\documentclass[preprint,review,10pt,times]{elsarticle}
\usepackage[margin=3cm]{geometry}

\usepackage{graphicx}
\usepackage{numprint}

\usepackage[T1]{fontenc}
\usepackage[english]{babel}

\usepackage[table]{xcolor}
\usepackage{enumerate}
\usepackage{lineno} 

\usepackage{amssymb}
\usepackage{amsmath}
\usepackage{amsthm}
\usepackage{scalerel,stackengine}
\usepackage{mathtools}
\usepackage{gensymb}
\usepackage{dsfont}
\usepackage{siunitx}
\usepackage{textcomp} 
\usepackage{xspace}

\usepackage{tabularx} 
\usepackage{booktabs} 
\usepackage{arydshln} 
\usepackage{multirow} 

\usepackage{caption}
\usepackage{subcaption}

\usepackage{ifthen,ifpdf}
\ifpdf
  \usepackage[hidelinks]{hyperref}
  \hypersetup{colorlinks=false,allcolors=blue}
  \usepackage{hypcap}
  \pdfpagewidth=\paperwidth
  \pdfpageheight=\paperheight
\fi

\usepackage{tikz} 
\usetikzlibrary{positioning}
\graphicspath{{figures/dbs/}}

\newcommand\crule[3][black]{\textcolor{#1}{\rule{#2}{#3}}}

\usepackage[many]{tcolorbox}
\usepackage{multicol}
\tcbuselibrary{raster}
\tcbuselibrary{skins}

\usepackage[draft]{listofsymbols}
\newcommand{\NewSym}[3][]{\newsym[#1]{#2}{\unexpanded{\unexpanded{#3}}}}

\newcommand*{\eg}{e.g.,\@\xspace}
\newcommand*{\ie}{i.e.,\@\xspace}
\newcommand*{\wrt}{w.r.t.\@\xspace}

\tcbset{boxouter/.style={halign title=flush center, fonttitle=\bfseries, colframe=black!70!white, colback=white!95!black, halign upper=left, valign upper=center}}
\tcbset{boxinner/.style={halign title=flush center, size=small, colframe=black!70!white, colback=white!95!white, colbacktitle=white!95!black, coltitle=black, halign=flush center, valign=top}}

\newcommand{\f}[1]{\mbox{$ #1 $}}

\renewcommand{\d}{{\,\mathrm  d}}

\newcommand{\I}[1]{{#1}^{\mathrm -1}} 

\newcommand{\norm}[1]{\Vert #1 \Vert}

\newcommand\equalhat{\mathrel{\stackon[1.5pt]{=}{\stretchto{%
				\scalerel*[\widthof{=}]{\wedge}{\rule{1ex}{3ex}}}{0.5ex}}}}

\renewcommand{\det}[1]{{\mathrm{det}}( #1 )}

\newcommand{\Sym}[1]{{\mathrm{Sym}}( #1 )}

\newcommand{\lb}{\left(} 
\newcommand{\rb}{\right)} 
\newcommand{\ls}{\{} 
\newcommand{\rs}{\}} 
\newcommand{\lr}{\left[} 
\newcommand{\rr}{\right]} 

\newcommand{\sop}{{\textit{SO}(3)}}

\newcommand{\fempty}[1]{{}}

\newcommand{\sty}[1]{\mbox{\boldmath $#1$}}
\newcommand{\styy}[1]{{\mathbb{#1}}}

\newcommand{\fa}{\sty{ a}}
\newcommand{\fb}{\sty{ b}}

\newcommand{\fd}{\sty{ d}}
\newcommand{\fe}{\sty{ e}}

\newcommand{\fh}{\sty{ h}}

\newcommand{\fn}{\sty{ n}}

\newcommand{\fp}{\sty{ p}}

\newcommand{\fu}{\sty{ u}}
\newcommand{\fv}{\sty{ v}}

\newcommand{\fx}{\sty{ x}}

\newcommand{\fz}{\sty{ z}}
\newcommand{\fzero}{\sty{ 0}}

\newcommand{\fA}{\sty{ A}}
\newcommand{\fB}{\sty{ B}}

\newcommand{\fD}{\sty{ D}}

\newcommand{\fI}{\sty{ I}}

\newcommand{\fR}{\sty{ R}}

\newcommand{\fW}{\sty{ W}}

\newcommand{\ffB}{\styy{ B}}
\newcommand{\ffC}{\styy{ C}}

\newcommand{\ffR}{\styy{ R}}
\newcommand{\ffS}{\styy{ S}}

\newcommand{\ftau}{\mbox{\boldmath $\tau$}}
\newcommand{\fsigma}{\mbox{\boldmath $\sigma$}}

\newcommand{\feps}{\mbox{\boldmath $\varepsilon $}}

\definecolor{boxrahmen}{gray}{0.0}
\definecolor{boxhintergrund}{gray}{0.999}

\definecolor{KITgreen}{RGB}{  0,150,130}
\definecolor{KITgreen70}{RGB}{ 76,181,167}
\definecolor{KITgreen50}{RGB}{127,202,192}
\definecolor{KITgreen30}{RGB}{178,223,217}
\definecolor{KITgreen15}{RGB}{217,239,236}
\definecolor{KITblue}{RGB}{70,100,170}
\definecolor{KITblue70}{RGB}{125,146,195}
\definecolor{KITblue50}{RGB}{162,177,212}
\definecolor{KITblue30}{RGB}{199,208,229}
\definecolor{KITblue15}{RGB}{227,232,242}
\definecolor{KITblack}{RGB}{0,0,0}
\definecolor{KITblack70}{RGB}{77,77,77}
\definecolor{KITblack50}{RGB}{128,128,128}
\definecolor{KITblack30}{RGB}{179,179,179}
\definecolor{KITblack15}{RGB}{217,217,217}
\definecolor{KITpalegreen}{RGB}{130,190,60}
\colorlet{KITpalegreen70}{KITpalegreen!70}
\colorlet{KITpalegreen50}{KITpalegreen!50}
\colorlet{KITpalegreen30}{KITpalegreen!30}
\colorlet{KITpalegreen15}{KITpalegreen!15}
\definecolor{KITyellow}{RGB}{250,230,20}
\colorlet{KITyellow70}{KITyellow!70}
\colorlet{KITyellow50}{KITyellow!50}
\colorlet{KITyellow30}{KITyellow!30}
\colorlet{KITyellow15}{KITyellow!15}
\definecolor{KITorange}{RGB}{220,160,30}
\colorlet{KITorange70}{KITorange!70}
\colorlet{KITorange50}{KITorange!50}
\colorlet{KITorange30}{KITorange!30}
\colorlet{KITorange15}{KITorange!15}
\definecolor{KITbrown}{RGB}{160,130,50}
\colorlet{KITbrown70}{KITbrown!70}
\colorlet{KITbrown50}{KITbrown!50}
\colorlet{KITbrown30}{KITbrown!30}
\colorlet{KITbrown15}{KITbrown!15}
\definecolor{KITred}{RGB}{160,30,40}
\colorlet{KITred70}{KITred!70}
\colorlet{KITred50}{KITred!50}
\colorlet{KITred30}{KITred!30}
\colorlet{KITred15}{KITred!15}
\definecolor{KITlilac}{RGB}{160,0,120}
\colorlet{KITlilac70}{KITlilac!70}
\colorlet{KITlilac50}{KITlilac!50}
\colorlet{KITlilac30}{KITlilac!30}
\colorlet{KITlilac15}{KITlilac!15}
\definecolor{KITcyan}{RGB}{80,170,230}
\colorlet{KITcyan70}{KITcyan!70}
\colorlet{KITcyan50}{KITcyan!50}
\colorlet{KITcyan30}{KITcyan!30}
\colorlet{KITcyan15}{KITcyan!15}
\definecolor{Dark2-A}{RGB}{27,158,119}
\definecolor{Dark2-B}{RGB}{217,95,2}
\definecolor{Dark2-C}{RGB}{117,112,179}
\definecolor{Dark2-D}{RGB}{231,41,138}

\newcommand{\stressdamageinitsub}[1]{\sigma_{0, #1}}
\newcommand{\effective}[1]{\bar{#1}}
\newcommand{\macauly}[1]{\langle #1 \rangle_{+}}
\newcommand{\set}[1]{\left\{ #1 \right\}}

\opensymdef
    \NewSym[Fiber volume fraction]{fvc}{f}
    \NewSym[Second order fiber orientation tensor]{oritensor}{\fA}
    \NewSym[Direction of a fiber or fiber bundle]{fiberdir}{\fp}
    \NewSym[Molding property state vector]{statevec}{\vec{s}}
    \NewSym[Mechanical feature vector]{featvec}{\vec{y}}
    \NewSym[Multivariate normal distribution]{normal}{\mathcal{N}}
    \NewSym[Mean]{meanvec}{\vec{\mu}}
    \NewSym[Covariance]{cov}{\Sigma}
    \NewSym[Standard deviation]{std}{\sigma}
    \NewSym[Mold friction stress]{fricstress}{\ftau}
    \NewSym[Hydrodynamic friction coefficient]{hydrofric}{\lambda}
    \NewSym[Hydrodynamic power-law coefficient]{hydropower}{m}
    \NewSym[Reference velocity]{refvelocity}{v_0}
    \NewSym[Reference viscosity]{refviscosity}{D_1}
    \NewSym[Transition shear rate]{transshear}{\dot{\gamma}_0}
    \NewSym[Power-law viscosity coefficient]{visccoef}{n}
    \NewSym[Temperature parameter]{visct}{T^*}
    \NewSym[Fitting parameter]{visca}{A_1}
    \NewSym[Fitting parameter]{viscb}{A_2}
    \NewSym[Matrix mass density]{density}{\rho}
    \NewSym[Thermal conductivity]{conductivity}{\kappa}
    \NewSym[Gap conductance]{conductance}{k_\textrm{T}}
    \NewSym[Specific heat capacity]{heatcapacity}{c_\textrm{p}}
    \NewSym[Velocity]{velocity}{\fv}
    \NewSym[Stress]{stress}{\sigma}
    \NewSym[Strain]{strain}{\varepsilon}
    \NewSym[Microscopic strain]{microstrain}{\varepsilon}
    \NewSym[Macroscopic strain]{macrostrain}{\effective{\microstrain}}
    \NewSym[Microscopic stress]{microstress}{\sigma}
    \NewSym[Macroscopic stress]{macrostress}{\effective{\sigma}}
    \NewSym[Microscopic strain tensor]{fmicrostrain}{\feps}
    \NewSym[Macroscopic strain tensor]{fmacrostrain}{\effective{\feps}}
    \NewSym[Microscopic stress tensor]{fmicrostress}{\fsigma}
    \NewSym[Macroscopic stress tensor]{fmacrostress}{\effective{\fsigma}}
    \NewSym[Gap conductance between mold and SMC]{gapcond}{K_T}
    \NewSym[Bundle cross section]{abundle}{A_\textrm{B}}
    \NewSym[Fiber orientation distribution function]{oridensity}{\Phi}
    \NewSym[Length of a bundle in a cell]{lengthincell}{\Delta l_{ai}}
    \NewSym[Total bundle length in a cell]{totallengthincell}{L_{a}}
    \NewSym[Bundle modulus]{bundlemodulus}{E}
    \NewSym[Maximum compression force]{maxforce}{F_\textrm{max}}
    \NewSym[Press speed]{pressspeed}{\dot{h}}
    \NewSym[Volume of a cell]{cellvolume}{V_a}
    \NewSym[Free energy density]{freeenergy}{\psi}
    \NewSym[Free energy potential]{energypotential}{w}
    \NewSym[Dissipation]{dissipation}{\phi}
    \NewSym[Dissipation potential]{dissipotential}{\Phi}
    \NewSym[Algorithmic potential]{condensedpotential}{\Psi}
    \NewSym[Hardening parameter]{hardparam}{H}
    \NewSym[Damage variable]{hardeningvar}{q} 
    \NewSym[Extraction tensor]{extract}{\ffB}
    \NewSym[Stress for damage initiation]{stressdamageinit}{\sigma_{0}} 
    \NewSym[Power-law damage exponent]{powerexponent}{m}
    \NewSym[Internal state variables]{fstatev}{\fz}
    \NewSym[Rotation operator]{rotOperator}{\fR}
    \NewSym[Gradient operator]{gradOperator}{\fD}
    \NewSym[Weight operator]{weightOperator}{\fW}
    \NewSym[Averaging operator]{averagingOperator}{\fB}
    \NewSym[Unit sphere]{unitsphere}{\mathcal{S}^2}
    \NewSym[Set of bundles crossing a cell]{bundlesincell}{\mathcal{B}_a}
    \NewSym[Laminate building block]{BB}{\mathcal{B}}
    \NewSym[Deep material network ]{DMNLIN}{\mathcal{DMN}^{\mathcal{L}}_{Y}}
\closesymdef

\journal{Composites Part B}

\begin{document}

\begin{frontmatter}



\title{A probabilistic virtual process chain to quantify process-induced uncertainties in Sheet Molding Compounds}

\author[inst1]{Nils Meyer}\ead{nils.meyer@kit.edu}
\author[inst2]{Sebastian Gajek}
\author[inst2]{Johannes G\"orthofer}
\author[inst4]{Andrew Hrymak}
\author[inst1]{Luise K\"arger}
\author[inst1,inst3]{Frank Henning}
\author[inst2]{Matti Schneider}
\author[inst2]{Thomas B\"ohlke}

\affiliation[inst1]{organization={Institute of Vehicle Systems Technology - Division Lightweight Technology, Karlsruhe Institute of Technology~(KIT)},
            addressline={Rintheimer Querallee~2}, 
            city={Karlsruhe},
            postcode={76131}, 
            state={Baden-W\"urttemberg},
            country={Germany}}

\affiliation[inst2]{organization={Institute of Engineering Mechanics - Chair for Continuum Mechanics, Karlsruhe Institute of Technology~(KIT)},
            addressline={Kaiserstra\ss e~10}, 
            city={Karlsruhe},
            postcode={76131}, 
            state={Baden-W\"urttemberg},
            country={Germany}}
            
\affiliation[inst3]{organization={Fraunhofer Institute for Chemical Technology~ICT},
            addressline={Joseph-von-Fraunhofer Stra\ss e~10}, 
            city={Pfinztal},
            postcode={76327}, 
            state={Baden-W\"urttemberg},
            country={Germany}}

\affiliation[inst4]{organization={Department of Chemical and Biochemical Engineering, University of Western Ontario~(UWO)},
            addressline={1151~Richmond Street}, 
            city={London},
            postcode={N6A 5B9}, 
            state={Ontario},
            country={Canada}}

\begin{abstract}
The manufacturing process of Sheet Molding Compound (SMC) influences the properties of a component in a non-deterministic fashion.
To predict this influence on the mechanical performance, we develop a virtual process chain acting as a digital twin for SMC specimens from compounding to failure. 
More specifically, we inform a structural simulation with individual fields for orientation and volume fraction computed from a direct bundle simulation of the manufacturing process.
The structural simulation employs an interpolated direct deep material network to upscale a tailored SMC damage model.
We evaluate hundreds of virtual specimens and conduct a probabilistic analysis of the mechanical performance.
We estimate the contribution to uncertainty originating from the process-induced inherent random microstructure and from varying initial SMC stack configurations. 
Our predicted results are in good agreement with experimental tensile tests and thermogravimetric analysis.
\end{abstract}

\begin{graphicalabstract}
\centering
\footnotesize
\begin{tikzpicture}
    [block/.style={draw,minimum width=#1,minimum height=2em},
    block/.default=10em,high/.style={minimum height=3em},
    node distance=1.5em and 5em,auto]
    \node (n0) {$\statevec_0$};
    \node[block=3em,fill=lightgray,rounded corners,right=1em of n0] (n1) {Compounding};
    \node[block=3em,fill=lightgray,rounded corners,right=8em of n1] (n2) {Molding};
    \node[block=3em,text=darkgray,draw=darkgray,below left= 1.5em and 0em of n2,rounded corners, align=center] (n2a) {
            \emph{Direct bundle simulation}\\
            \includegraphics[width=8em]{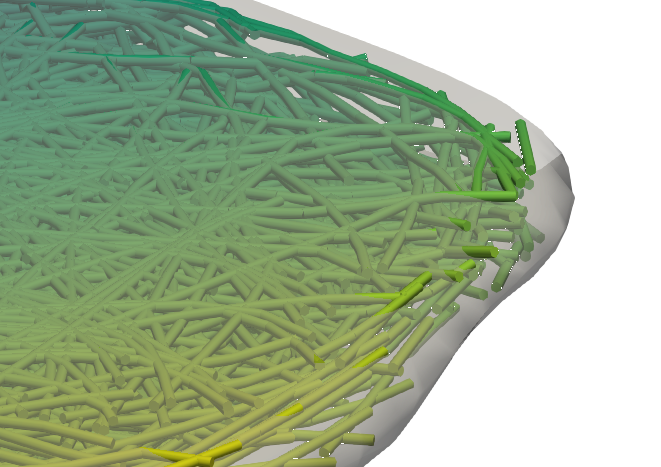}
        };
    \node[block=3em,fill=lightgray,rounded corners,right=8em of n2] (n3) {Loading};
    \node[block=3em,fill=lightgray,rounded corners,right=8em of n3] (n4) {Database};
    \node[block=3em,text=darkgray,draw=darkgray,below=of n3,rounded corners, align=center] (n3a) {
            \emph{Deep Material Network}\\
            \includegraphics[trim=100 0 130 0,clip, width=7em]{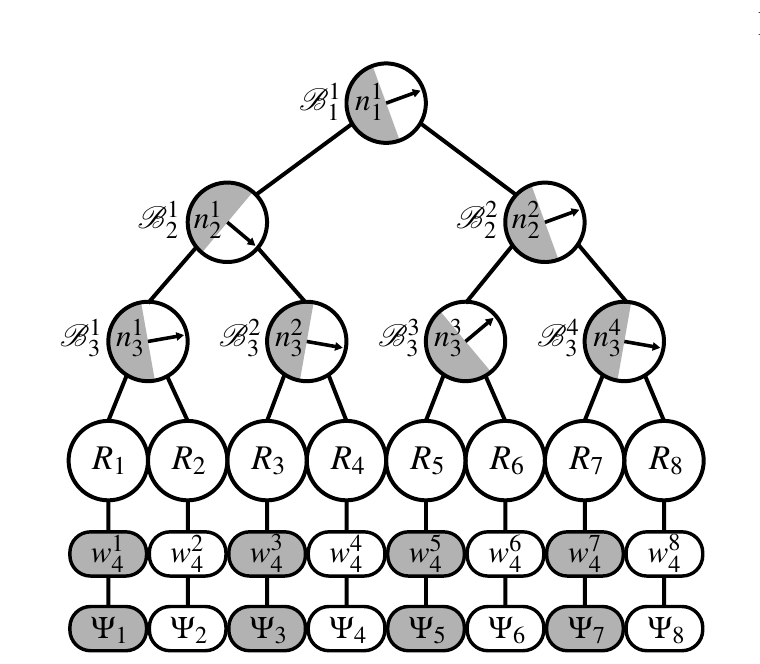}
        };
    \node[block=3em,text=darkgray,draw=darkgray,below left= 0.7em and -4em of n3a,rounded corners, align=center] (n3b) {
        \emph{SMC microstructures}\\
        \includegraphics[width=13em]{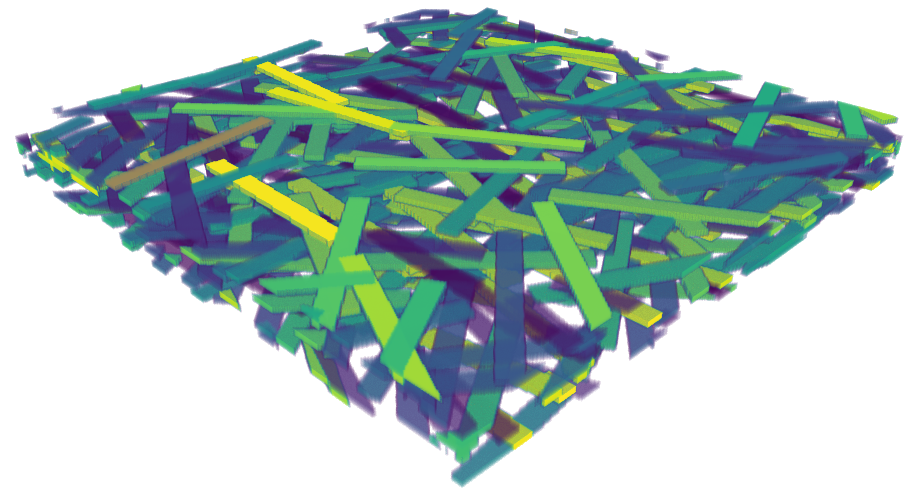}
    };
    \node[block=3em,text=darkgray,draw=darkgray,below right=0.7em and -2em of n3a,rounded corners,align=center] (n3c) {
        \emph{Damage model}\\
        \includegraphics[width=7em]{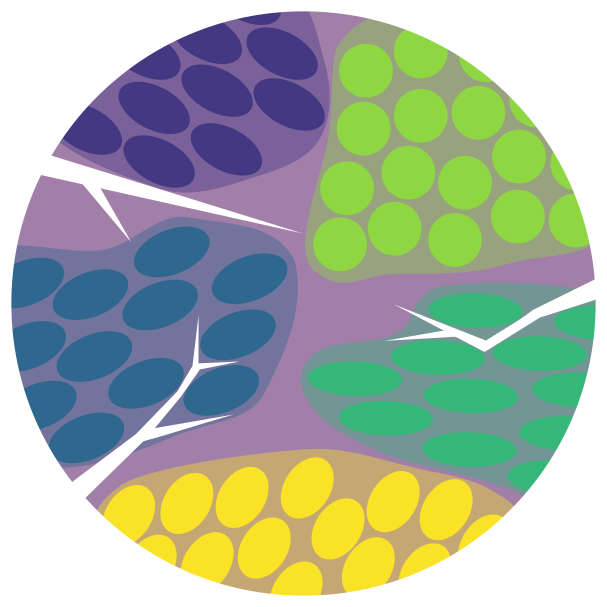}
    };
    \node[block=3em,text=darkgray,draw=darkgray,below right=0.7em and -1em of n4,rounded corners, align=center] (n4a) {
            \emph{Uncertainty evaluation}\\
            \includegraphics[width=7em]{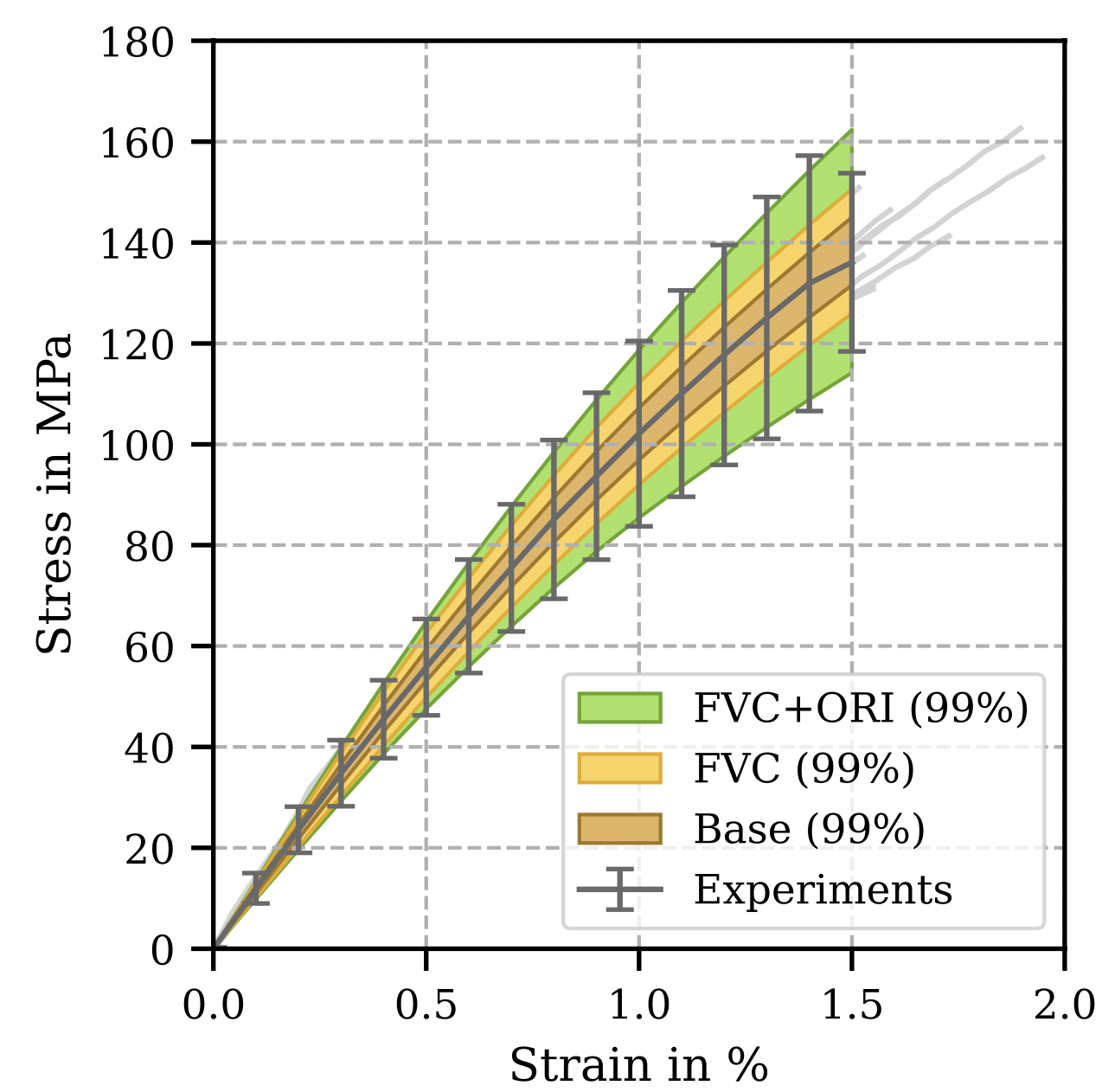}
        };
    \draw[-latex, draw=darkgray] (n0) -- (n1);
    \draw[-latex, draw=darkgray] (n1) -- (n2) node[midway,align=center]{ \includegraphics[width=5em]{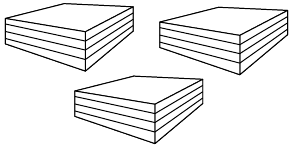}\\
    $\statevec_\textrm{C} \sim \normal(\meanvec_0, \cov_\textrm{C})$}; 
    \draw[-latex, draw=darkgray] (n2) -- (n3) node[midway,align=center]{\includegraphics[width=5em]{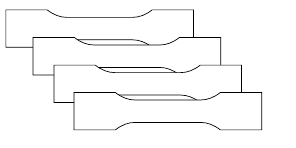}\\
    $\statevec_\textrm{M}  \sim \normal(\meanvec_0, \cov_\textrm{M})$};
    \draw[-latex, draw=darkgray] (n3) -- (n4)
    node[midway, align=center]{\includegraphics[width=5em]{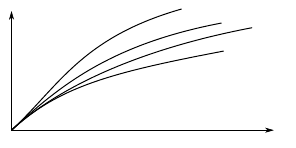}\\
    $\featvec_\textrm{L} \sim \normal(\meanvec_\textrm{L}, \cov_\textrm{L})$};
    \draw[-latex, draw=darkgray] (n2a) -- (n2);
    \draw[-latex, draw=darkgray] (n3a) -- (n3);
    \draw[-latex, draw=darkgray] (n3b) -- (n3a);
    \draw[-latex, draw=darkgray] (n3c) -- (n3a);
    \draw[-latex, draw=darkgray] (n4) -- (n4a);
\end{tikzpicture}
\end{graphicalabstract}



\end{frontmatter}

\newpage


\section{Introduction}
\label{sec:introduction}

\subsection{State of the art}

Sheet Molding Compound (SMC) is a discontinuous fiber reinforced polymer material that is available in the form of sheets which are pre-impregnated with a thermoset resin. 
The manufacturing process leads to a random planar orientation distribution of bundled fibers in the prepreg sheets.
These sheets are cut and stacked to form an initial charge that is subsequently molded to a part in a compression molding process. 
The resulting SMC parts have superior mechanical properties in comparison to injection molded parts due to their higher fiber length while maintaining the ability to fill complex geometrical features due to the flow process during compression molding. 
However, the manufacturing process leads to a spatially varying fiber configuration, \ie an inhomogeneous fiber volume fraction and anisotropic fiber orientation, which determines the properties of a resulting part~\cite{Schemmann2015parameter, Schemmann2018parameter}.

The fiber configuration found in SMC components is a complex three-dimensional arrangement of fiber bundles. 
As an efficient description, we may define appropriate microstructure characteristics to describe the properties within a region, such as the fiber volume fraction $\fvc$ and the (second-order) fiber orientation tensor $\oritensor$~\cite{Advani.1987}. 
For curved SMC fiber bundles, we use a length-weighted second-order fiber orientation tensor 
\begin{equation}
    \oritensor = \frac{1}{L} \sum_{i=1}^N \int_0^{L_i} \fiberdir_i(s) \otimes \fiberdir_i(s) \textrm{d}s 
    \quad
    \text{with}
    \quad
    L = \sum_{i=1}^N L_i
    \label{eq:oritensor}
\end{equation}
for $N$ fibers, each of which is described with a varying direction $\fiberdir_i(s) \in \ffR^3$ along its arc length $s \in [0, L_i]$  \cite{Schneider2022}.  

Virtual process chains for composite materials are used to address the coupling between manufacturing, the resulting microstructure and the mechanical performance in an early virtual product development stage.
These virtual process chains integrate manufacturing simulations to inform structural simulations about the fiber orientation state after molding~\cite{Karger2015, Buck2015, Gorthofer2019}.
While previous virtual process chains increased accuracy of structural simulations in a deterministic fashion, typically they are not able to quantify uncertainty of the predicted processing effects. 
Indeed, in SMC components, even for simple geometries, there is a significant variation in terms of the underlying microstructure, described by the fiber volume fraction $\fvc$ and the fiber orientation tensor $\oritensor$, between individual realizations due to the randomness of the prepreg manufacturing process~\cite{Kia1991, Bretz2021, Rothenhaeusler2022}. Accounting for such effects is imperative to fully exploit the lightweight potential of composite materials by reducing safety factors.

The effect of a stochastic microstructure on mechanical performance of SMC has been addressed for example by Chen et al.~\cite{Chen2018generator} in a multiscale framework using a stochastic microstructure reconstruction algorithm and a Kriging model for scale bridging. 
However, the authors did not consider propagation of uncertainties from the molding simulation and used a macroscopic process simulation with fixed fiber volume fraction instead of directly simulating the motion of individual fiber bundles.
Mansour et al.~\cite{Mansour2019} analyzed the inherent randomness and size dependent properties in thin fiber networks constituting the microstructure of paper. 
They use stochastic volume elements to derive a statistically equivalent stochastic effective constitutive model for paper from direct micromechanical simulations.
Their model is based on correlated random fields for modeling the spatially distributed strength and strain fields, which are constructed from multivariate kernels. 
However, the damage mechanisms and microstructures in paper are vastly different to polymer composites and their work is limited to isotropic orientation distributions.
Sommer et al.~\cite{Sommer2020} investigated the variable properties of stochastic prepreg platelet molded composites (PPMC). 
They analyze the process-structure-property relationships for the stochastic microstructure by experimental methods and by linking a smoothed particle hydrodynamics (SPH) flow simulation with a progressive failure model.
However, they simulate only few realizations and assume constant fiber volume fraction (no resin pockets).

\subsection{Contribution}

In this work, we propose a framework for estimating the stochastic outcomes of a virtual SMC process chain. To address uncertainty in SMC components, we enhance our previous virtual process chain~\cite{Gorthofer2019} in three areas:
\begin{enumerate}
    \item We utilize a more capable direct process simulation that represents fiber bundles directly by truss elements which interact with the matrix through hydrodynamic forces~\cite{Meyer2020, Meyer2021, Meyer2021Diss, Meyer2022, Rothenhaeusler2022}. Contrary to macroscopic simulation models, the former allows us to compute multiple realizations of a molding process with different stochastic bundle structures.
    Further, the improved process model allows to evaluate a spatially varying fiber volume fraction field in addition to the fiber orientation field.
    \item We utilize a convex and anisotropic damage model that takes the characteristic bundle microstructure of SMC composites~\cite{Gorthofer2020} into account via dedicated extraction tensors~\cite{Gorthofer2022damage}. Our framework enables to combine different damage functions in a modular way to capture any anisotropic damage evolution. Motivated by Puck's laminate criteria, we developed different extraction tenors for specific damage cases present in the SMC composite. Directly operating on the compliance tensor as primary damage variable in combination with a set of physically meaningful damage parameters yields a well-posed model with which we can accurately capture the damage evolution on the microscale~\cite{Gorthofer2022failure}. 
    \item For conducting two-scale simulations of SMC components, we use the framework of direct deep material networks~\cite{Gajek2020, Gajek2021, Gajek2021PAMM, Gajek2022} which we augment by a fiber orientation and fiber volume fraction interpolation scheme. In contrast to previous works~\cite{Gajek2021}, this approach allows to accurately resolve the spatially varying fiber volume fraction as well as the spatially varying fiber orientation in a component scale simulation.
\end{enumerate}

These individual contributions, \ie a direct process simulation approach, an accurate constitutive model for anisotropic damage in SMC and a versatile upscaling approach, form an enhanced virtual process chain. 
Notably, our enhanced process chain accounts for both fiber orientation and fiber volume fraction by a direct evaluation of the bundle structure on a structural simulation mesh.
The efficient implementation and upscaling enables us to compute hundreds of macroscopic virtual specimens with individual spatially varying fiber configurations. 
We vary the initial stack configuration in terms of the overall fiber volume fraction and overall fiber orientation while keeping material parameters and boundary conditions constant.
For each variation, we compute multiple realizations of SMC components and analyze these to understand sources of process-induced uncertainties. 
Eventually, this allows a further reduction in safety factors to exploit the full lightweight potential of SMC parts.

The virtual process chain is described in Section~\ref{sec:methods} for the three individual contributions \emph{compression molding simulation}, \emph{damage modeling} and \emph{upscaling}.
In Section~\ref{sec:application}, we apply the methods to establish a database of virtual tensile tests for a plate molding process. 
In Section~\ref{sec:results}, we evaluate the results in this database to estimate uncertainties in the mechanical response depending on specimen size and prepreg properties.
Finally, we compare our predictions to the scattering observed in previously reported experiments~\cite{Trauth2020Diss}.

\section{A virtual process chain for Sheet Molding Compounds}
\label{sec:methods}

\subsection{Concept}
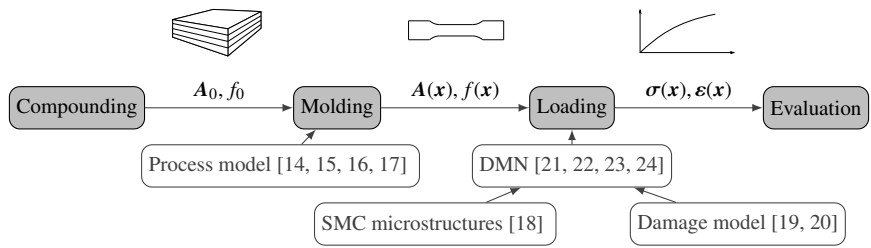
\begin{figure*}[htbp]
\centering
\footnotesize
    \begin{tikzpicture}
        [block/.style={draw,minimum width=#1,minimum height=2em},
        block/.default=10em,high/.style={minimum height=3em},
        node distance=0.7em and 7em,auto]
        \node[block=3em,fill=lightgray,rounded corners] (n1) {Compounding};
        \node[block=3em,fill=lightgray,rounded corners,right=of n1] (n2) {Molding};
        \node[block=3em,text=darkgray,draw=darkgray,below left= 0.7em and -6em of n2,rounded corners, align=center] (n2a) {
            Process model~\cite{Meyer2020, Meyer2021, Meyer2021Diss, Meyer2022}
        };
        \node[block=3em,fill=lightgray,rounded corners,right=of n2] (n3) {Loading};
        \node[block=3em,text=darkgray,draw=darkgray,below= of n3,rounded corners] (n3a) {
            DMN~\cite{Gajek2020, Gajek2021, Gajek2021PAMM, Gajek2022}
        };
        \node[block=3em,text=darkgray,draw=darkgray,below left= 0.7em and -4em of n3a,rounded corners, align=center] (n3b) {
            SMC microstructures~\cite{Gorthofer2020}
        };
        \node[block=3em,text=darkgray,draw=darkgray,below right=0.7em and -2em of n3a,rounded corners] (n3c) {
            Damage model~\cite{Gorthofer2022damage, Gorthofer2022failure}
        };
        \node[block=3em,fill=lightgray,rounded corners,right=of n3] (n4) {Evaluation};
        \draw[-latex, draw=darkgray] (n1) -- (n2) node[midway,align=center]{ \includegraphics[width=5em]{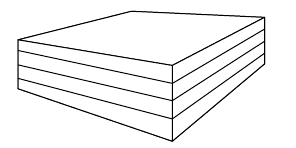}\\
        $\oritensor_0, \fvc_0$}; 
        \draw[-latex, draw=darkgray] (n2) -- (n3) node[midway,align=center]{ \includegraphics[width=5em]{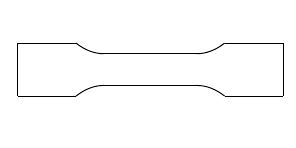}\\
        $\oritensor(\fx), f(\fx)$};
        \draw[-latex, draw=darkgray] (n3) -- (n4) node[midway,align=center]{
        \includegraphics[width=5em]{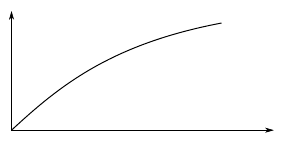}\\ $\fsigma(\fx), \feps(\fx)$};
        \draw[-latex, draw=darkgray] (n2a) -- (n2);
        \draw[-latex, draw=darkgray] (n3a) -- (n3);
        \draw[-latex, draw=darkgray] (n3b) -- (n3a);
        \draw[-latex, draw=darkgray] (n3c) -- (n3a);
    \end{tikzpicture}
    \caption{Schematic representation of the virtual SMC process chain for a single realization with nominal initial fiber orientation state $\oritensor_0$ and nominal initial fiber volume fraction $\fvc_0$}
    \label{fig:vpc_schematic}
\end{figure*}

We design a virtual process chain which is a continuous digital representation of a specimen's life from semi-finished raw materials up to its end of life due to failure in a testing rig. 
Such a virtual process chain of an SMC specimen is schematically illustrated in Figure~\ref{fig:vpc_schematic}.

The starting point of the virtual SMC process chain is the configuration of a prepreg sheet stack, which is placed at a dedicated position in a mold. The sheets in this stack have an initial fiber configuration described by the second-order fiber orientation tensor $\fA_0$ (see equation~\eqref{eq:oritensor}) and the initial fiber volume fraction $\fvc_0$. A process simulation computes the change of the initial fiber configuration during a compression molding process and predicts the locally distributed orientation states $\fA(\fx)$ and fiber volume fractions $f(\fx)$ after the manufacturing process. These fields serve as an input to a structural simulation to compute the tensile stress $\stress$, the tensile strain $\strain$ and the damage in the specimen due to loading.



\subsection{Compression molding simulation} \label{sec:fund_molding}

\paragraph{Traditional macroscopic approaches}

A compression molding simulation of SMC describes the mold filling process starting from an initial SMC stack until the mold is completely filled. 
The process may be decomposed into three steps~\cite{Odenberger.2004}: An initial \emph{squish phase} with a complex flow front and a release of entrapped air in the stack, a stable \emph{plug-flow} with an extensional flow of SMC and a final \emph{boiling phase} after complete mold closing. 
The plug-flow is the dominant kinematic flow mechanism during SMC compression molding.
The plug-flow kinematic was first shown by Barone and Caulk~\cite{Barone.1985}, who suggested that the heated resin near mold walls forms a thin lubrication layer that is best described by hydrodynamic friction models~\cite{Barone.1986}.
Calibrated models for the hydrodynamic friction in SMC plug-flows and more advanced friction models for it are available~\cite{Abrams.2003, Dumont.2007, Hohberg2017}.
As the flow phase is short compared to the time scale of curing and heat transfer, simulation models often simplify the process model to be isothermal without curing, such that the heated lubrication layer is only accounted for through the friction model between mold and SMC~\cite{Dumont.2007, Hohberg2017}. 
The fiber orientation is typically described by an evolution equation for the second-order fiber orientation tensor, which was introduced by Advani and Tucker~\cite{Advani.1987} as an efficient numerical treatment for spatially varying fiber-orientation distributions and is based on Jeffery's equation~\cite{Jeffery.1922}.
There are approaches to account for varying fiber volume content during processing~\cite{Dumont.2005, Perez.2017, Perez2019}, but most macroscopic compression molding simulations assume a constant fiber volume fraction.

Instead of solving effective measures like fiber orientation tensors or fiber volume fractions, one may directly simulate the motion of individual fiber bundles during the compression molding process~\cite{Meyer2020, Meyer2021}. 
The key here is the observation that fiber bundles often stay intact during molding~\cite{Le.2008, Guiraud.2012, Motaghi.2019}, which allows us to treat hundreds of fibers as individual bundle instances.
This reduces the computational effort significantly and allows for a direct simulation of all flexible bundles in entire parts of industrial complexity with up to \numprint{250000} bundles \cite{Meyer2021Diss}.

\paragraph{A direct bundle simulation approach for the molding process}

The process simulation model solves conservation of mass, momentum and internal energy in a Coupled Eulerian Lagrangian (CEL) framework~\cite{Benson.1992}, where each Eulerian element is attributed an Eulerian volume fraction $e \in [0,1]$.
The motion of SMC in the Eulerian phase is driven by molds that are modeled as rigid bodies and are in contact with the reconstructed SMC surface~\cite{Benson.2004}.

The core idea of the direct bundle simulation approach is the treatment of fiber bundles as one dimensional flexible objects represented by truss elements (\ie one dimensional elements that transfer axial forces only) which interact with a viscous matrix via a body force field in the CEL momentum equation~\cite{Meyer2020}.
Initially, these truss elements are generated in the region of the SMC stack by drawing random directions from a uniform distribution on a unit sphere, which are then projected to a planar isotropic orientation state $\fA_0$, shifted randomly within the stack and cut if extending outside the stack (see Section \ref{sec:stack_generation} for details and \cite{Gorthofer2020} for a comprehensive overview).
This procedure results in a realization of a random process similar to the sheet manufacturing and introduces a length distribution due to cut fiber bundles at the edges of the stack similar to the physical cutting process of sheets. 

During the flow process, the matrix is modeled as a compressible viscous material with an equation of state fitted to compaction trials and a Cross-WLF viscosity model~\cite{Meyer2022}.
Each truss element experiences a hydrodynamic drag force from the surrounding matrix that is proportional to the relative velocity of the surrounding matrix~\cite{Meyer2020}. 
The relative velocity is computed from weighted neighboring elements and an opposed body force field is applied to those neighbors ensuring a two-way coupled anisotropic flow~\cite{Meyer2020}.

The tangential hydrodynamic mold friction causes a shear stress traction vector at the mold surface $\fricstress \in \ffR^2$ that follows a power-law model 
\begin{equation}
    \fricstress = -\hydrofric \left( \frac{\norm{ \velocity_\textrm{s} }}{\refvelocity}\right)^{\hydropower-1} \velocity_\textrm{s},
\end{equation}
where $\hydrofric \in \ffR_{>0}$ is a hydrodynamic friction coefficient, $\hydropower \in [0,1]$ is a power-law coefficient, $\refvelocity \in \ffR_{>0}$ is an arbitrary reference velocity for non-dimensionalization, and $\velocity_\textrm{s} \in \ffR^2$ is the slip velocity at the mold surface~\cite{Dumont.2007, Hohberg.2017}.
For the thermal contact we assume that the heat flux is proportional to the difference between SMC temperature and the mold temperature with a gap conductance $\gapcond \in \ffR_{>0}$~\cite{Meyer2022}. 

\paragraph{Evaluation of process simulation results}
\label{sec:evaluation_mesoscale_process}

When using a direct bundle simulation, the fiber-bundle statistics are readily computable in post-processing.
The fiber-bundle statistics may be transferred to virtual specimens with data about fiber orientation as well as fiber volume fraction. 
In contrast to mapping methods, which may introduce interpolation errors to the model~\cite{Krauss2022}, we evaluate the bundle configuration directly in each cell of the target mesh and thus ensure coherent and natural correlations between cells.
The fiber volume fraction in cell $a$ is approximated as 
\begin{equation}
    \fvc_a = \abundle \sum_{i \in \bundlesincell} \frac{\lengthincell}{\cellvolume},
\end{equation}
where $\bundlesincell$ contains all truss elements crossing cell $a$ and $\abundle$ denotes the cross section area of a bundle. 
The length of bundle element $i$ within a cell $a$ is denoted $\lengthincell$ and the cell volume is $\cellvolume$. 
Similarly, the second-order discrete fiber orientation tensor is computed as 
\begin{equation}
    \oritensor_a = \frac{1}{\totallengthincell} \sum_{i \in \bundlesincell} \lengthincell \fiberdir_i \otimes \fiberdir_i 
    \quad 
    \textrm{with}
    \quad 
    \totallengthincell = \sum_{i \in \bundlesincell} \lengthincell
\end{equation}
and  the bundle orientation $\fiberdir_i$.

\subsection{Structural damage modeling}
\label{sec:fund_damage}

\paragraph{Current state of research}

Damage modeling is devoted to the continuum description of progressive stiffness degradation on the macroscale due to the growth of defects such as voids or microcracks on a lower length scale~\cite{Krajcinovic1984,Lemaitre1986}. Two general approaches have prevailed, phenomenological models in combination with suitable kinetic laws operating on the macroscale~\cite{Krajcinovic1989,Hansen1994} and micromechanics-based models directly taking into account the damage mechanisms on the microscale~\cite{Fitoussi1996b,Guo1997,Gorthofer2022failure}. The latter allow for an easier handling of microstructural stochastics such as uncertainty in the elastic properties of fiber reinforced concrete~\cite{Liu2020}, random loading in fatigue processes~\cite{Franko2017}, progressive fiber breakage~\cite{Ju2016,Wu2017}, interfacial effects and strength~\cite{Schemmann2018damage,Chen2018generator} or localized microcracks~\cite{Li2020}.

Depending on the state of anisotropy induced by progressive damage, different damage model approaches are used: Scalar-valued damage approaches to describe isotropic stiffness degradation, as e.\,g., used to model steel-fiber reinforced concrete~\cite{Moradi2020}, fiber reinforced plastic composite plies~\cite{Sharma2020}, or notched epoxy resin specimens~\cite{Rahimi2020}. Second-order damage tensors are used to model orthotropic stiffness degradation due to damage evolution~\cite{Murakami1981}. Especially in the field of composite research, these approaches are frequently applied, e.\,g., to describe composite fabrics and laminated panels~\cite{Wei2020}, composite laminates~\cite{Okabe2018,Onodera2020} or ceramic-matrix composites~\cite{Alabdullah2020}. At least a fourth-order damage tensor is necessary to capture any anisotropic stiffness degradation~\cite[section 4]{Krajcinovic1989}. Indeed, an approach operating on the stiffness or compliance tensor as primary damage variable seems natural. Some applications include modeling of concrete~\cite{OrtizPopov1982,Ortiz1985,Yazdani1990} and elasto-plasticity coupled to damage~\cite{Simo1987,Ju1989}.

When used to describe softening material behavior, local damage model formulations generally become ill-posed due to localization effects~\cite{Lemaitre1986}, which lead to strongly mesh-dependent results~\cite{DeBorst1996}. Solution approaches include non-local formulations~\cite{Belytschko1986,Bazant1991} such as gradient enhancements~\cite{Brunig2005,AbuAl-Rub2009,Junker2021}, convolution with a tapering function~\cite{PijaudierCabot1987} or an augmentation via an elliptic differential operator~\cite{Aifantis1984}. Other approaches, inter alia, apply relaxation techniques to the local damage formulation as a countermeasure~\cite{Balzani2012,Schmidt2016,Schwarz2021}. For a summary on ill-posedness and regularization methods, we refer to Forest et al.~\cite{Forest2004}.

As long as the material behavior is dominated by a hardening regime, locally formulated damage material models remain well-posed. An approach that reflects this entire class of hardening-type materials and is able to capture any anisotropic stiffness degradation due to damage evolution was introduced by Görthofer et al.~\cite{Gorthofer2022damage}. We apply the model to SMC using specific adaptions to capture matrix and bundle damage.

\paragraph{A convex damage model for hardening-type materials}

We formulate the model in the setting of generalized standard materials~\cite{GSM} with a free energy density~\f{\freeenergy: \Sym{3} \times S \times Q \rightarrow \ffR} which is comprised of an elastic part and a part related to damage. The elastic part is defined on the space of symmetric \f{3\times 3} strain tensors~\f{\feps\in\Sym{3}}. Furthermore, the model directly operates on the space of symmetric and positive definite compliance tensors~$\ffS \in S=\ls\ffS\in\Sym{\Sym{3}}  \, | \, \ftau\cdot\ffS\lr\ftau\rr>0 \quad \forall \, \tau\in\Sym{3}\backslash \ls 0\rs\rs$ as the primary damage variable and a set of general variables~\f{\hardeningvar \in Q} describing the shape and size of the damage surfaces. Whereas the domain \f{Q} may in general be abstract, we will utilize damage-activation functions (see equation~\eqref{eq:damage:activation}) so that \f{Q\equiv\ffR^M} with \f{M} being the number of different damage cases. With the classical linear elastic energy density and a power-law ansatz for the damage part, we introduce the free energy density~\f{\freeenergy} as
\begin{equation}
    \freeenergy(\feps,\ffS,\hardeningvar)  = \dfrac{1}{2}\feps\cdot\I\ffS\lr\feps\rr + \sum_{i=1}^M \dfrac{\hardparam_i}{\powerexponent_i+1}\hardeningvar_i^{\powerexponent_i+1},
\end{equation}
with hardening parameters~\f{\hardparam_i\in\ffR_{>0}} and exponents~\f{\powerexponent_i\in\ffR_{>0}}.  Suitably, we introduce the force potential~\f{\dissipotential^*} in terms of~\f{M} convex damage-activation functions $g$ that bound the elastic regime in analogy to associated elasto-plastic models. Directly integrating the driving forces for the compliance~\f{\ffS} and the damage variables~\f{\hardeningvar}, we formulate the force potential~$\dissipotential^*: \Sym{3} \times Q \rightarrow \ffR \cup \set{+\infty} $ in its simplified version
\begin{equation}
    \dissipotential^*\lb\fsigma,\hardeningvar\rb = \begin{cases}
    0, & g_i\lb\fsigma,\hardeningvar_i\rb \leq 0, \quad \forall \, i=1,\ldots,M, \\
    +\infty, & \text{else}, \end{cases}
\end{equation}
The damage-activation functions (see equation~\eqref{eq:damage:activation}) take the current stress state into account. For damage to evolve, this stress state has to exceed a damage-activation threshold~\f{\stressdamageinitsub{i}\in\ffR_{>0}} in combination with a part accounting for the onset of damage
\begin{equation}
    g_i\lb\fsigma,\hardeningvar_i\rb = \norm{\extract_i\lr\fsigma\rr}^2 - \stressdamageinitsub{i}^2 - \hardparam_i^2\hardeningvar_i^{\powerexponent_i}.
    \label{eq:damage:activation}
\end{equation}
Case-specific stresses are extracted via dedicated extraction tensors~\f{\extract_i} that can be tailored to the application at hand. Via Biot's dual equation~\cite{ConvexAnalysis} we determine the evolution equations for the internal variables and the associated Karush-Kuhn-Tucker (KKT) conditions~\cite{Karush1939,Kuhn1951}. Eventually, a proper reformulation yields the compliance for any state of damage captured by the associated damage variables
\begin{equation}
    \ffS = \ffS_0 + 2\sum_{i=1}^M \dfrac{\hardeningvar_i}{\hardparam_i}\extract_i^2,
\end{equation}
where~\f{\ffS_0} is the initial compliance. The corresponding KKT conditions have the form
\begin{equation}
    g_i\lb\fsigma,\hardeningvar_i\rb \leq 0, \quad \dot{\hardeningvar}_i \geq 0, \quad \dot{\hardeningvar}_i g_i\lb\fsigma,\hardeningvar_i\rb = 0, \quad i=1,\ldots,M.
\end{equation}
The model is thermodynamically consistent and satisfies Wulfhinghoff's  damage growth criterion~\cite{Wulfinghoff2017}. Furthermore, it can be applied to any hardening-type damage material. An efficient predictor-corrector framework in analogy to problem settings in elasto-plasticity allows for an efficient computation. For a detailed overview on the model, the reader is referred to the original publication~\cite{Gorthofer2022damage}.

\paragraph{Extraction tensors to describe damage in SMC composites}

A variety of concurring experimental investigations~\cite{Meraghni1996,Larbi2006,Trauth2017,Schottl2020} show matrix and bundle damage to be the dominant damage mechanisms in SMC composites. To capture damage onset in the matrix, we implement the extraction tensor
\begin{equation}
    \extract_M = \dfrac{1}{3}\fI\otimes\fI,
    \label{eq:extraction:matrix}
\end{equation}
accounting for damage due to dilatation~\cite{Gorthofer2022failure}. Motivated by Puck's criteria for laminates~\cite{Puck2002,Knops2008}, we derive extraction tensors for bundle damage that capture normal stresses perpendicular to the bundle direction and shear stresses in bundle direction~\cite{Gorthofer2022damage,Gorthofer2022failure}. Using a maximum stress approach based on a pencil glide ansatz~\cite{Krawietz1981,Krawietz2001}, we implement the associated extraction tensors as
\begin{equation}
    \extract_{B,N} = \dfrac{\sqrt{2}}{2}\lb\fe_2^{\otimes2}+\fe_3^{\otimes2}\rb^{\otimes2} + \dfrac{\sqrt{2}}{4}\lb\fe_2^{\otimes2}-\fe_3^{\otimes2}\rb^{\otimes2} + \lb\fe_2\otimes_{\sf S}\fe_3\rb^{\otimes2}
    \label{eq:extraction:bundle:normal}
\end{equation}
and
\begin{equation}
    \extract_{B,S} = \lb\fe_1\otimes_{\sf S}\fe_2\rb^{\otimes2} + \lb\fe_1\otimes_{\sf S}\fe_3\rb^{\otimes2},
    \label{eq:extraction:bundle:shear}
\end{equation}
with orthonormal basic vectors~\f{\fe_1}, \f{\fe_2} and \f{\fe_3}. For better clarity, we use the abbreviations \f{\fa^{\otimes n}=\fa\otimes\fa\dots\fa} ($n$~repetitions) and \f{\fa\otimes_{\sf S}\fb = \lb\fa\otimes\fb+\fb\otimes\fa\rb/2} for any vectors~\f{\fa,\fb\in\ffR^3}.

As each damage-activation function~$g_i$ is comprised of one extraction tensor~\f{\extract_i} and three damage parameters (\f{\stressdamageinitsub{i}}, \f{\hardparam_i}, \f{\powerexponent_i}), we have to determine a set of $9$ parameters to describe the anisotropic damage evolution of our SMC composite.

\subsection{An efficient upscaling approach using deep material networks} \label{sec:fund_DMN}

\paragraph{Current state of research}

Two-scale simulation approaches are a powerful tool for the analysis of microstructured materials. If the microstructure inhomogeneities fluctuate on a scale much smaller than the actual component, homogenization techniques may be used to obtain effective material models. The former emerge by solving a partial differential equation, the cell problem of first order homogenization. Such an effective model accounts for the physical mechanisms of the microstructure and the nonlinear behavior of the constituents in a macroscopic simulation.

The approach of solving the cell problem on a finite element (FE) model of the underlying microstructure as well as in every Gauss point of a macroscopic finite element simulation is commonly referred to as the $\textrm{FE}^2$ method~\cite{Renard1987, Smit1998,Feyel1999}. Similar in spirit, but relying on a fast Fourier-transform (FFT) based micromechanics solver~\cite{MoulinecSuquet1994,MoulinecSuquet1998,Schneider2021Review}, the FE-FFT method~\cite{Spahn2014,Kochmann2016} was introduced. Both the $\textrm{FE}^2$ and FE-FFT method provide excellent accuracy which comes at the expense of considerable computational costs, limiting their practicality for problems of industrial complexity. As a remedy, model order reduction techniques, \ie the transformation field analysis (TFA)~\cite{TFADvorakBenveniste1992, TFA2, TFA3}, the self-consistent clustering analysis (SCA)~\cite{SCALiu2016,SCA_CVP_2017,SCA_softening_2018} or the non-uniform transformation field analysis (NTFA)~\cite{NTFA1} seek to approximate the solution of the cell problem. However, the weak approximation qualities of piece-wise uniform functions give rise to a slow convergence rate of the TFA and SCA in terms of the number of used clusters~\cite{TFA4, SCA_convergence}. This problem is mitigated in the NTFA by relying upon non-uniform basis functions for the inelastic strains. However, the difficulty is then transferred to prescribing suitable evolution equations~\cite{NTFA_viscoelastic1,NTFA_viscoelastic2}.

Furthermore, data-driven approaches relying on feed-forward neural networks~\cite{NasserJadid1997, Penumadu1999, Srinivasu2012} or recurrent neural networks~\cite{Mozaffar2019, Koeppe2019, Gorji2020} discard with micromechanics and seek to approximate the effective stress-strain relationship of the microstructure and constituents. Nevertheless, as powerful as these data-driven approaches are, accounting for inherent physical properties, preservation of thermodynamic consistency and stress-strain monotonicity, especially far away from the training domain, may be difficult.

Liu and co-workers~\cite{LiuDMN2018, LiuDMN2019} introduced deep material networks (DMNs), a micromechanics inspired data-driven modeling approach relying on nested laminates, which serve as high-fidelity surrogate models for full-field simulations on microstructures with inelastic constituents. Gajek et al.~\cite{Gajek2020} introduced direct DMNs with a reduced number of fitting parameters to be identified also allowing for an efficient implementation in a two-scale context~\cite{Gajek2021, Gajek2021PAMM, Gajek2022}. In contrast to approximating the effective stress-strain relationship, DMNs seek to replace the underlying microstructure by a surrogate model which dispenses with a physical representation but retains positive characteristics, \ie preservation of fundamental micromechanical bounds, preservation of thermodynamical consistency and monotonicity or fulfillment of the Hill-Mandel condition, see Gajek et al.~\cite{Gajek2020} for a discussion. Furthermore, the associated cell problem can be solved efficiently and can easily be extended to account for other physical effects, \eg to account for thermodynamical coupling~\cite{Gajek2022} in a two-scale setting.

\paragraph{Micro-oriented direct deep material networks}

\begin{figure*}
	\centering
	\includegraphics[width=0.80\textwidth]{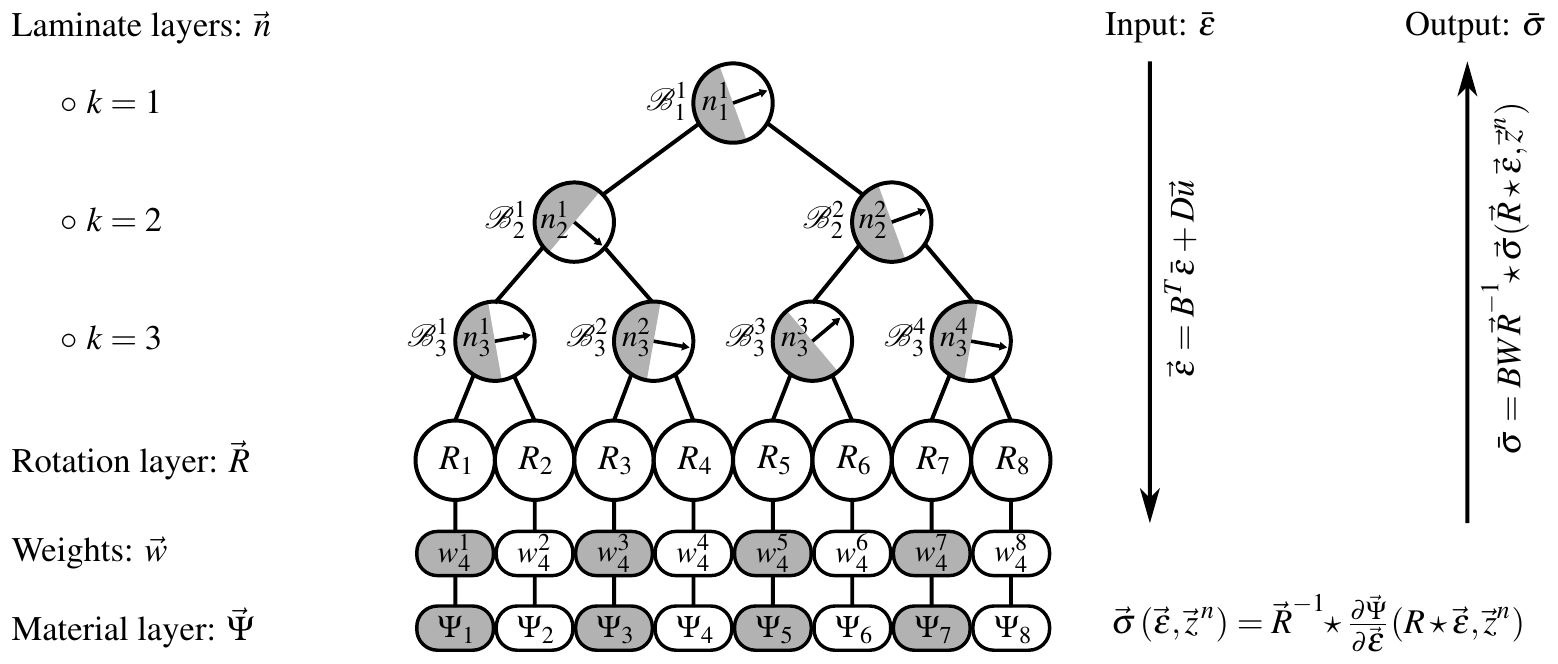}
	\caption{A two-phase direct DMN of depth 3}
	\label{fig:DMN}
\end{figure*}

In their original formulation, two-phase direct DMNs~\cite{Gajek2020} were introduced as perfect, ordered, rooted binary trees of laminates. More precisely, every node of the binary tree is given by a two-phase, rank-one laminate building block $\BB^i_k$ ($k$ indexes the depth and $i$ designates the horizontal position), see Figure~\ref{fig:DMN}. However, this formulation is restricted to microstructures without micro-oriented phases, \ie considering anisotropic SMC bundles as second phase beside the isotropic unsaturated polyester polyurethane hybrid (UPPH) matrix (see B\"ucheler~\cite{Bucheler2018} for information on the material system) is not possible. Thus, we augment the direct DMN framework~\cite{Gajek2020, Gajek2021, Gajek2021PAMM, Gajek2022} with an additional rotation layer at the bottom of the binary tree, enabling the treatment of micro-oriented problems.

On a more formal level, we consider a micro-oriented direct DMN of two phases in three spatial dimensions and of depth $K$ to consist of the following.
\begin{enumerate}
	\item A vector $\vec{\fn} = [\fn^1_K, \fn^2_K, \dots, \fn^2_1, \fn^2_2, \fn^1_1 ] \in \mathcal{N} := (\ffR^3)^{2^K - 1}$ comprising the lamination directions of all laminate building blocks inserted in a reversed breadth-first ordering. 
	\item A vector of non-negative weights $\vec{w} = [ w^1_{K+1}, w^2_{K+1}, \dots, w^{2^K}_{K+1} ] \in \mathcal{W} := \ffR^{2^K}_{\geq 0}$, summing to unity, which are used to parameterize the volume fractions of all laminate blocks, see Gajek et al.~\cite{Gajek2020} or Liu et al.~\cite{LiuDMN2018} for more details.
	\item A vector of rotation matrices $\vec{\rotOperator} = [ \rotOperator_1, \dots, \rotOperator_{2^K} ] \in \mathcal{R} := \sop^{2^K}$ specifying the material orientation.
\end{enumerate}
The vector of lamination directions $\vec{\fn}$, the vector of weights $\vec{w}$ and the vector of rotation matrices $\vec{\rotOperator}$ uniquely determine the direct DMN and serve as the fitting parameters of the surrogate model. Most notably, the DMN is trained on linear elastic training data exclusively. Subsequently, after parameter identification, the DMN is used to extrapolate to the nonlinear regime with astonishing accuracy, see Gajek et al.~\cite{Gajek2020} for more background.

The parameters ($\vec{\fn}, \vec{w}, \vec{\rotOperator}$) only depend on the geometric composition of the underlying microstructure the model is fitted on and are independent of the constituents. However, for many material classes including SMC, the geometric composition of the microstructure (typically described by suitable microstructure characteristics) fluctuates significantly on the macroscopic scale, an effect usually induced by the manufacturing process of the composite. As a consequence, a plethora of parameter sets~($\vec{\fn}, \vec{w}, \vec{\rotOperator}$), typically one for every Gauss point of the macroscopic simulation, needs to be identified in order to employ the DMN surrogate model in a two-scale simulation. As a remedy, the interpolation of the parameters ($\vec{\fn}, \vec{w}, \vec{\rotOperator}$) has been proposed~\cite{Gajek2021, Liu2019interpVF, Liu2020interpOri}, which assumes that ($\vec{\fn}, \vec{w}, \vec{\rotOperator}$) depend continuously on the relevant microstructure characteristics such that only a single parameter identification process is necessary.

\paragraph{Interpolating direct deep material networks} \label{sec:DMN_interpolation}
For SMC, the relevant microstructure characteristics are given by the fiber volume fraction and the second-order fiber orientation tensor. In principle, the fiber volume fraction $f \in \left[0, 1\right]$ ranges between zero and one. Typically, the maximum fiber content is capped well bellow one, mainly due to geometric considerations. By general covariance considerations, two fiber orientation states which differ only by an orthogonal transformation should give rise to effective material responses which differ only by this orthogonal transformation \cite{Gajek2021}. Consequently, restricting to planar fiber orientation states, we may parameterize essentially different fiber orientation states by the following second-order fiber orientation tensor
\begin{equation}
	\fA \equalhat \textrm{diag}(a, 1 - a, 0) \in \ffR^{3 \times 3}
\end{equation}
which only depends on a single parameter $a \in \left[0.5, 1\right]$ (assuming that the principal fiber orientations coincide with the axes of the underlying SMC microstructure). For instance, for $a = 0.5$, we recover a planar orientation state whereas for $a = 1$ a unidirectional state is observed.

To obtain a surrogate model admissible for any pair $(f, a)$, we define the (interpolated) parameter vector $\vec{\fp}(f, a) = [ \vec{\fn}(a), \vec{w}(f), \vec{\rotOperator}(a) ]$. Here, we assume that the directions of lamination $\vec{\fn}(a)$ and the rotation matrices $\vec{\rotOperator}(a)$ depend linearly on the fiber orientation parameter $a$ and are independent of the fiber volume fraction $f$, see Gajek et al~\cite{Gajek2021} for a suitable parametrization. Furthermore, we assume that the weights $\vec{w}(f)$ are parameterized by an (affine) linear function of the fiber volume fraction $f$ and expressed in terms of the unconstrained weights $\vec{v} \in \ffR^{2^K}$ (to ensure non-negativity, see also Gajek et al.~\cite{Gajek2020} or Liu et al.~\cite{LiuDMN2018}), \ie 
\begin{equation}
	w^{2i}_{K+1}(f) = f \ \macauly{v_{2i}} \quad \textrm{and} \quad w^{2i-1}_{K+1}(f) = (1 - f) \ \macauly{v_{2i-1}}
\end{equation}
holds, together with the consistency conditions
\begin{equation}
	\sum_{i=1}^{2^{K-1}} \macauly{v_{2i}} = 1 \quad \textrm{and} \quad \sum_{i=1}^{2^{K-1}} \macauly{v_{2i-1}} = 1.
\end{equation}
Here, $\macauly{\cdot}: \ffR \rightarrow \ffR_{\geq 0}$, $x \mapsto \max(0, x)$, denotes the Macaulay bracket, commonly known as the ReLU activation function in machine learning.

\paragraph{Upscaling with deep material networks}

After the parameter identification, for a fixed fiber orientation $a$ and a fixed volume fraction $f$,  given nonlinear laws for the phases and a time discretization by an implicit Euler method, the DMN might be thought of as the mapping $\fh_{a, f}: \Sym{3} \times \effective{\mathcal{Z}} \rightarrow \Sym{3}$, 
\begin{equation}
    \fmacrostress^{n+1} = \fh_{a, f} (\fmacrostrain^{n+1}, \vec{\fstatev}^{\, n})
\end{equation}
which computes an effective stress increment $\fmacrostress^{n+1}$ in dependence of the given macroscopic strain increment $\fmacrostrain^{n+1}$ and given vector of internal variables $\vec{\fstatev}^{\, n}$ of the last converged time step. This function can be implemented rather efficiently, enabling the analysis of industrial-scale composite components in a two-scale setting which resolves the fluctuation of $f$ and $a$ on the macroscopic scale. Details on the implementation are summarized in \ref{sec:appendix:DMN_implementation}.

\section{Application to a plaque with tensile specimens}
\label{sec:application}

\subsection{Experimental investigations} \label{sec:experimental_investigations}

\begin{figure}[ht]
    \centering
    \begin{minipage}[t]{7.5cm}
        \includegraphics{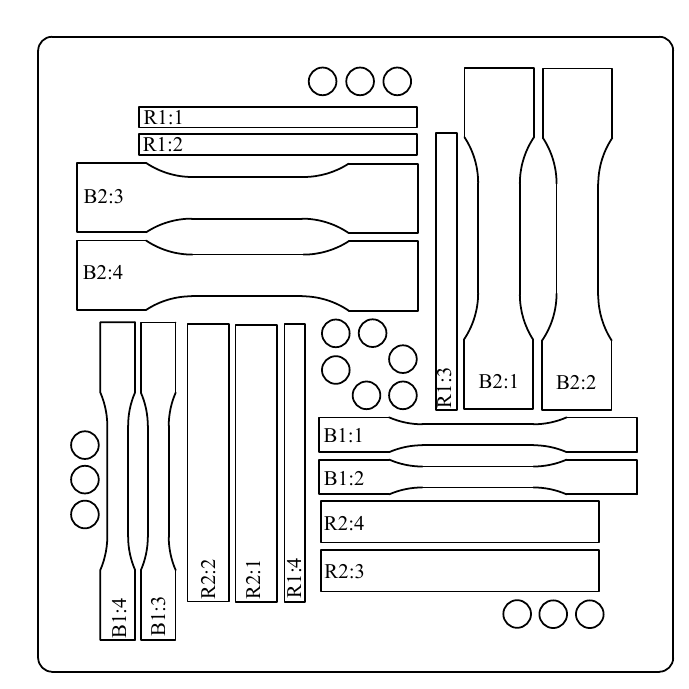}
        \caption{Locations of specimens extracted from molded plates}
        \label{fig:specimen_locations}
    \end{minipage}%
    \hspace{2mm}
    \begin{minipage}[t]{7.5cm}
        \includegraphics[width=0.9\textwidth]{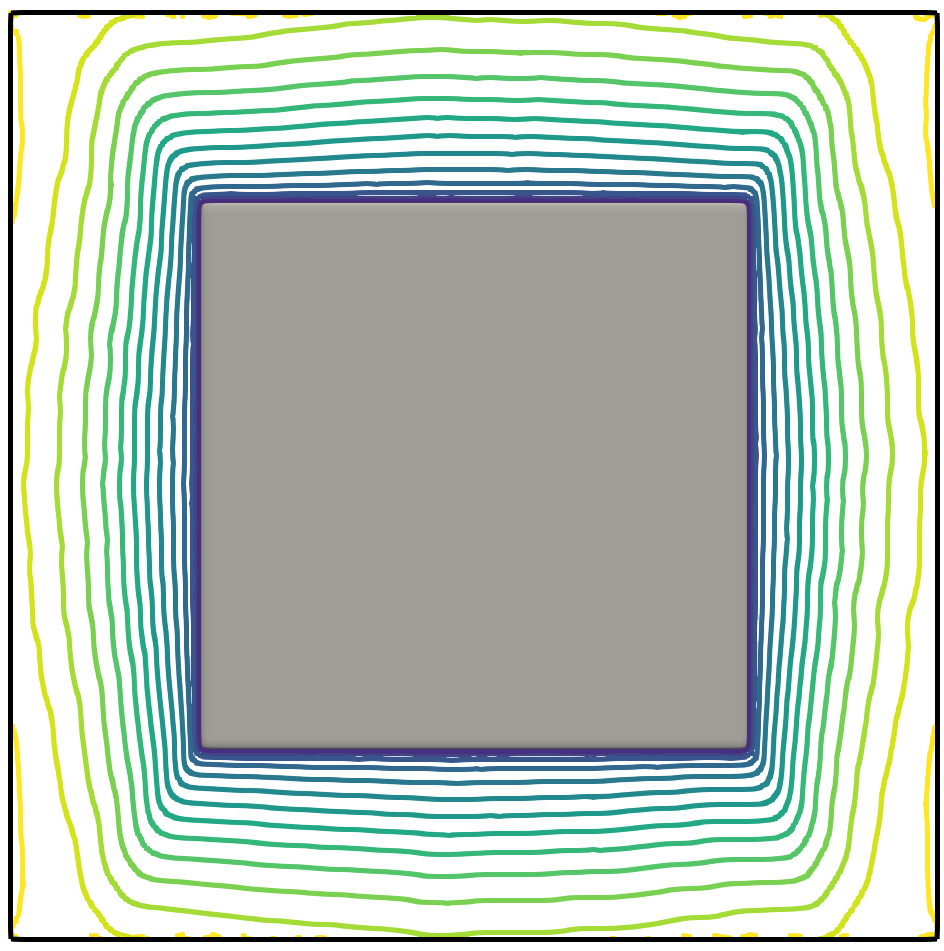}
        \caption{Initial prepreg stack (gray) and contour lines of the filled region at equally spaced time steps}
        \label{fig:fill_contours}
    \end{minipage}
\end{figure}

The experimental results used for comparison in this work are taken from Trauth~\cite{Trauth2020Diss}. 
An unsaturated polyester polyurethane hybrid (UPPH) resin system introduced by B\"ucheler~\cite{Bucheler2018} filled with a nominal E-glass fiber volume fraction of \SI{26}{\percent} was used to manufacture SMC plates (\SI{458}{\milli\meter} $\times$ \SI{458}{\milli\meter}). 
The prepreg stacks were placed in the center of the mold with an initial coverage of the molding area of \SI{35}{\percent}.
After compression molding, five types of specimens were extracted from the plates according to the cutting plan illustrated in Figure~\ref{fig:specimen_locations}.
The specimen types include circular TGA samples with \SI{25}{\milli\meter} diameter, rectangular specimens with \SI{15}{\milli\meter} width (\emph{R1}) and \SI{30}{\milli\meter} width (\emph{R2}) as well as dog-bone specimens with \SI{15}{\milli\meter} width (\emph{B1}) and \SI{30}{\milli\meter} width (\emph{B2}).
The specimens for mechanical testing (R1, R2, B1, B2) were subjected to tensile loading in a universal testing machine and loaded until failure. 
The strain was recorded with an optical system in a gauge section of \SI{70}{\milli\meter} $\times$ \SI{10}{\milli\meter} at the center of the specimens.

\subsection{Compression molding simulation}
The compression molding simulation is a transient thermo-mechanical simulation computed in a Coupled Eulerian Lagrangian framework in SIMULIA Abaqus 2021. 
We model the UPPH matrix as a purely viscous fluid with a Cross-WLF-like equation 
\begin{equation}
    \eta (T, \dot{\gamma}) = \frac{\eta_0(T)}{1+\left(\frac{\dot{\gamma}}{\dot{\gamma}_0}\right)^{1-n}}
    \quad
    \textrm{with}
    \quad
    \eta_0(T) = D_1 \exp{\lb\frac{-\alpha_1(T - T^*)}{\alpha_2 + (T - T^*)}\rb},
    \label{eq:cross_wlf}
\end{equation}
where $T$ is the temperature and $\dot{\gamma}=\sqrt{2\fD':\fD'}$ is the scalar shear rate computed from the deviatoric part of the symmetric strain rate tensor $\fD'$.
The parameters $n$, $\dot{\gamma}_0$, $T^*$, $D_1$, $\alpha_1$, $\alpha_2$ are fitted to experimental plate-plate rheometry data~\cite{Meyer2022}. 
The compressible behavior of the compound, due to air pockets in the prepreg stack, was characterized via compaction trials and we interpolate the measured relation between pressure and volumetric strain during the simulation~\cite{Meyer2022}.
The transverse thermal parameters of the compound are determined by fitting the solution of a one-dimensional heat equation to the temperature history of multiple temperature sensors embedded in an SMC stack, which was placed in a hot mold~\cite{Meyer2022}.
Molds are represented by isothermal (\SI{145}{\celsius}) rigid body shells which interact with the surface of SMC via normal contact and hydrodynamic tangential friction.
The hydrodynamic mold friction parameters were determined by press rheometry using a compression molding tool equipped with multiple pressure sensors~\cite{Meyer2022}.
A summary of parameters for the compression molding simulation is given in Table \ref{tab:dbs_parameters}.

The Eulerian domain is meshed with $670.000$ elements of Abaqus type \emph{EC3D8RT} (\ie linear hexaeders with reduced integration and a temperature degree of freedom) with \SI{1}{\milli\meter} thickness and \SI{2.5}{\milli\meter} in-plane dimensions.
The matrix volume fraction in elements is assigned according to the \SI{270}{\milli\meter} $\times$ \SI{270}{\milli\meter} $\times$ \SI{12}{\milli\meter} sized stack that is positioned at the center of the mold. 
The initial stack temperature is set to \SI{25}{\celsius} and bundles with \SI{25}{\milli\meter} length are generated in the stack.
The bundles are meshed with truss elements of \SI{2.5}{\milli\meter} length and bundles leaving the domain are clipped. 
The nominal fiber volume content $f_0 \in [0,1]$ and a nominal planar second-order fiber orientation tensor $\fA_0 \equalhat \mathrm{diag}(a_0, 1-a_0,  0) \in \ffR^{3 \times 3}$ with $a_0 \in [0.5,1]$ in the initial stacks follow Table \ref{tab:stack_realizations} with four different configurations \emph{A} to \emph{D}.

We compute four different realizations for each configuration with identical nominal properties and compute the mold filling process for each plate realization, \ie sixteen compression molding simulations in total.
The computational setups and computation times are described in \ref{sec:dbs_computational_setup}.

\begin{table}[ht!]
    \centering
    \footnotesize
    \begin{tabular}{lcccc}
        \toprule
         & Nominal volume fraction $f_0$ & Nominal orientation $a_0$ & Truss element count & Color Code\\
        \midrule
        A & \SI{22.5}{\percent} & 0.5 & 1.8 M & \crule[Dark2-A]{6mm}{3mm}\\
        B & \SI{26.0}{\percent} & 0.5 & 2.1 M & \crule[Dark2-B]{6mm}{3mm}\\
        C & \SI{29.0}{\percent} & 0.5 & 2.4 M & \crule[Dark2-C]{6mm}{3mm}\\
        D & \SI{26.0}{\percent} & 0.6 & 2.1 M & \crule[Dark2-D]{6mm}{3mm}\\
        \bottomrule
    \end{tabular}
\caption{Investigated initial stack configurations}
\label{tab:stack_realizations}
\end{table}

During molding, the bottom mold remains at rest, while the upper mold is closed by a press controller with a constant velocity $\pressspeed = \SI{3}{\milli\meter\per\second}$ until a maximum compression force $\maxforce = \SI{6}{\mega\newton}$ is reached. Subsequently the compression force remains constant.
The contour lines in Figure~\ref{fig:fill_contours} show the progress of a compression molding simulation with equally spaced time steps between lines. 
The gray area in the center indicates the initial prepreg stack placement. The illustration shows a simulation result of configuration \emph{D} with a slight horizontal orientation preference.
The compression molding simulation takes the anisotropy into account and predicts faster flow perpendicular to the preferred fiber orientation.

\subsection{Structural material parameters}
\label{sec:identified_material_parameters}

The isotropic elastic phase properties of the E-glass fibers and the UPPH matrix resin system~\cite{Bucheler2018} are listed in Table~\ref{tab:elastic:properties}. 
UPPH was characterized by Trauth, see~\cite[Section 6.3.1]{Trauth2020Diss}. 
Using FFT-based full-field homogenization as well as a Mori-Tanaka mean-field approach~\cite{Mori1973}, we compute the elastic properties of a representative bundle containing $225$ unidirectional aligned fibers. 
Each fiber has a diameter of about \f{13.5} \textmu m and a length of about \f{\SI{25.4}{\milli\meter}}. 
Based on given \textmu CT scan analyses~\cite{Schottl2020b,Schottl2021}, we assume the volume fraction of fibers within a bundle to be \f{\SI{70}{\percent}}. 
The homogenized, transversely isotropic properties of a bundle are summarized in Table~\ref{tab:elastic:properties}. 

We identify the parameters describing damage evolution in matrix and bundles (compare Section~\ref{sec:fund_damage}) via a Bayesian optimization~\cite{Bolstad2016} approach with Gaussian regression~\cite{Williams2006} as presented in~\cite{Gorthofer2022failure}. 
For the SMC at hand, we fix all power-law exponents~\f{\powerexponent} to unity, leaving a set of three damage-activation thresholds~\f{\stressdamageinit} and three hardening parameters~\f{\hardparam} to be identified. 
The identified parameters that minimize the cost-function comparing the experimental and simulation results via a least-square error measure are listed in Table~\ref{tab:damage:parameters}.

\subsection{Training of the DMN surrogate model}

\begin{figure*}[ht]
	\centering
	\includegraphics[width=0.95\textwidth]{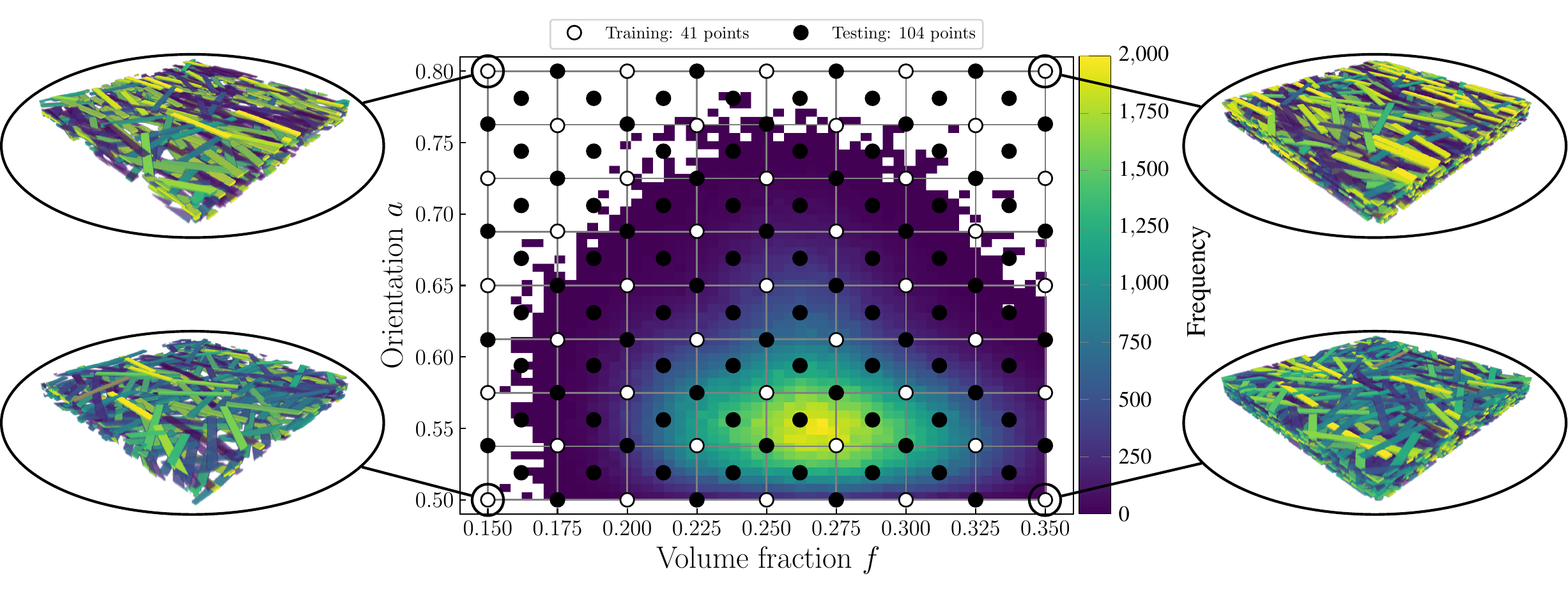}
	\caption{Discretization of the space of admissible fiber volume fraction and fiber orientation and frequencies indicating the prevalence of these after the compression molding simulation}
	\label{fig:DMN_discretization_frequencies}
\end{figure*}

As discussed in Section~\ref{sec:DMN_interpolation}, the space of admissible fiber volume fractions and fiber orientations is given by a continuum $(f, a) \in [0, 1] \times [0.5, 1.0]$ in two dimensions. Before sampling the training data for the linear elastic training, we seek a suitable discretization of our continuum domain. In order to obtain reasonable bounds for these parameters (to keep the computational costs manageable), we analyze the realized fiber volume fractions and fiber orientations occurring as a result of the mold filling simulations. Figure~\ref{fig:DMN_discretization_frequencies} summarizes this information in a heat map representing the frequency of observed fiber orientations and volume fractions.

Please note that local fiber volume fractions of up to \SI{40}{\percent} are predicted by the mold filling simulation. However, fiber volume fractions of \SI{35}{\percent} correspond to bundle fractions of over \SI{60}{\percent} which is close to the theoretical maximum of our random sequential addition (RSA) based SMC microstructure generator~\cite{Gorthofer2020}. For this reason, we restrict the maximum fiber volume fraction to \SI{35}{\percent}. As this limit is exceeded in only about \SI{0.5}{\percent} of all cells, we do not expect a significant impact on the macroscale simulations. In addition, we observe a minimum fiber volume fraction of well above \SI{15}{\percent}, hence we consider $f \in \left[0.15, 0.35\right]$ as suitable. Furthermore, we observe that (close to) unidirectional fiber orientation states do not occur at all. Thus, it suffices to consider fiber orientations $a \in [0.5, 0.8]$ during training as indicated by Figure~\ref{fig:DMN_discretization_frequencies}.

The white dots in Figure~\ref{fig:DMN_discretization_frequencies} represent the $41$ tuples $\left\{\left(f^s, a^s\right)\right\}^{41}_{s=1}$ of our discretized space. We generate an artificial SMC microstructure~\cite{Gorthofer2020} for any of those tuples. In the next step, we sample $N_{\textrm{s}} = 1230$ pairs of stiffnesses $(\ffC^s_1$, $\ffC^s_2)$ as explained in Gajek et al.~\cite{Gajek2020}, assign each stiffness tuple to one of the generated SMC microstructures in a cyclic fashion, \ie $(f^s, a^s) \mapsto (f^{(s - 1) \!\! \mod 41 + 1}, a^{(s - 1) \!\! \mod 41 + 1})$, and compute the associated effective stiffness $\effective{\ffC}^s$ by means of an FFT-based computational micromechanics code as described in Schneider~\cite{Schneider2021Review}.

\begin{figure}[ht]
	\centering
	\begin{subfigure}[t]{0.45\textwidth}
		\includegraphics[width=\textwidth]{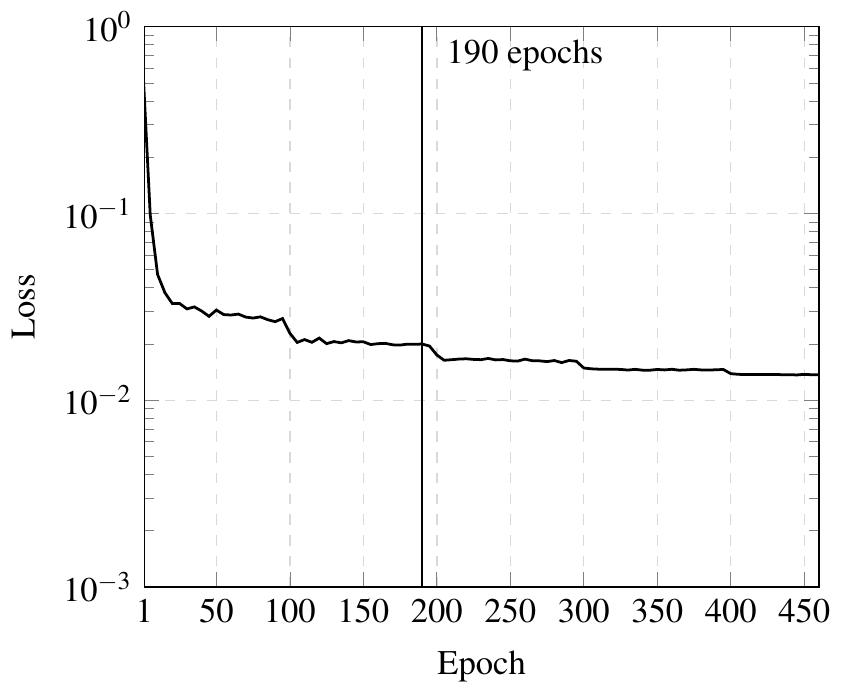}
		\caption{Loss vs.\ epochs}
		\label{fig:training_loss}
	\end{subfigure}
	\begin{subfigure}[t]{0.45\textwidth}
		\includegraphics[width=\textwidth]{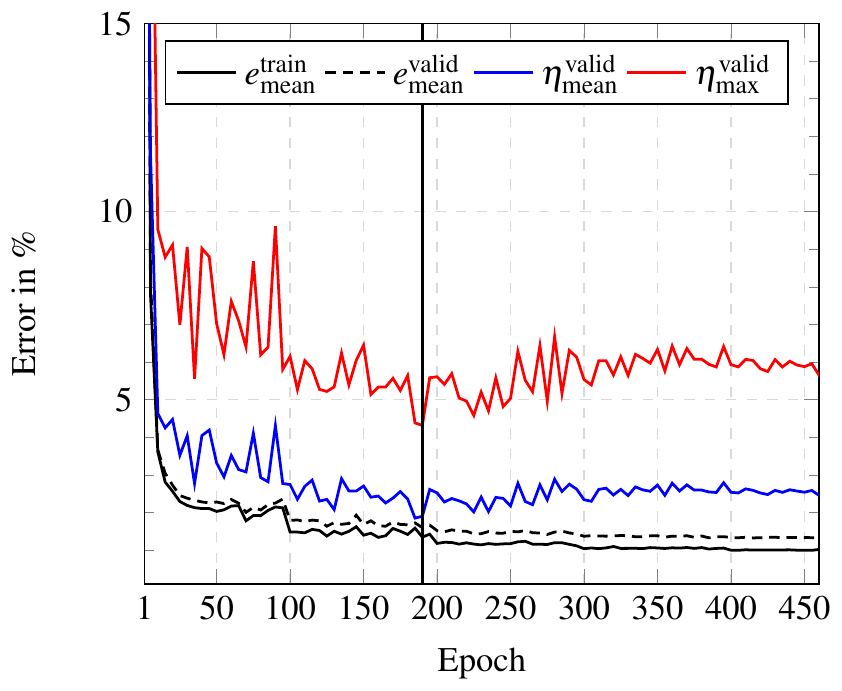}
		\caption{Training and validation errors vs.\ epochs}
		\label{fig:training_error}
	\end{subfigure}
	\caption{Loss function (a) and model performance during training (b)}
	\label{fig:training}
\end{figure}

We randomly split the pre-computed training data $\left\{ (\ffC^s_1, \ffC^s_2, \effective{\ffC}^s, f^{s}, a^{s}) \right\}^{N_{\textrm{s}}}_{s=1}$ into a training and validation set comprising $80\%$ and $20\%$ of the samples, respectively. The DMN is trained based on the loss function
\begin{equation}
	J(\vec{\fp}) = \sum^{N_{\textrm{s}}}_{s=1} J_s(\vec{\fp}) + J_p(\vec{\fp}) \rightarrow \min_{\vec{\fp}}
\end{equation}
where the first part
\begin{equation}
	J_s(\vec{\fp}) = \frac{1}{N_{\textrm{s}}} \frac{\norm{\DMNLIN(\ffC^s_1, \ffC^s_2, \vec{\fp}(f^s, a^s)) - \effective{\ffC}^s}_1}{\norm{\effective{\ffC}^s}_1}
\end{equation}
measures the proximity of the pre-computed effective stiffness $\effective{\ffC}^s$ to the DMN's effective stiffness $\effective{\ffC}^s_{\textrm{DMN}} = \DMNLIN{}(\ffC^s_1, \ffC^s_2, \vec{\fp}(f^s, a^s))$. The penalty term
\begin{equation}
	J_p(\vec{\fp}) =  \lambda \left(\sum_{i=1}^{2^{K-1}} \macauly{v_{2i}} - 1\right)^2 + \lambda \left(\sum_{i=1}^{2^{K-1}} \macauly{v_{2i-1}} - 1\right)^2
\end{equation}
with the penalty parameter $\lambda = 1000$ serves as a regularizer and enforces that the weights sum to unity and the DMN is consistent \wrt the given volume fraction $f$, \ie the equations
\begin{equation}
	\sum_{i=1}^{2^K} w^i_{K+1} = 1, \quad \sum_{i=1}^{2^{K-1}} w^{2i}_{K+1} = f \quad \textrm{and} \quad \sum_{i=1}^{2^{K-1}} w^{2i-1}_{K+1} = 1 - f
\end{equation}
hold. To assess the accuracy of the fit, we define the sample-wise mean training $e^{\textrm{train}}_{\textrm{mean}}$ and validation $e^{\textrm{valid}}_{\textrm{mean}}$ error via
\begin{equation}
	e_\textrm{mean} = \sum^{N_{\textrm{s}}}_{s=1} \frac{\norm{\DMNLIN(\ffC^s_1, \ffC^s_2, \vec{\fp}(f^s, a^s)) - \effective{\ffC}^s}_1}{\norm{\effective{\ffC}^s}_1},
\end{equation}
where $N_{\textrm{s}}$ denotes the number of elements in the training and validation sets, respectively. To significantly reduce the number of training epochs and to improve the results in the nonlinear regime, we employ an early-stopping technique as proposed by Dey et al.~\cite{DeyBosh2022}. For this purpose, we use the identified material parameters of the UPPH matrix and the bundles summarized in Section~\ref{sec:identified_material_parameters} and simulate three unidirectional strain loadings~\cite{Kabel2016}
\begin{equation} \label{eq:DMN_validation_loading}
	\fmicrostrain = \microstrain \ \fd \otimes \fd \quad \textrm{for} \quad \fd \equalhat \left[ \cos(\alpha), \sin(\alpha), 0 \right] \in S^2 \quad \textrm{and} \quad \alpha \in \ls 0\degree, 45\degree, 90\degree \rs
\end{equation}
with a strain amplitude of $\microstrain = \SI{4}{\percent}$ for each of the $41$ generated SMC microstructures. We consider this generated nonlinear data as the basis of the employed early-stopping technique. To quantify the deviation, we compute the nonlinear mean and maximum validation errors via
\begin{equation} \label{eq:DMN_nonlinear_errors}
	\quad \eta_\textrm{mean} = \max_{s \in \set{1, \ldots, N_{\textrm{s}}}} \frac{1}{T} \int_{0}^{T} \eta_s(t) \d t \quad \textrm{and} \quad \eta_\textrm{max} = \max_{s \in \set{1, \ldots, N_{\textrm{s}}}} \max_{t \in [0, T]} \eta_s(t) \quad \textrm{with} \quad \eta_s(t) = \frac{\norm{\fmacrostress^{\textrm{DMN}}_{\! s}(t) - \fmacrostress^{\textrm{FFT}}_{\! s}(t)}_1}{\underset{\tilde{t} \in [0, T]}{\max} \norm{\fmacrostress^{\textrm{FFT}}_{\! s}(\tilde{t})}_1}
\end{equation}
and track them every five epochs. In Figure~\ref{fig:training} the training progress is shown. The stepwise reduction of the loss function observed in Figure~\ref{fig:training_loss} results from the used learning rate modulation which reduces the learning rate by a factor of two for every $100$ epochs starting from an initial learning rate of $\alpha_{\textrm{ini}} = 1.5 \cdot 10^{-2}$. A closer look at the elastic training $e^{\textrm{train}}_{\textrm{mean}}$ and validation $e^{\textrm{valid}}_{\textrm{mean}}$ errors in Figure~\ref{fig:training_error} shows that there is no significant model overfitting \wrt the linear elastic training data. However, the maximum validation error $\eta^{\textrm{valid}}_{\textrm{max}}$ takes its minimum of $\SI{4.32}{\percent}$ at $190$ epochs and increases thereafter. The early-stopping approach stops the training after another $300$ epochs as the nonlinear errors have not improved for the subsequent $60$ steps. The best model (at $190$ epochs) is then stored.

To ensure that the DMN generalizes well for all tuples $(\fvc, a) \in [0.15, 0.35] \times [0.5, 0.8]$, we evaluate the DMN on an additional test set, \ie microstructures the DMN has not seen before. To be more precise, for each black point in Figure~\ref{fig:DMN_discretization_frequencies}, \ie $104$ microstructures in total, we generate an artificial SMC microstructure, perform three virtual uniaxial extension tests each (compare equation~\eqref{eq:DMN_validation_loading}), and compute the corresponding nonlinear mean and maximum test errors~\eqref{eq:DMN_nonlinear_errors}. The results are summarized in Table~\ref{tab:DMN_errors}. The nonlinear errors evaluated on the test set are only slightly increased compared to the training set. The maximum nonlinear test error (as well as the nonlinear validation error) are well below $\SI{5}{\percent}$, \ie in the range of engineering requirements, for all considered fiber volume fractions and fiber orientations. Further information on the model validation can be found in \ref{sec:appendix:DMN_validation} and the computational setup and the computational costs are presented in \ref{sec:structure_computational_setup}.

\begin{table}
    \centering
    \footnotesize
    \begin{tabular}{c c c c c c}
        \hline
        $e^{\textrm{train}}_{\textrm{mean}}$ & $e^{\textrm{valid}}_{\textrm{mean}}$ & $\eta^{\textrm{valid}}_{\textrm{mean}}$ & $\eta^{\textrm{valid}}_{\textrm{max}}$ & $\eta^{\textrm{test}}_{\textrm{mean}}$ & $\eta^{\textrm{test}}_{\textrm{max}}$ \\
        \hline
        $\SI{1.35}{\percent}$ & $\SI{1.60}{\percent}$ & $\SI{1.90}{\percent}$ & $\SI{4.32}{\percent}$ & $\SI{2.20}{\percent}$ & $\SI{4.71}{\percent}$\\
        \hline  
    \end{tabular}
    \caption{Model performance after training}
    \label{tab:DMN_errors}
\end{table}

\subsection{Structural simulation}
\label{sec:structural_simulation}

\begin{figure*}
	\centering
	\begin{subfigure}[t]{0.49\textwidth}
		\includegraphics[width=\textwidth]{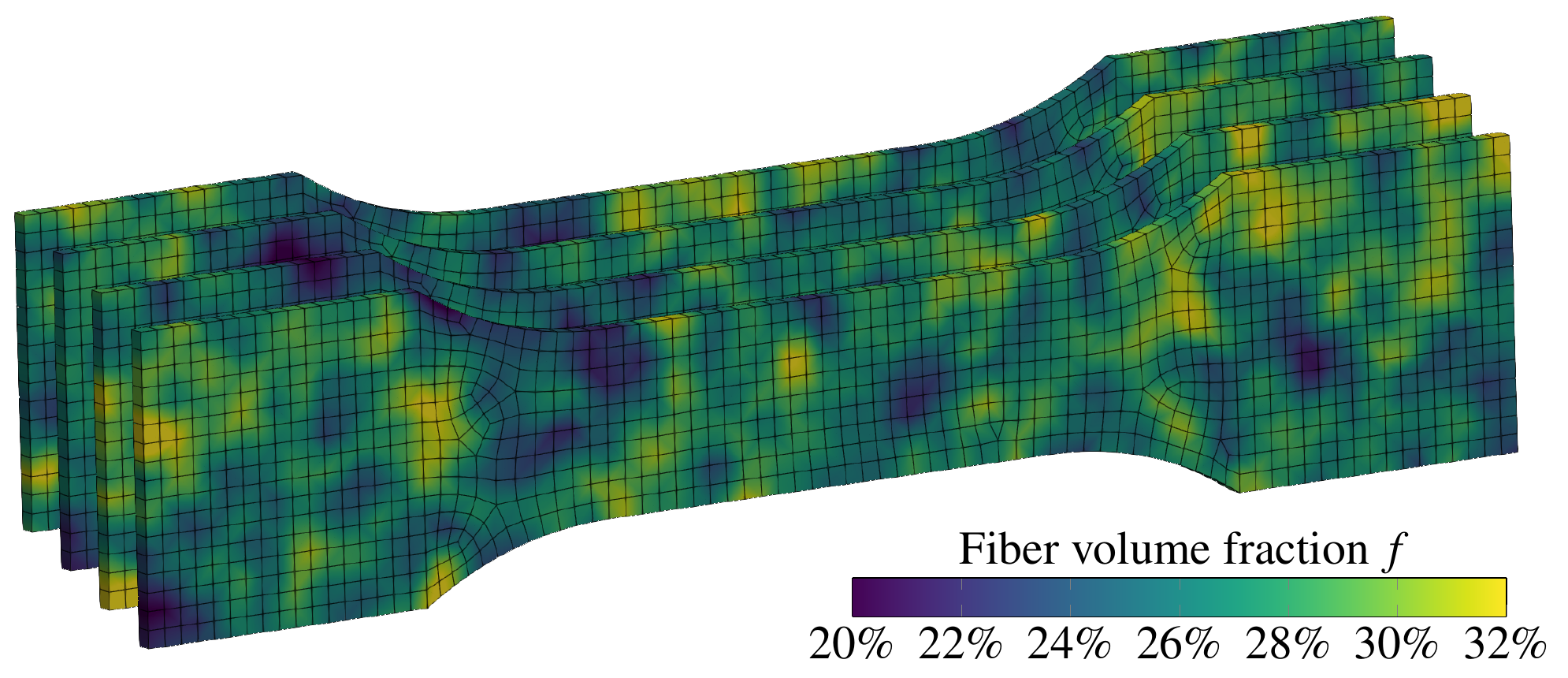}
		\caption{Mapped fiber volume fraction}
	\end{subfigure}
	\begin{subfigure}[t]{0.49\textwidth}
		\includegraphics[width=\textwidth]{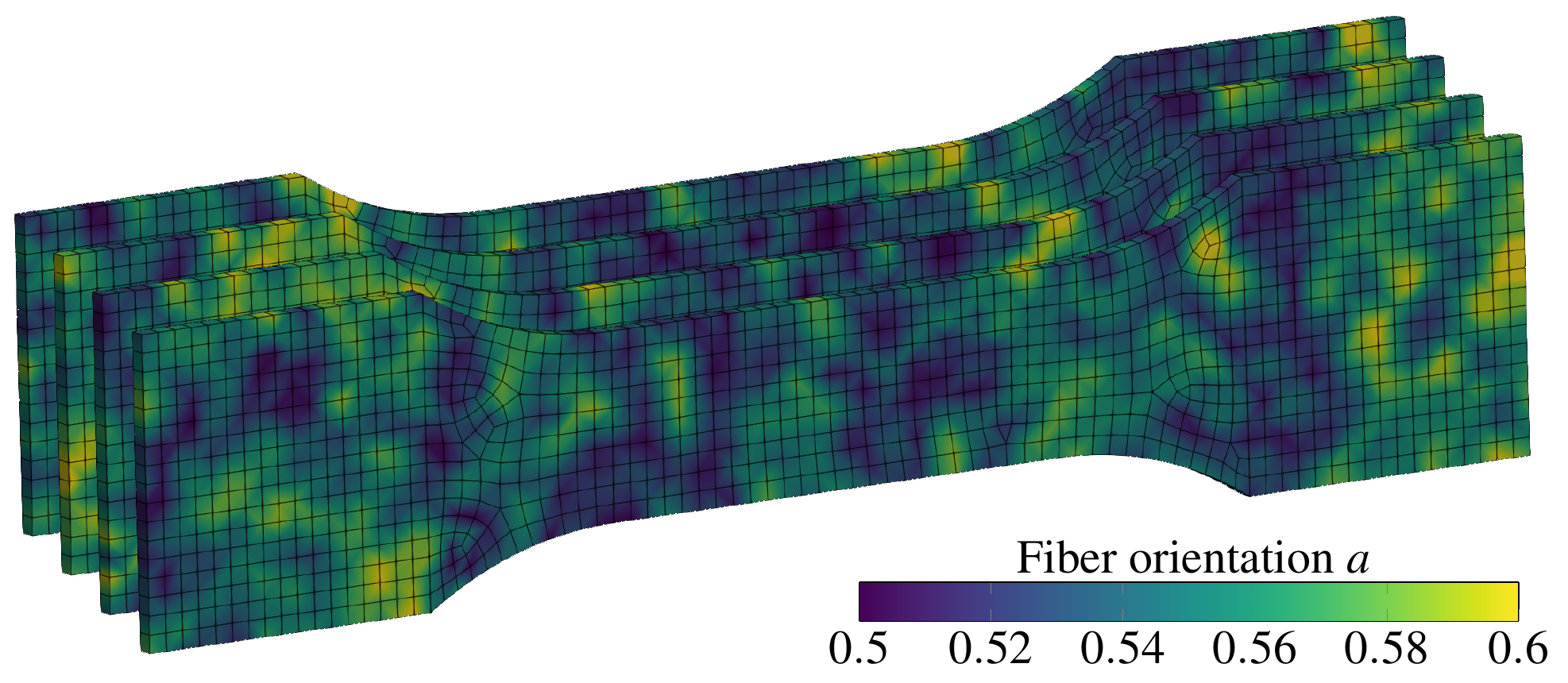}
		\caption{Mapped fiber orientation}
	\end{subfigure}
	\caption{Fiber volume fraction (left) and fiber orientation (right) property fields for different realizations at the same specimen position with identical nominal properties}
	\label{fig:fvc_mapped_stacked}
\end{figure*}

Each compression molding simulation result is transferred to multiple FE simulation models with fiber orientation and fiber volume fraction data. 
To be more precise, all $16$ mechanical specimens shown in Figure~\ref{fig:specimen_locations} are meshed with \emph{C3D8} Abaqus elements with \SI{3}{\milli\meter} edge length and we apply the procedure described in Section~\ref{sec:evaluation_mesoscale_process} to virtually cut specimens from the plates. We set the element-wise material orientations to the eigensystem of $\oritensor$, which reduces the necessary field information to $\fvc$ and $a$. 
These fields are interpolated to the nodes of the structural mesh and serve as a predefined field for the subsequent structural simulations (see Figure~\ref{fig:fvc_mapped_stacked} for an illustration of the mapped data). Contrary to physical plates, we may rotate the virtual plates, virtually cut the specimens and map the aforementioned data. Thus, we generate 1024 unique FE models (64 per plate) with individual distributed property fields.
Figure~\ref{fig:fvc_mapped_stacked} shows several realizations generated in the described way for the specimen shape B$2$ extracted at identical specimen positions (compare Figure~\ref{fig:specimen_locations}). Indeed, as observed in Figure~\ref{fig:fvc_mapped_stacked}, the mold filling simulation predicts highly fluctuating fiber volume fractions and fiber orientations even for identical nominal properties $\fvc_0$ and $a_0$.

\begin{figure*}
    \centering
	\includegraphics[width=0.8\textwidth]{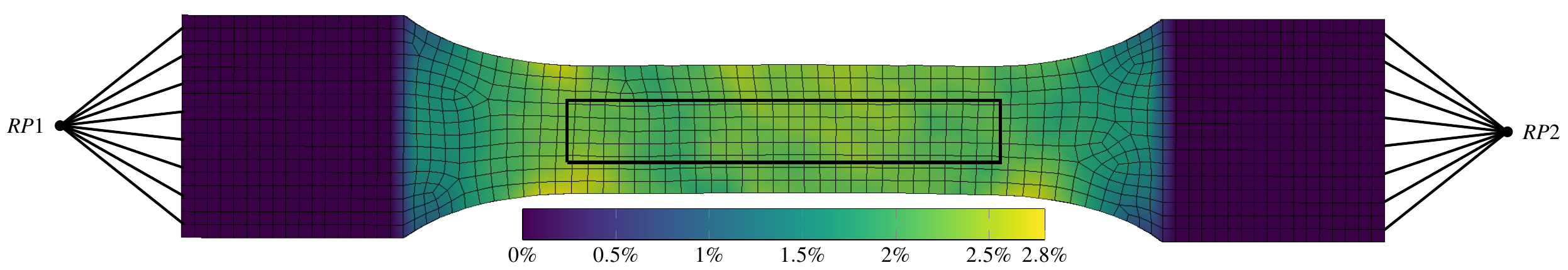}
	\caption{Distribution of the strain $\varepsilon_{11}$ component as a result of the applied load in conjunction with the gauge used for averaging the strain}
    \label{fig:sim_boundary_conditions}
\end{figure*}

For the structural simulations, a unidirectional elongation of $\SI{3}{\milli\meter}$ is applied via the two reference points RP$1$ and RP$2$, which are coupled to the specimen arms, see Figure~\ref{fig:sim_boundary_conditions}. In every Gauss point, a direct DMN is integrated implicitly to incorporate the fiber volume fraction and fiber orientation into the structural simulation.

\begin{figure*}
    \centering
    \begin{minipage}[t]{7.5cm}
        \includegraphics[width=\textwidth]{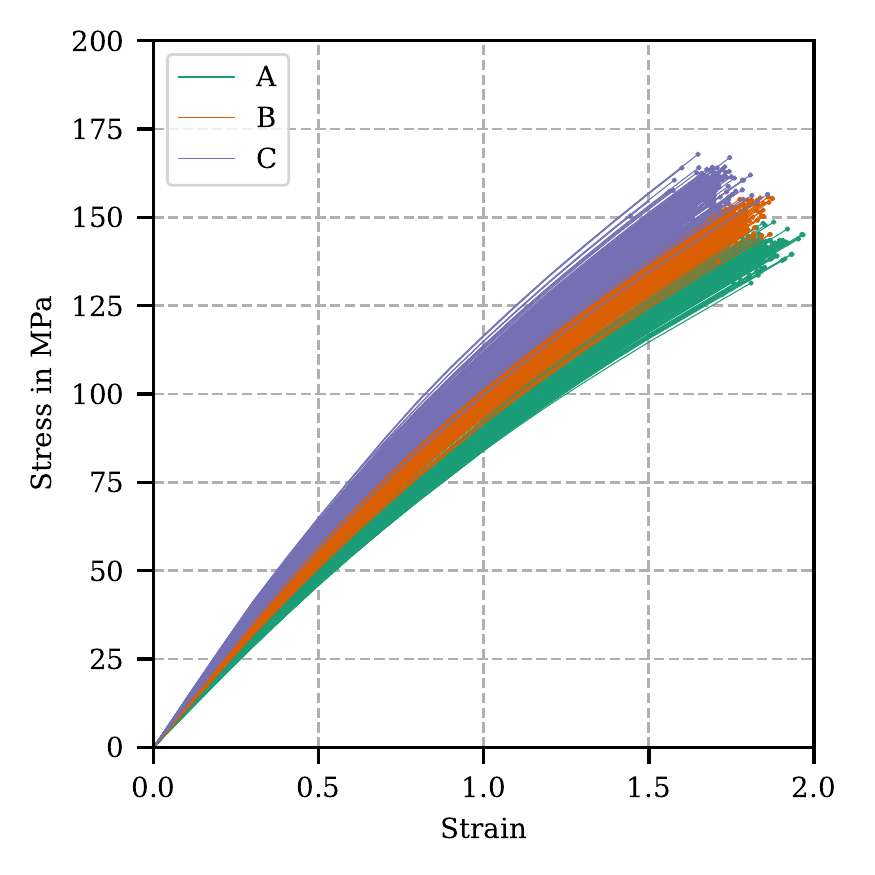}
        \caption{Stress strain curves of all virtual specimens (B$1$, B$2$, R$1$, R$2$) in configuration~A ($f=22.5\%$), configuration~B ($f=26.0\%$) and configuration~C ($f=29.0\%$) with planar isotropic initial stacks}
        \label{fig:simABC}
    \end{minipage}%
    \hspace{2mm}
    \begin{minipage}[t]{7.5cm}
        \includegraphics[width=\textwidth]{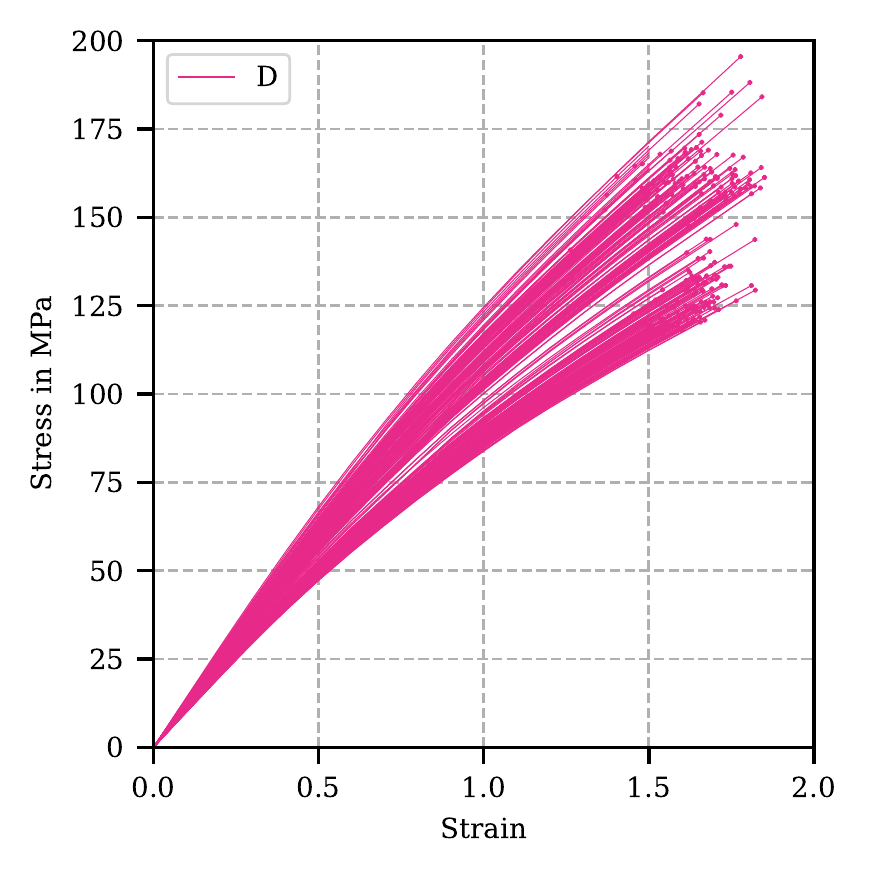}
        \caption{Stress strain curves of all virtual specimens (B$1$, B$2$, R$1$, R$2$) in configuration~D ($f=26.0\%$) with an initial stack that has an orientation preference ($a_0=0.6$). There are two branches for specimens with preferred orientation in load direction and transverse to the load direction}
        \label{fig:simD}
    \end{minipage}
\end{figure*}

For the strain measurement, we follow the experimental setup as described in Section~\ref{sec:experimental_investigations} and record the strain via averaging over the shown gauge section with dimensions \SI{70}{\milli\meter} $\times$ \SI{10}{\milli\meter}. The stress is computed by tracking the reaction force at RP$2$ which is averaged over the cross section. As SMC composites are dominated by brittle failure~\cite{Trauth2020Diss}, it suffices to predict crack initiation in order to determine the global failure of the specimen. For this purpose, we assume that crack initiation only occurs in the UPPH matrix. We determine crack initiation when the phase mean of the damage variable $q_M$ describing the isotropic matrix damage exceeds a threshold, \ie $\langle q_{\textrm{M}} \rangle_{\textrm{M}} > \SI{1.3}{\percent}$, in any element of the macroscopic simulation. This threshold is calibrated based on Trauth's~\cite{Trauth2020Diss} experimental results. The stress at crack initiation represents the strength and the corresponding strain is the failure strain. The resulting stress-strain curves of all individual virtual specimens (types B$1$, B$2$, R$1$ and R$2$) are shown in Figure~\ref{fig:simABC} and Figure~\ref{fig:simD} and saved in a database for further analysis.

\section{Uncertainty evaluation, validation and discussion}
\label{sec:results}
\subsection{Probabilistic evaluation of the virtual process chain}
\begin{figure*}[htbp]
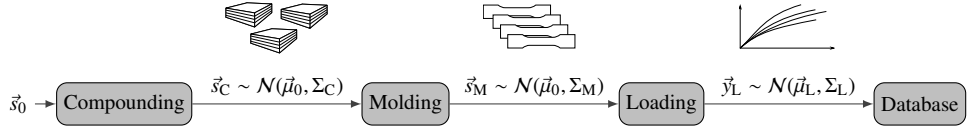

    \centering
    \footnotesize
    \begin{tikzpicture}
        [block/.style={draw,minimum width=#1,minimum height=2em},
        block/.default=10em,high/.style={minimum height=3em},
        node distance=0.7em and 5em,auto]
        \node (n0) {$\statevec_0$};
        \node[block=3em,fill=lightgray,rounded corners,right=1em of n0] (n1) {Compounding};
        \node[block=3em,fill=lightgray,rounded corners,right=8em of n1] (n2) {Molding};
        \node[block=3em,fill=lightgray,rounded corners,right=8em of n2] (n3) {Loading};
        \node[block=3em,fill=lightgray,rounded corners,right=8em of n3] (n4) {Database};
        \draw[-latex, draw=darkgray] (n0) -- (n1);
        \draw[-latex, draw=darkgray] (n1) -- (n2) node[midway,align=center]{ \includegraphics[width=5em]{figures/schematic/stack_multi.pdf}\\
        $\statevec_\textrm{C} \sim \mathcal{N}(\vec{\mu}_0, \Sigma_\textrm{C})$}; 
        \draw[-latex, draw=darkgray] (n2) -- (n3) node[midway,align=center]{\includegraphics[width=5em]{figures/schematic/specimen_multi.pdf}\\
        $\statevec_\textrm{M}  \sim \mathcal{N}(\vec{\mu}_0, \Sigma_\textrm{M})$};
        \draw[-latex, draw=darkgray] (n3) -- (n4)
        node[midway, align=center]{\includegraphics[width=5em]{figures/schematic/curve_multi.pdf}\\
        $\featvec_\textrm{L} \sim\mathcal{N}(\vec{\mu}_\textrm{L}, \Sigma_\textrm{L})$};
    \end{tikzpicture}
    \caption{Schematic representation of the probabilistic virtual process chain. The nominal fiber bundle configuration state $\statevec_0 = [f_0, a_0]$ is prescribed, while the state after compounding $\statevec_\textrm{C}$ and after molding $\statevec_\textrm{M}$ follows multivariate Gaussian distributions.
    Finally, the mechanical behavior including damage is also described by a feature vector $\featvec_\textrm{L}$ that follows a multivariate Gaussian distribution.}
    \label{fig:pvpc_schematic}
\end{figure*}

Due to the different initial microstructures resulting from variations in the impregnation process and its propagation through molding, each specimen has different property fields, \ie fiber volume fraction and fiber orientation, that affect its behavior under loading. 
Figure~\ref{fig:pvpc_schematic} outlays a probabilistic virtual process chain that interprets the multitude of specimen realizations as a Gaussian process.
Hence, we assume the uncertainty to follow a multivariate Gaussian distribution $\vec{x} \sim \mathcal{N}(\vec{\mu}, \Sigma)$ for a random variable $\vec{x}$ with a probability density distribution $p: \ffR^k \rightarrow \ffR$ defined as 
\begin{equation}
    p(\vec{x}) = \frac{1}{\sqrt{(2\pi)^k\det{\Sigma}}}
        \exp{\left(-\frac{1}{2}(\vec{x}-\vec{\mu})^\top\Sigma^{-1}(\vec{x}-\vec{\mu})\right)},
\end{equation}
where $k$ is the dimensionality of the random variable $\vec{x}$ (see Section \ref{sec:discussion} for a discussion on the applicability of a Gaussian distribution). 
The mean and the positive semi-definite covariance matrix of $\vec{x}$ are denoted as \f{\vec{\mu} \in \ffR^k} and \f{\Sigma \in \ffR^{k \times k}}.
The nominal initial state of a stack is given by an initial state vector \f{\statevec_0 = [f_0, a_0]}.

\paragraph{Compounding} The compounding process introduces uncertainty to the SMC prepreg stacks, as stacks consist of sheets from different sections of the prepreg coil.
Properties along this coil may vary due to a preferred fiber orientation or varying fiber volume fraction during production.

For the compounding process, the distribution is given by $\mathcal{N}(\vec{\mu}_0, \Sigma_\textrm{C})$ assuming that the production process has a mean fulfilling the nominal requirement ($\vec{\mu}_0 = \statevec_0$) and a covariance matrix $\Sigma_\textrm{C} \in \ffR^{2 \times 2}$. 
The covariance $\Sigma_\textrm{C} = \mathrm{diag}(\sigma_\textrm{C,f}^2, \sigma_\textrm{C,a}^2)$ is a machine-specific parameter set and we assume no correlation between the state variables as there are no off-diagonal terms in the covariance matrix $\Sigma_\textrm{C}$.

\paragraph{Molding} We assume that both, compression molding process and the extraction of specimens, increase the uncertainty, whereas the mean remains unchanged.
This assumption is specific to the symmetric centered plate molding application considered here and simplifies the evaluation.
By linearity of expectation, the covariance can then be written as
\begin{equation}
    \Sigma_\textrm{M} = \mathrm{diag}
        \left(
            \sigma_\textrm{C,f}^2+\sigma_\textrm{M,f}^2, \sigma_\textrm{C,a}^2+\sigma_\textrm{M,a}^2
        \right), 
        \quad 
        \Sigma_\textrm{M} \in \ffR^{2 \times 2}, 
\end{equation}
where $\sigma_\textrm{M,f}$ and $\sigma_\textrm{M,a}$ are the additional standard deviations introduced by molding and extraction for fiber volume fraction and orientation, respectively. 
The values of these parameters depend on the geometry and the molding process parameters. 
We will estimate these parameters from multiple realizations of the virtual process chain in Section \ref{sec:results:lenghtscale}.

\paragraph{Loading} Finally, the uncertainty of the average state in specimens $\statevec_\textrm{M} \in \ffR^{2}$ translates to uncertainties of the features upon loading $\featvec_\textrm{L} \in \ffR^N$.
The feature vector $\featvec_\textrm{L} \in \ffR^N$ may contain $N$ loading features such as Young's modulus, strength or stresses at specific strains imposed to the specimens.
For a linear combination 
\begin{equation}
    \featvec_\textrm{L}= M \statevec_\textrm{M}
\end{equation}
with $M \in \ffR^{N \times 2}$, we may compute the covariance matrix of the $N$ loading features by linear propagation of uncertainty
\begin{equation}
    \Sigma_\textrm{L} = M \Sigma_\textrm{M}  M^\top.
    \label{eq:error_propagtion}
\end{equation}
Similar to $\sigma_\textrm{M,f}$ and $\sigma_\textrm{M,a}$, the feature-extraction matrix $M$ depends on the geometry and material. 
It is also estimated from multiple evaluations of the virtual process chain in Section \ref{sec:results:contributions}.

\subsection{The effect of size on uncertainty}
\label{sec:results:lenghtscale}

\begin{figure*}[ht]
    \includegraphics[width=\textwidth]{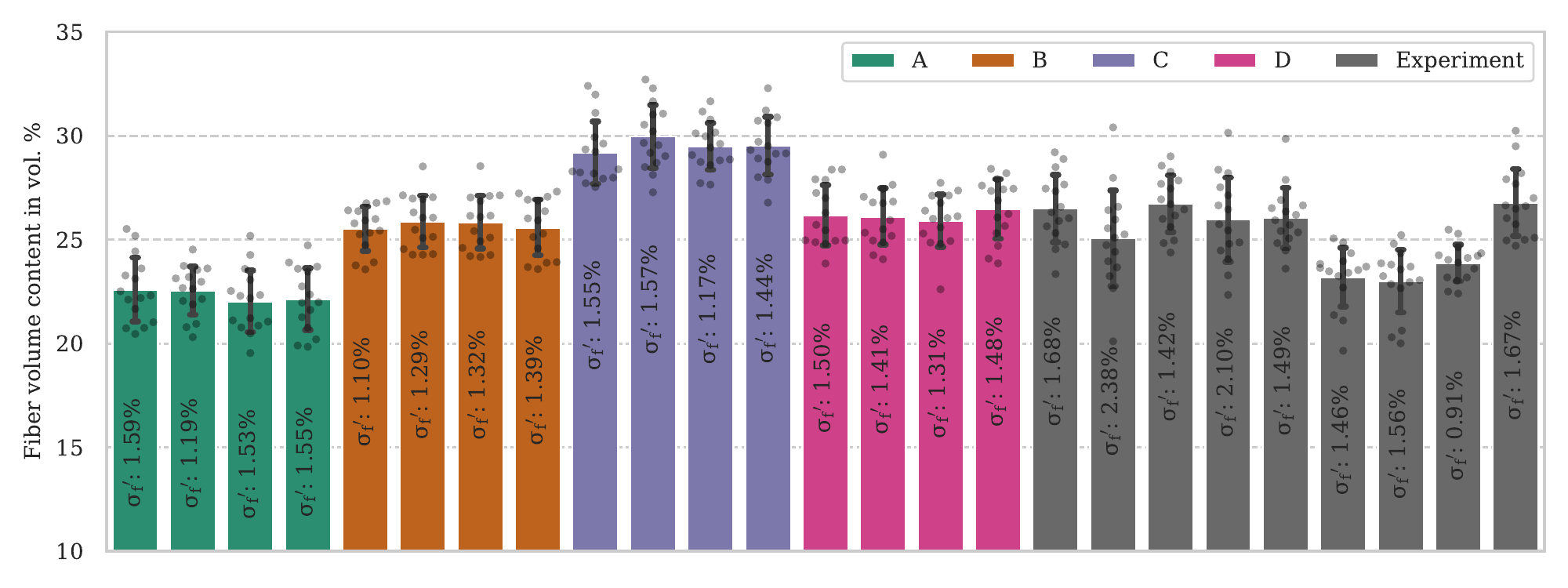}
    \caption{Simulated fiber volume fraction evaluated at TGA specimen positions and experimentally evaluated TGA specimens by Trauth~\cite{Trauth2020Diss} (gray)}
    \label{fig:sim_tga}
\end{figure*}

First, we verify that the predicted fiber configuration accurately predicts scatter of fiber volume fraction across the plates. 
Figure~\ref{fig:sim_tga} summarizes the evaluation of all $15$ round TGA samples for each realization. 
The standard deviation of fiber volume fraction within a plate $\hat{\sigma}_f$ ranges from \SI{1.10}{\percent} to \SI{1.59}{\percent} in the simulation models.
This is in good agreement with the experimental data obtained by Trauth~\cite{Trauth2020Diss} which ranges from \SI{0.91}{\percent} to \SI{2.38}{\percent}. 
However, the experiments also feature scatter between individual plates.
Assuming that the mean fiber volume content of each plate is representative for a plate, we estimate the standard deviation of the compounding machine to be $\sigma_{C,f}=\SI{1.40}{\percent}$.

The uncertainty also depends on the specimen size, as larger specimens tend to average over larger fractions of the fiber configuration. 
We describe the specimen size by a characteristic length $L$, which is computed as $L=V^{1/3}$ from the specimen volume $V$. We presume that the standard deviation of fiber volume fraction $\sigma_{M,f}$ and of orientation preference $\sigma_{M,a}$ after molding are roughly inversely proportional to the characteristic specimen size. More precisely, we presume an $L^{-1}$ scaling of the standard deviations, which is typical for processes with non-negligible boundary-layer errors (cutting specimens from the plates), see, \eg Schneider~\cite{SchneiderOtto2022}.
A linear approximation is given by 
\begin{equation}
    \sigma_{M,f}(L) \approx \frac{\SI{1}{\percent}}{\SI{0.058}{\per\milli\meter}L} 
    \quad 
    \textrm{and}
    \quad 
    \sigma_{M,a}(L) \approx \frac{1}{\SI{4.184}{\per\milli\meter}L}, 
    \label{eq:size_approximation}
\end{equation}
which is obtained from the standard deviation of differently sized regions evaluated on the full plate (see Figure~\ref{fig:lengthscale} in \ref{sec:appendix:size_dependence}).
The regions are obtained by evaluating the fiber configuration on a full plate with dimensions \SI{400}{\milli\meter} $\times$ \SI{400}{\milli\meter} $\times$ \SI{3}{\milli\meter}, which is meshed with \SI{3}{\milli\meter} edge length. 
The different sizes are then realized by grouping elements to square-shaped subsets of different sizes.

The approximations provide a simple estimate for the intrinsic uncertainty of properties after plate molding. 
Even for a perfectly controlled process with absolutely identical material parameters of all constituents, we will observe this level of uncertainty due to the random microstructure and location.

\subsection{Contributions to uncertain mechanical properties}
\label{sec:results:contributions}
To evaluate the propagation of uncertain processing effects to mechanical loading, we define a feature vector $\featvec_L \in \ffR^{18}$ containing Young's modulus, strength, failure strain and stresses from \SI{0.1}{\percent} to \SI{1.5}{\percent} with a stepsize of \SI{0.1}{\percent}.
Using all realizations of the virtual process chain $\mathcal{Y}$, we minimize the component-wise squared residuals 
\begin{equation}
    \sum_{i \in \mathcal{Y}} \left(M\statevec_{M,i} - \featvec_{L,i} \right)^2 \rightarrow \min_{M}
\end{equation}
to obtain the feature-extraction matrix $M$. 
With this strategy, we compute the covariance matrix of the loading curve $\Sigma_L$ for the following three different cases of covariance matrices for the molded state $\Sigma_M$ using equation~\eqref{eq:error_propagtion}: 
\begin{description}
    \item[Base] A perfect compounding process and perfect molding process with uncertainty only originating from different microstructure realizations
    \begin{equation}
        \Sigma_\textrm{M}^\textrm{Base} = \mathrm{diag}\left(\sigma_\textrm{M,f}^2(L),\sigma_\textrm{M,a}^2(L) \right)
    \end{equation}
    \item[FVF] A compounding process with a standard deviation of $\sigma_{C,f}=\SI{1.40}{\percent}$ for the fiber volume fraction (FVF) in stacks
    \begin{equation}
        \Sigma_\textrm{M}^\textrm{FVF} = \mathrm{diag}\left(\sigma_{C,f}^2 + \sigma_\textrm{M,f}^2(L),\sigma_\textrm{M,a}^2(L) \right)        
    \end{equation}
    \item[FVF+ORI] A compounding process with an additional standard deviation $\sigma_{C,a}=0.05$ of the orientation state (ORI) in stacks
    \begin{equation}
        \Sigma_\textrm{M}^\textrm{FVF+ORI} = \mathrm{diag}\left(\sigma_\textrm{C,f}^2+\sigma_\textrm{M,f}^2(L), \sigma_\textrm{C,a}^2+\sigma_\textrm{M,a}^2(L)\right)        
    \end{equation}
\end{description}

\begin{figure}[ht]
	\centering
	\begin{subfigure}[t]{7.5cm}
		\includegraphics[width=\linewidth]{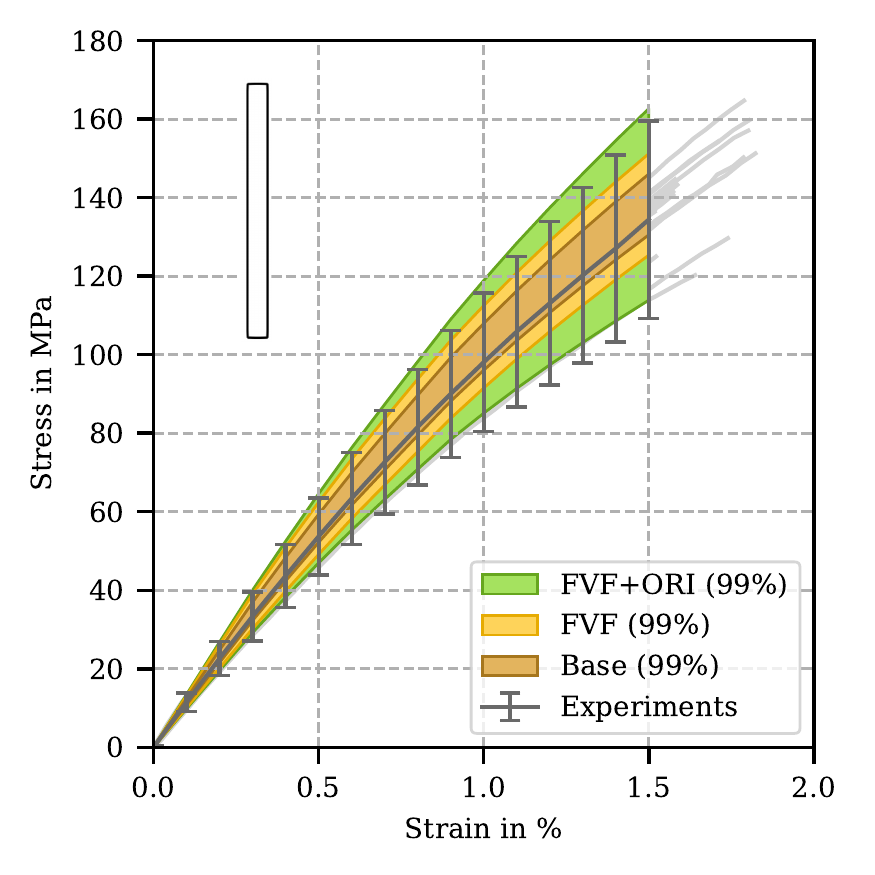}
		\caption{Specimen type R1}
	\end{subfigure}
	\begin{subfigure}[t]{7.5cm}
		\includegraphics[width=\linewidth]{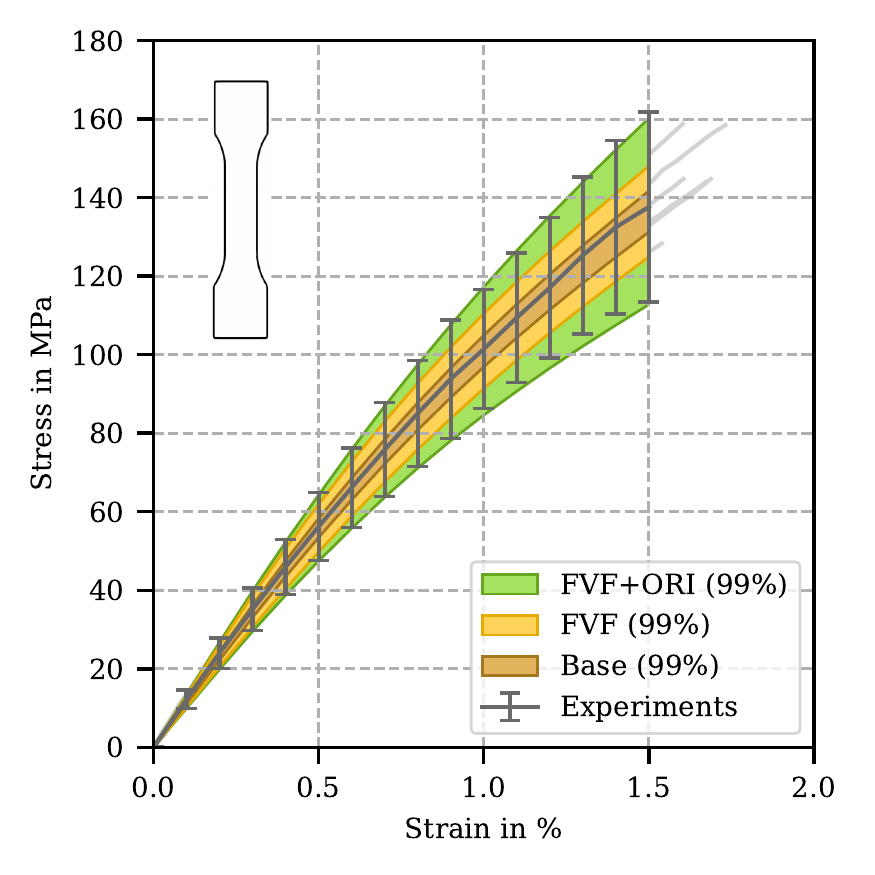}
		\caption{Specimen type B2}
	\end{subfigure}
	\caption{Stress-strain relations for the smallest and largest specimens}
	\label{fig:stress_strain}
\end{figure}

Evaluating a multitude of realizations of the virtual process permits us to estimate uncertainties associated to the loading curves of specimens from SMC plates. 
Subsequently, we simulate loading of a single homogeneous specimen and add an uncertainty estimation. 
Figure~\ref{fig:stress_strain} illustrates results of this procedure with colored areas indicating the $\pm 3 \sigma$ confidence interval containing approximately \SI{99}{\percent} of all realizations for the four specimen types (R$1$, R$2$, B$1$, B$2$) investigated here. 
The base uncertainty (brown in Figure~\ref{fig:stress_strain}) is slightly larger for the smaller specimen type R$1$ with \SI{15}{\milli\meter} width in the loading area. 
This contribution accounts only for a fraction of the total uncertainty observed in experiments (gray with error bars indicating a $\pm 3 \sigma$ confidence interval in Figure~\ref{fig:stress_strain}). 
Accounting for a variance of the fiber volume fraction between stacks results in an increase of predicted uncertainty (the area displayed in yellow in Figure~\ref{fig:stress_strain}) with a contribution similar to the basic microstructure-induced uncertainty. 
Small deviations of orientation in the stack propagate significantly through the molding process and mechanical loading, such that a deviation of $\sigma_{C,a}=0.05$ could explain the entire remaining contribution matching experimental error bars.  

\begin{figure}
	\centering
	\begin{subfigure}[t]{7.5cm}
		\includegraphics[width=\linewidth]{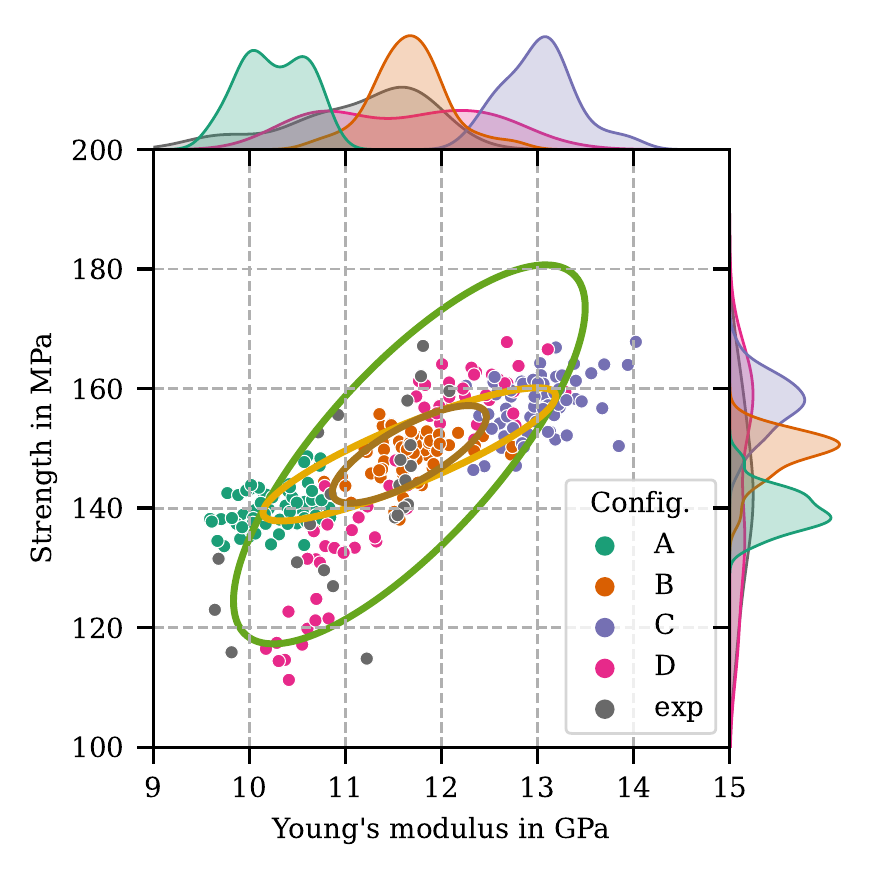}
		\caption{Specimen type R1}
	\end{subfigure}
	\begin{subfigure}[t]{7.5cm}
		\includegraphics[width=\linewidth]{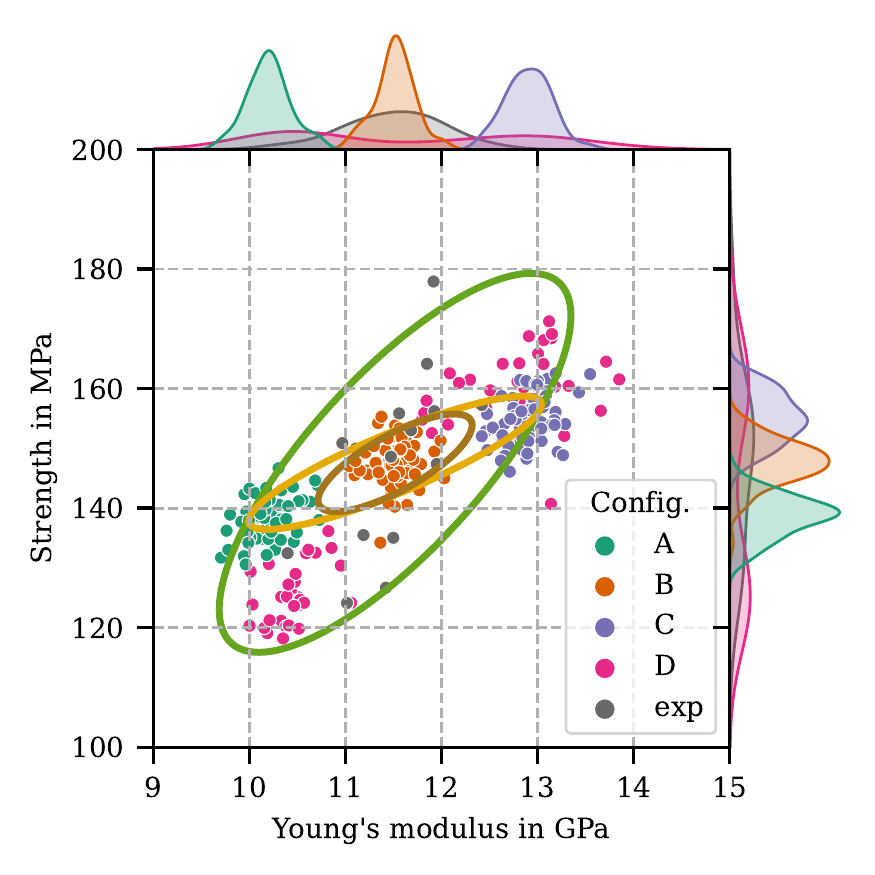}
		\caption{Specimen type R2}
	\end{subfigure}
	\begin{subfigure}[t]{7.5cm}
		\includegraphics[width=\linewidth]{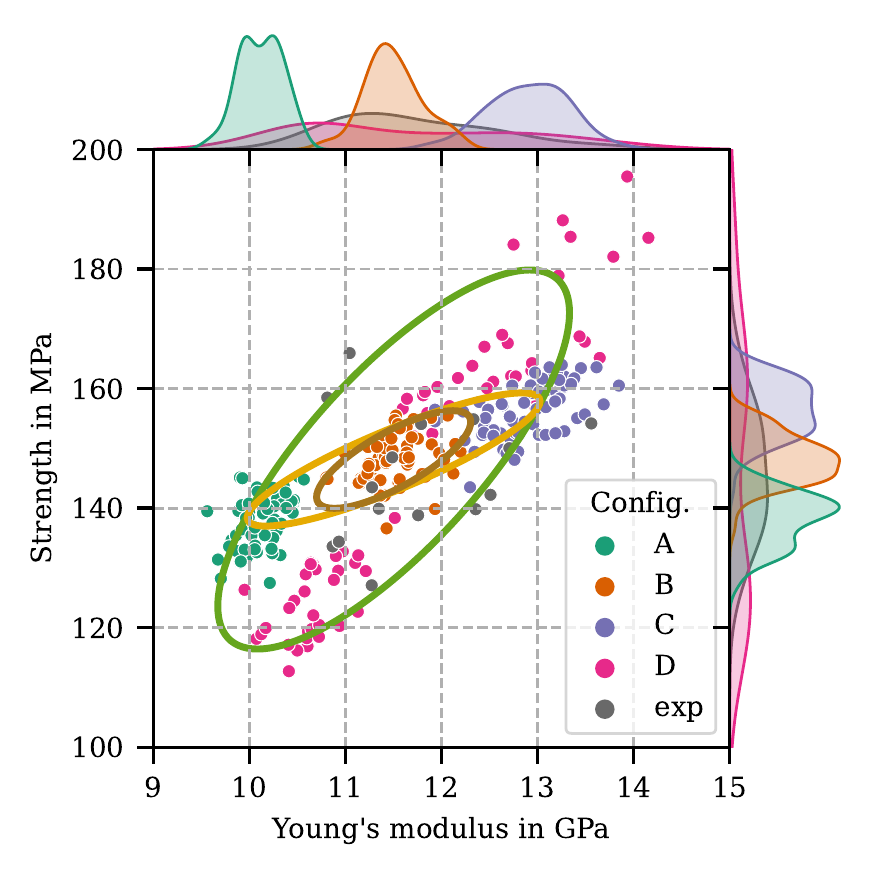}
		\caption{Specimen type B1}
	\end{subfigure}
	\begin{subfigure}[t]{7.5cm}
		\includegraphics[width=\linewidth]{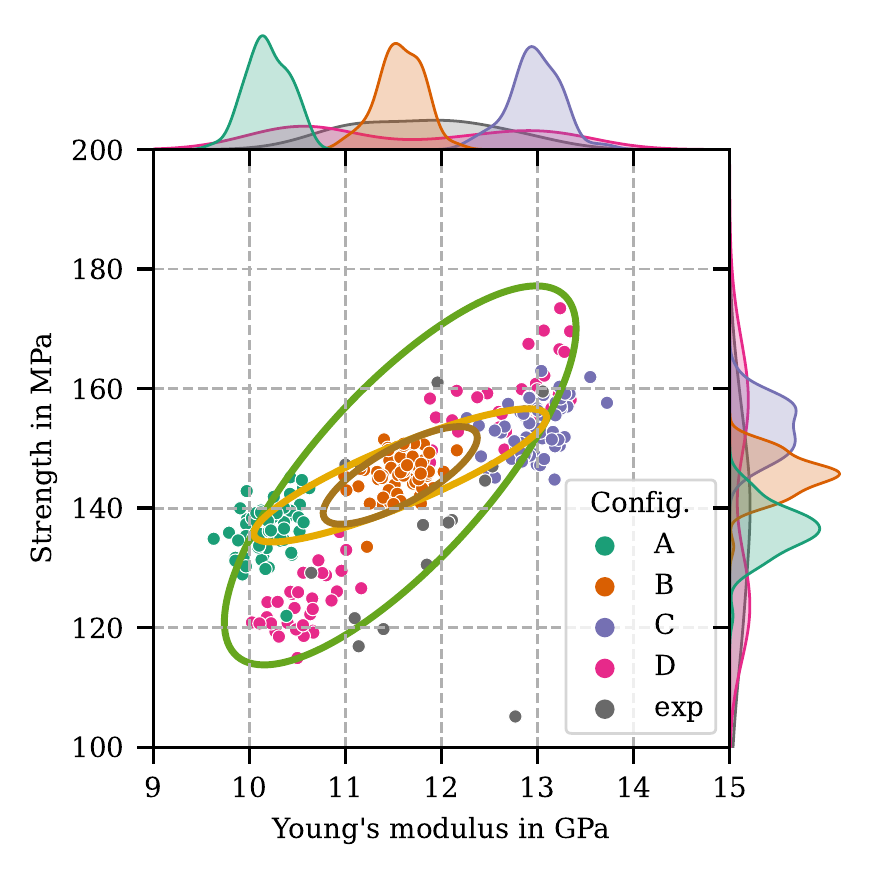}
		\caption{Specimen type B2}
	\end{subfigure}
	\caption{Correlations between strength and Young's modulus for simulated cases (colored points), experiments (gray points) and estimations of \SI{99}{\percent} confidence intervals (ellipses).}
	\label{fig:strenght_young_correlation}
\end{figure}

Additionally, evaluating the probabilistic virtual process chain allows to estimate the error propagation affecting strength, failure strain and Young's modulus including their correlation. 
Figure~\ref{fig:strenght_young_correlation} shows experimental results for strength and Young's modulus obtained by Trauth~\cite{Trauth2020Diss} as gray points.  

The simulation results for \SI{22.5}{\percent}, \SI{26}{\percent} and \SI{29}{\percent} nominal fiber volume fraction (configurations A, B and C) are shown as green, orange and purple points, respectively. 
The position of these clusters indicates an increase of strength and Young's modulus with increasing nominal fiber volume fraction. 
The simulation results for \SI{26}{\percent} nominal fiber volume fraction and an orientation preference $a_0=0.6$ (configuration D) results in two separate clusters illustrated by pink points separating those specimens tested in preferred fiber direction and those transverse to the preferred direction.
The increase of strength and Young's modulus with increasing fiber orientation follows a different relation than the one with changing fiber volume fraction.
For a preferred fiber orientation in loading direction, fibers carry more load and fatal matrix damage occurs at higher stress levels compared to a specimen without orientation preference that has the same initial Young's modulus due to a higher fiber volume fraction.

The \SI{99}{\percent} confidence interval for the contribution for the \emph{Base} case (brown ellipse), the \emph{FVC} case (yellow ellipse) and the \emph{FVC+ORI} case (green ellipse) are also shown in Figure~\ref{fig:strenght_young_correlation}. 
Even though the covariance of the molding state $\cov_M$ is uncorrelated, the propagation to the covariance of the loading curve $\cov_L$ includes the correlation and rotates the ellipses if the ratio of variance due to fiber volume fraction and orientation changes. 
Evaluating the basic uncertainty recovers simulation results with nominal stack properties. 
Adding the variance of fiber volume fraction between stacks to the model stretches the ellipse such that its first principal axis aligns with the simulated clusters for different fiber volume fractions (yellow in Figure~\ref{fig:strenght_young_correlation}). 
Adding additional uncertainty from the orientation rotates the ellipse to align with the clusters from different preferred orientation states and increases the uncertain domain significantly.
The experimental results are predominantly located within this ellipse, however isolated points are found further away. 
These are likely influenced by uncertainties outside the scope of this model, such as crack initiation during specimen clamping.

\subsection{Discussion of model simplifications}
\label{sec:discussion}

In this work, we focus on planar fiber orientation distributions and thus simplify the parametrization of the orientation state to a single variable $a$.
This is reasonable, as the out-of-plane component never exceeds a value of $A_\textrm{zz}=1.5\%$ in our application.
However, advanced structural SMC applications often deviate from the planar state. 
The proposed procedure could be enhanced for such cases with an advanced generator for three-dimensional SMC bundle structures as well as an extension of the orientation parametrization and interpolation scheme.

The proposed inverse relation of standard deviation and characteristic  length of specimens is a rough estimation without any claim of general validity.
The standard deviation plotted in Figure~\ref{fig:lengthscale} is an average value influenced by a location effect.

The uncertainty evaluation models property distributions as multivariate Gaussians.
However, the properties are limited to a certain domain (e.g. fiber volume fraction $f \in [0,1]$) and the Gaussian probability distribution would predict a small probability outside the admissible domain. 
This could be addressed with truncated Gaussian distributions, but in the application here, the probability outside the admissible range is so marginal, that we prefer regular Gaussians for simplicity.

The machine specific orientation uncertainty in compounded stacks is only estimated and not measured.
The value could be smaller, thus leaving the remaining uncertainty of experimental results to other effects such as varying material properties or crack initiation upon clamping to name a few.
These additional uncertainties may also be the reason why some experimental results in Figure~\ref{fig:strenght_young_correlation} are located outside the predicted confidence ellipses.

\section{Conclusions and outlook}
The developed virtual process chain for SMC is a complete digital representation of specimen lives from prepreg manufacturing to failure upon testing. 
The applied direct process simulation computes outcomes of an SMC compression molding process with different initial realizations of a  fiber bundle stack.
Virtual specimens with individual fields for fiber orientation and fiber volume fraction are informed with these molding results for structural simulation.
The structural macroscale simulation is based on a multiscale damage model, which is upscaled via a micro-oriented deep material network using appropiate interpolation techniques.
The resulting database of virtual samples is evaluated with a Gaussian process model and compared to experimental results.

There is an inherent size dependent base uncertainty, which can be quantified approximately by a inverse proportional relation to the characteristic length.
This base uncertainty cannot be improved by more accurate processing, more careful testing or more samples - it is intrinsic to this type of composite.
Additional uncertainty is introduced by propagation of uncertainties from the SMC prepreg stack. 
This propagated uncertainty represents a significant contribution to the total uncertainty associated with the investigated SMC specimens.
Characterizing the fiber orientation and fiber volume fraction during SMC prepreg production is an interesting field for future work due to its significant contribution to uncertainty.
Manufacturers of SMC stacks may reduce the uncertainty by shuffling sheets in a stack in such a way that the variance is minimized.

The work at hand demonstrates that it is imperative to have powerful upscaling techniques available in order to conduct two-scale simulations of microstructured materials. We show that it is essential to consider the locally varying microstructure characteristics in a macroscopic simulation in order to capture the probabilistic process chain with good accuracy. The outstanding efficiency, the high fidelity and reliability and the possibility of extension by interpolation schemes make direct DMN a powerful piece of technology for multiscale simulations. The coupled evaluations are not limited to the simple specimens shown here. 
The proposed simulation framework can be applied to entire parts by automated computation of multiple virtual realizations. 
Quantifying the uncertainty of such parts leverages the full light weighting potential by omitting over-dimensioned structures that account for uncertainty by rough safety factors.


\section*{Acknowledgements}
NM, SG and JG thank M. Bartkowiak and A. Trauth for providing detailed insight into their experimental procedures, observations and knowledge regarding our considered SMC specimens.
The research documented in this manuscript has been funded by the Deutsche Forschungsgemeinschaft (DFG, German Research Foundation), project number 255730231, within the International Research Training Group “Integrated engineering of continuous-discontinuous long fiber reinforced polymer structures“ (GRK 2078).
The work also benefited from discussion with DFG project MeproSi, project number 464119659, which aims at a probabilistic CAE chain for injection molded parts, considering uncertainties in microstructure, process conditions and material properties.
The support by the German Research Foundation (DFG) is gratefully acknowledged.

\section*{Declaration of competing interest}
The authors declare that they have no known competing financial interests or personal relationships that could have appeared to influence the work reported in this publication.

\section*{CRediT authorship contribution statement}
NM, SG, JG, LK and TB were responsible for the development of the methodology presented in this publication. NM, AH, LK, and FH developed the direct bundle compression molding approach. SG, MS and TB established the deep material network. JG, MS, AH and TB developed the anisotropic damage model. Conceptualization and establishment of the probabilistic process chain was taken over by NM, SG and JG. Validation, investigation of the results, formal analysis and the subsequent visualization were performed by NM, SG and JG. Resources were provided by LK, FH, MS and TB. The original manuscript draft was written by NM, SG and JG and extensively reviewed and edited by all authors. The research project was administrated by NM, SG and JG. Funding was acquired by LK, MS, FH and TB. The research was supervised by LK, MS, AH, FH and TB.

\appendix


\clearpage
\section{Generation of initial SMC stack}
\label{sec:stack_generation}

\begin{figure}[!ht]
  \centering
  \footnotesize
  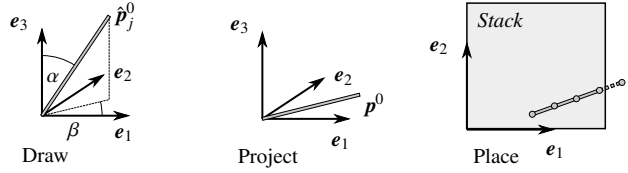
  \caption{Illustration of a single bundle generation in the initial SMC stack}
  \label{fig:stack_generation}
\end{figure}

The direct bundle simulation requires to generate fiber bundles in the initial stack that represent the initial fiber bundle configuration before molding. To draw and place the bundles we follow the procedure presented by G\"orthofer et al.~\cite{Gorthofer2020}. The corresponding placement procedure for a single bundle is illustrated in Figure~\ref{fig:stack_generation} and is repeated until the total fiber volume fraction equals the prescribed SMC fiber volume fraction.

To place a single bundle in the initial stack, we draw a primer direction according to
\begin{equation}
    \hat{\fiberdir}^0 =
        \sin(\alpha)\cos(\beta) \fe_1
        +\sin(\alpha)\sin(\beta)\fe_2
        +\cos(\alpha) \fe_3,
\end{equation}
where $\alpha \in \mathcal{U}(-1, 1)$ and $\beta \in \mathcal{U}(0, 2\pi)$ where $\mathcal{U}(a,b)$ denotes a uniform probability distribution on the interval $[a,b]$.
Then, we map this primer direction to a prescribed initial planar fiber orientation state $\oritensor^0$ by
\begin{equation}
    \fiberdir^0 = \oritensor^0\hat{\fiberdir}^0
\end{equation}
and normalize it to unity.
We generate a fiber bundle comprised of several one dimensional elements with this direction and randomly translate it within the rectangular domain of the initial stack by a shift vector $\Delta \hat{\fx} \in \mathcal{U}(\Delta \fx_\textrm{min},\Delta \fx_\textrm{max})$, where $\Delta \fx_\text{min}$ and $\Delta \fx_\text{max}$ denote the minimum and maximum allowed shifts such that at least one node of the bundle resides within the initial stack domain.
All elements outside the initial stack are deleted, such that bundles close to the edge of the stack are shorter. 
This introduces a length distribution of bundles with \SI{12}{\percent} of fiber bundles being shorter than the nominal length of \SI{25}{\milli\meter} in the investigated application here.
The fiber bundles may overlap each other initially, which is either resolved by adjusting node positions in an initial overlap adjustment or, if that is not possible, by storing the offsets for contacts between bundles and resolving it during the flow process.
The procedure yields fiber bundles following a central Gaussian distribution~\cite{ACG} that captures the sought orientation state to high accuracy~\cite{Gorthofer2020}.

\section{Efficient implementation of micro-oriented direct deep material networks}
\label{sec:appendix:DMN_implementation}

For a detailed summary of the implementation as a user-defined subroutine, we refer to Gajek et al.~\cite{Gajek2021} for the purely mechanical and to Gajek et al.~\cite{Gajek2022} for the thermomechanical case. Please note that the mere difference in implementation between the work at hand and the aforementioned works is the additional rotation layer which rotates the computed stresses and strains (and algorithmic tangents) of the materials.

We assume that the parameter vector $\vec{\fp}$ is given as the result of a suitable interpolation scheme $(f, a) \mapsto \vec{\fp}(f, a)$ and fixed. First, we introduce the averaging operator ${\averagingOperator: \Sym{3}^{2^K} \rightarrow \Sym{3}}$, a $2^K$-fold copy of the identity on $\Sym{3}$ and the symmetrized gradient operator $\gradOperator: (\ffR^3)^{2^K-1} \rightarrow \Sym{3}^{2^K}$, which depends on the vector of lamination directions $\vec{\fn}$ and the vector of weights $\vec{w}$ and which encodes the DMN's topology into a single linear mapping, see Gajek et al.~\cite{Gajek2020} or Dey et al.~\cite{DeyBosh2022} for the specific structure of the gradient operator $\gradOperator$. Then, the vector of compatible strains $\vec{\fmicrostrain} \in \Sym{3}^{2^K}$ of the DMN admits the representation
\begin{equation}
	\vec{\fmicrostrain} = \averagingOperator^T \fmacrostrain + \gradOperator \vec{\fu}
\end{equation}
where $\fmacrostrain \in \Sym{3}$ designates the macrostrain (increment) and $\vec{\fu} \in (\ffR^3)^{(2^K-1)}$ stands for the vector of (unknown) displacements. Next, we define the vector of incremental algorithmic potentials $\vec{\condensedpotential} = \left[ \condensedpotential_1, \dots, \condensedpotential_{2^K} \right]$, alternating between the incremental potentials of both phases, \ie
\begin{equation}
	\condensedpotential_i = \left\{
	\begin{array}{l r}
		\condensedpotential_1, \quad i \textrm{ odd},\\
		\condensedpotential_2, \quad i \textrm{ even},
	\end{array}
	\right.
\end{equation}
see Gajek et al.~\cite{Gajek2020, Gajek2021, Gajek2022} for more information. The incremental algorithmic potential ${\condensedpotential_i: \Sym{3} \times \mathcal{Z}_i \rightarrow \ffR}$ of the $i$-th phase is the result of a time discretization by the implicit Euler method
\begin{equation}
	\condensedpotential_i\left(\fmicrostrain, \fstatev^n_i\right) = \inf_{\fstatev^{n+1}_i \in \mathcal{Z}_i}\left(\freeenergy_i\left(\fmicrostrain, \fstatev^{n+1}_i\right) + \triangle t \, \dissipation_i\left(\frac{\fstatev^{n+1}_i - \fstatev^n_i}{\triangle t}\right) \right),
\end{equation}
where $\fstatev^n_i$ designates the internal variables of the last converged time step. The microscopic vector of stresses is defined via
\begin{equation}
	\vec{\fmicrostress}\left(\vec{\fmicrostrain}, \vec{\fstatev}^{\, n} \right) = \vec{\fR}^{-1} \!\! \star \frac{\partial \vec{\condensedpotential}}{\partial \vec{\fmicrostrain}}(\fR \star \vec{\fmicrostrain}, \vec{\fstatev}^{\, n}) \quad \textrm{with} \quad 	\frac{\partial \vec{\condensedpotential}}{\partial \vec{\fmicrostrain}} (\cdot, \vec{\fstatev}^{\, n} ) = \left[ \frac{\partial \condensedpotential_1}{\partial \vec{\fmicrostrain}}\left( \cdot, \fstatev^n_1 \right), \dots, \frac{\partial \condensedpotential_{2^K}}{\partial \vec{\fmicrostrain}}\left( \cdot, \fstatev^n_{2^K} \right)  \right],
\end{equation}
where $\vec{\fstatev}^{\, n} = [\fstatev_1, \ldots, \fstatev^{2^K}] \in \bar{\mathcal{Z}}$ denotes the vector of internal variables and the operator $\vec{\rotOperator} \, \star: ~\Sym{3}^{2^K} \rightarrow \Sym{3}^{2^K}$ encodes a forward rotation of strains
\begin{equation}
	\vec{\fmicrostrain} \mapsto \vec{\rotOperator} \star \vec{\fmicrostrain} = [\rotOperator_1^T \fmicrostrain_1 \rotOperator_1, \dots, \rotOperator_{2^K}^T \fmicrostrain_{2^K} \rotOperator_{2^K}]
\end{equation}
and $\vec{\rotOperator}^{-1} \! \star: ~\Sym{3}^{2^K} \rightarrow \Sym{3}^{2^K}$ denotes the corresponding backward rotation of stresses
\begin{equation}
	\vec{\fmicrostress} \mapsto \vec{\rotOperator}^{-1} \! \star \vec{\fmicrostress} = [\rotOperator_1 \fmicrostress_1 \rotOperator_1^T, \dots, \rotOperator_{2^K} \fmicrostress_{2^K} \rotOperator_{2^K}^T],
\end{equation}
both expressed in terms of the vector of rotation matrices $\vec{\rotOperator} = [ \rotOperator_1, \dots, \rotOperator_{2^K} ] \in \mathcal{R}$.
Then, for a prescribed total macrostrain increment $\fmacrostrain$, we seek the vector of displacements $\vec{\fu}$, which solves the balance of linear momentum
\begin{equation}
	\gradOperator^T \weightOperator \vec{\rotOperator}^{-1} \!\! \star \frac{\partial \vec{\condensedpotential}}{\partial \vec{\fmicrostrain}}  ( \vec{\rotOperator} \star ( \averagingOperator^T \fmacrostrain + \gradOperator \vec{\fu}), \vec{\fstatev}^{\, n} ) = \fzero, 
\end{equation}
where $\weightOperator \in \Sym{3}^{2^K} \rightarrow \Sym{3}^{2^K}$ designates the weighting matrix, a diagonal matrix comprising the weights $\vec{w}$ for which 
\begin{equation}
	\weightOperator = \operatorname{diag}\left(w^1_{K+1}, \ldots, w^{2^K}_{K+1}\right)
\end{equation}
holds. In a subsequent step, the effective stress $\fmacrostress$ is computed by averaging the phase stresses via
\begin{equation}
	\fmacrostress = \averagingOperator \weightOperator \vec{\rotOperator}^{-1} \!\! \star \vec{\fmicrostress} ( \vec{\rotOperator} \star ( \averagingOperator^T \fmacrostrain + \gradOperator \vec{\fu}), \vec{\fstatev}^{\, n}).
\end{equation}

\section{Model validation}
\label{sec:appendix:DMN_validation}

\begin{figure}
	\centering
	\begin{subfigure}{0.40\textwidth}
		\centering
		\begin{subfigure}{\textwidth}
			\centering
			\hspace{2mm}
			\includegraphics[height=0.7cm]{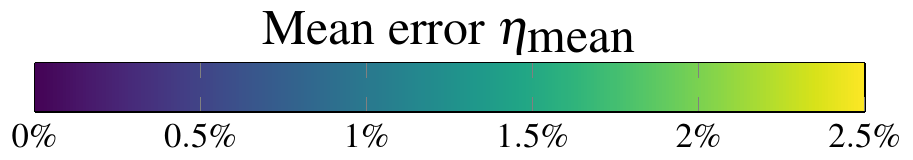}
		\end{subfigure}
		\centering
		\begin{subfigure}{\textwidth}
			\includegraphics[width=\textwidth]{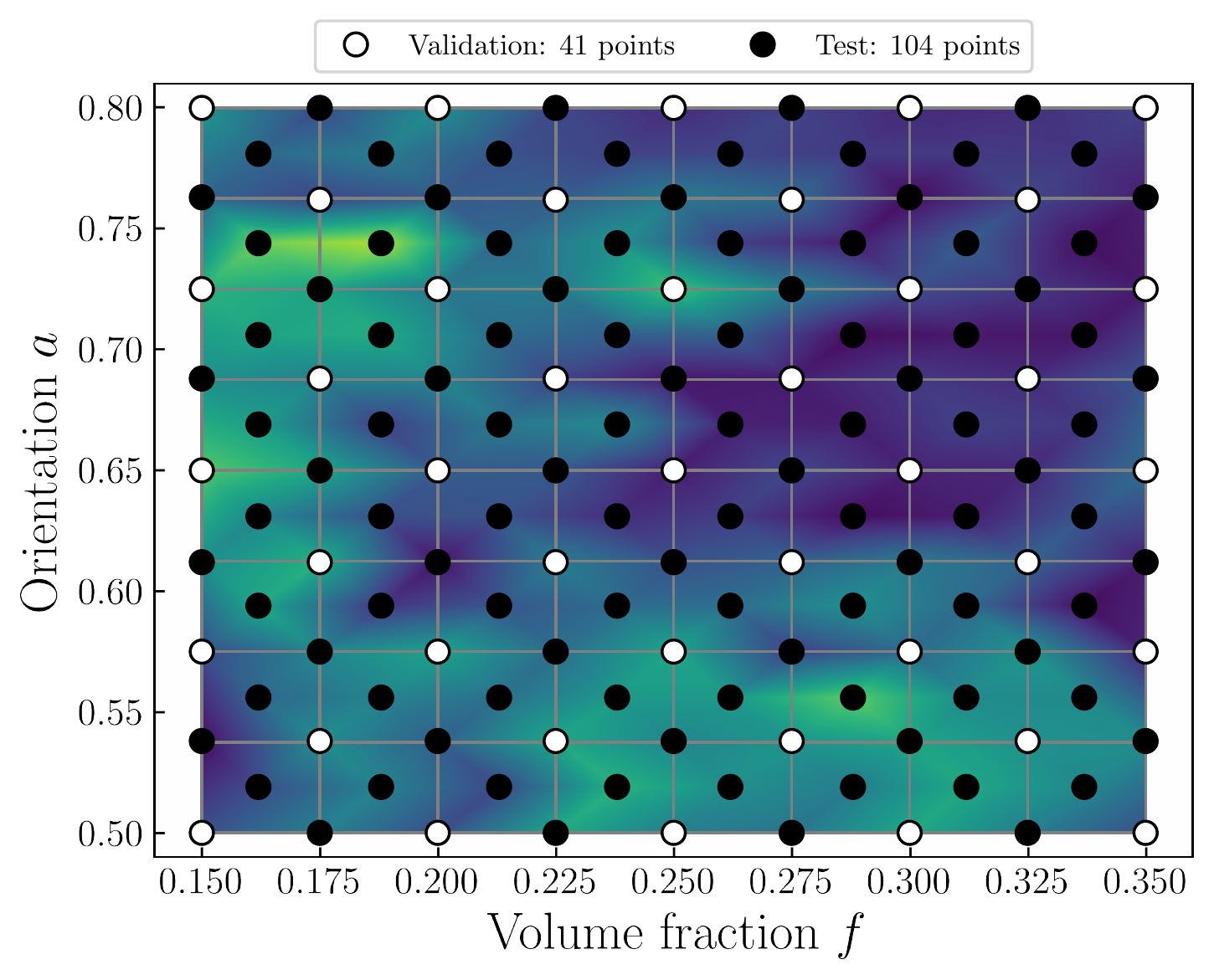}
			\caption{Mean nonlinear validation and test errors}
			\label{fig:DMN_error_countour_mean}
		\end{subfigure}
	\end{subfigure}
	\begin{subfigure}{0.40\textwidth}
		\centering
		\begin{subfigure}{\textwidth}
			\centering
			\hspace{2mm}
			\includegraphics[height=0.7cm]{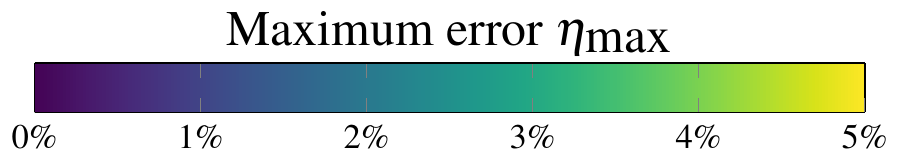}
		\end{subfigure}
		\centering
		\begin{subfigure}{\textwidth}
			\includegraphics[width=\textwidth]{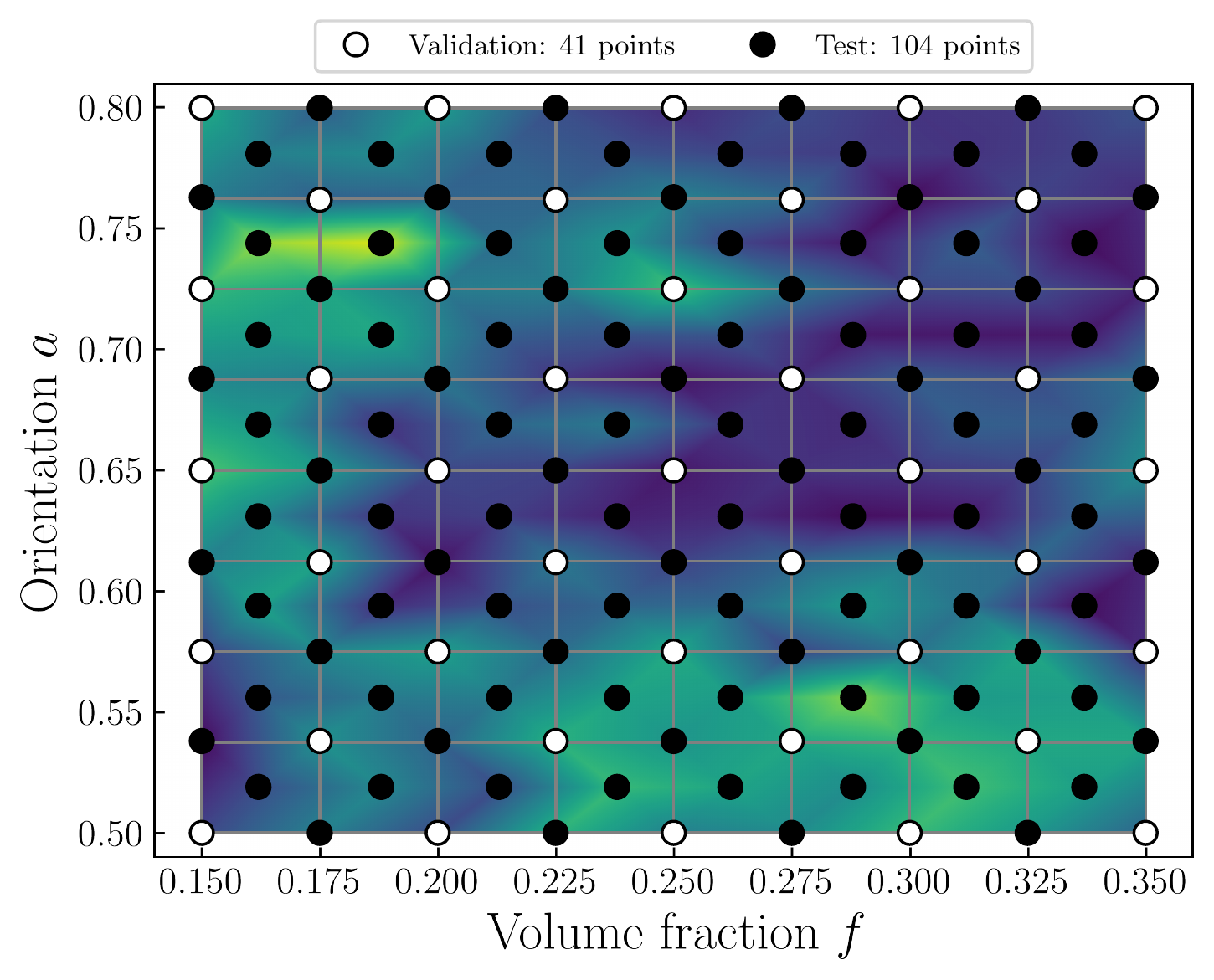}
			\caption{Maximum nonlinear validation and test errors}
			\label{fig:DMN_error_countour_max}
		\end{subfigure}
	\end{subfigure}
	\caption{Distribution of mean and maximum nonlinear validation and test errors on the space of admissible fiber volume fractions and fiber orientations}
	\label{fig:DMN_error_countour}
\end{figure}

To investigate the DMN's approximation capabilities, Figure~\ref{fig:DMN_error_countour_mean} and \ref{fig:DMN_error_countour_max} summarize the mean and maximum errors over the space of admissible fiber volume fractions and fiber orientations. We observe that both errors are strongly correlated, \ie large mean errors imply large maximum errors and vice versa. Furthermore, both errors fluctuate noticeably without indicating any distinguished dependence on certain fiber volume fractions or fiber orientations. For all considered microstructure realizations, even the ones the DMN was not trained but only interpolated on, the DMN gives nonlinear maximum errors well below $\SI{5}{\percent}$. 

\begin{figure}
	\begin{subfigure}{\textwidth}
		\centering
		\includegraphics[height=0.5cm]{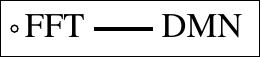}
		\includegraphics[height=0.5cm]{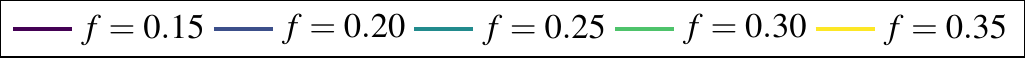}
		\includegraphics[height=0.5cm]{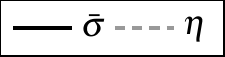}
	\end{subfigure}
	\begin{subfigure}{0.32\textwidth}
		\includegraphics[width=\textwidth]{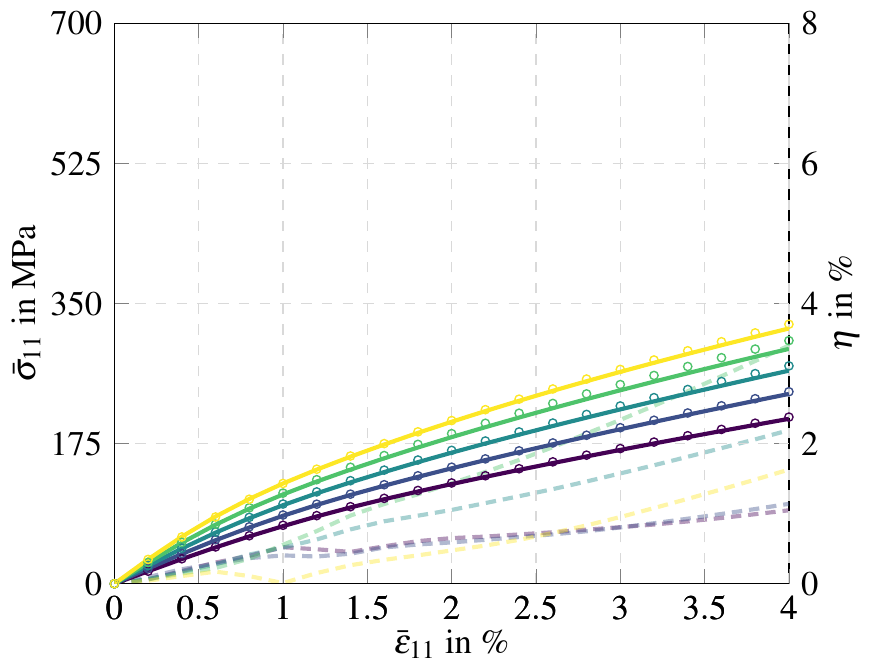}
		\caption{$a = 0.50$}
		\label{fig:DMN_stress_strain_curves_050}
	\end{subfigure}
	\begin{subfigure}{0.32\textwidth}
		\includegraphics[width=\textwidth]{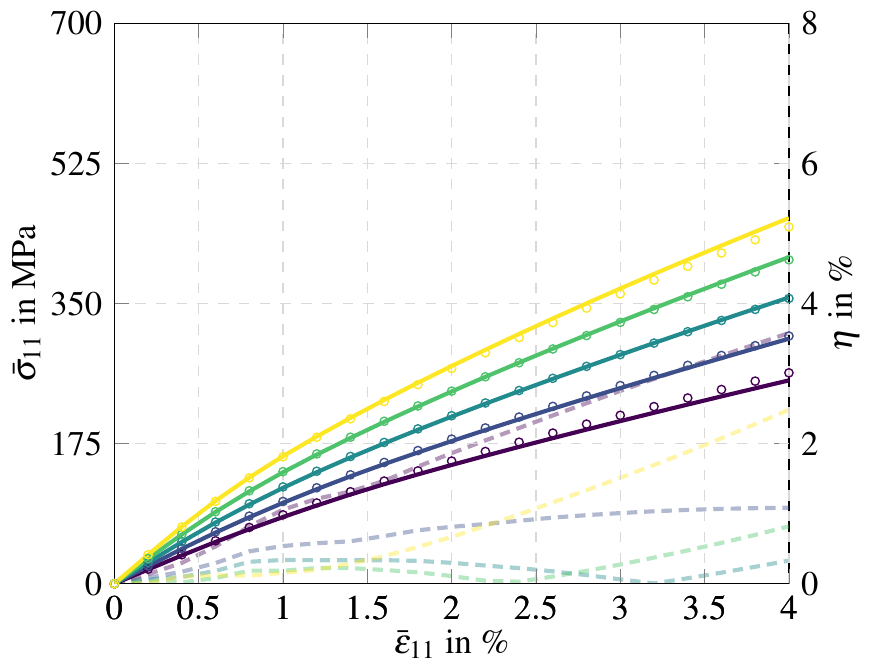}
		\caption{$a = 0.65$}
		\label{fig:DMN_stress_strain_curves_065}
	\end{subfigure}
	\begin{subfigure}{0.32\textwidth}
		\includegraphics[width=\textwidth]{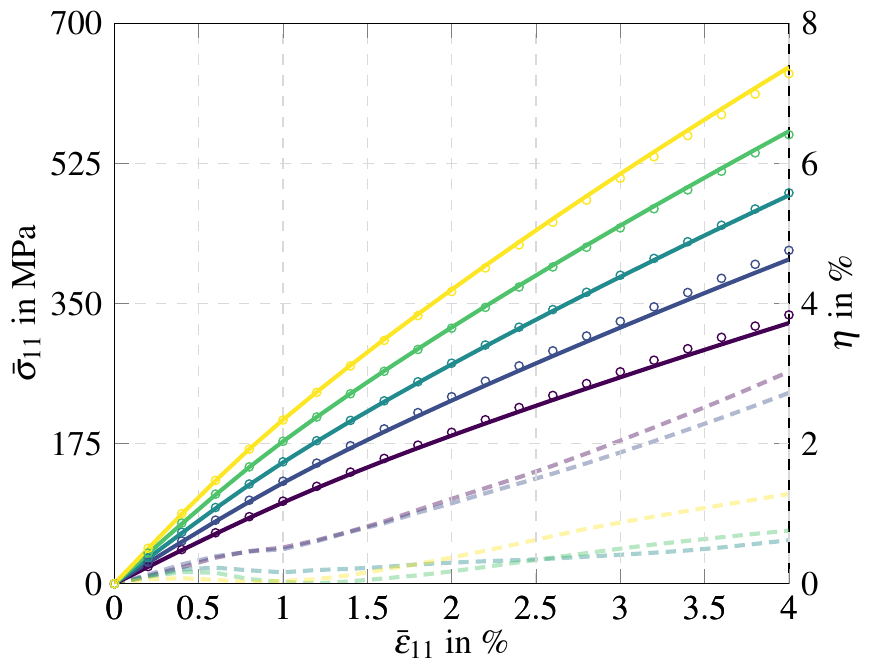}
		\caption{$a = 0.80$}
		\label{fig:DMN_stress_strain_curves_080}
	\end{subfigure}
	\caption{Comparison of full-field solution and meta model for a variety of fiber orientations and fiber volume fractions}
	\label{fig:DMN_stress_strain_curves}
\end{figure}

Figure~\ref{fig:DMN_stress_strain_curves} gives an impression on how the computed nonlinear errors listed in Table~\ref{tab:DMN_errors} translate into actual stress-strain curves. Illustrated are the predicted effective stress $\macrostress_{11}$ as well as the nonlinear error $\eta$~\eqref{eq:DMN_nonlinear_errors} for an uniaxial extension in the $11$-direction computed in $40$ equidistant time steps and a macroscopic strain of $\macrostrain = \SI{4}{\percent}$. We report the results for the planar fiber orientation $a = 0.5$ in Figure~\ref{fig:DMN_stress_strain_curves_050} as well as for the more aligned cases of $a = 0.65$ and $a = 0.80$ in Figures~\ref{fig:DMN_stress_strain_curves_065} and \ref{fig:DMN_stress_strain_curves_080} separately. Furthermore, we vary the fiber volume fraction from $f=0.15$ to $f=0.35$ in five equidistant steps.

We observe that the predicted effective stresses (in the $11$-direction) depends significantly on the fiber volume fraction as well as the fiber orientation. For all considered cases, the DMN gives an excellent prediction with nonlinear errors well below $\SI{4}{\percent}$. 

\clearpage
\section{Determination of unit cell properties}
\label{sec:appendix:unit:cell}

\begin{figure}[!ht]
    \centering
    \includegraphics[height=0.022\textheight]{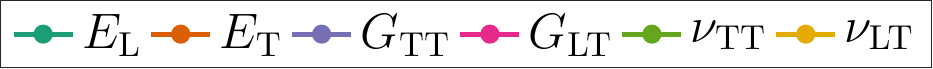}\vspace*{1ex}\\
    \begin{subfigure}[b]{0.32\textwidth}
        \centering
        \includegraphics[width=\textwidth, trim= 0 25 0 0, clip]{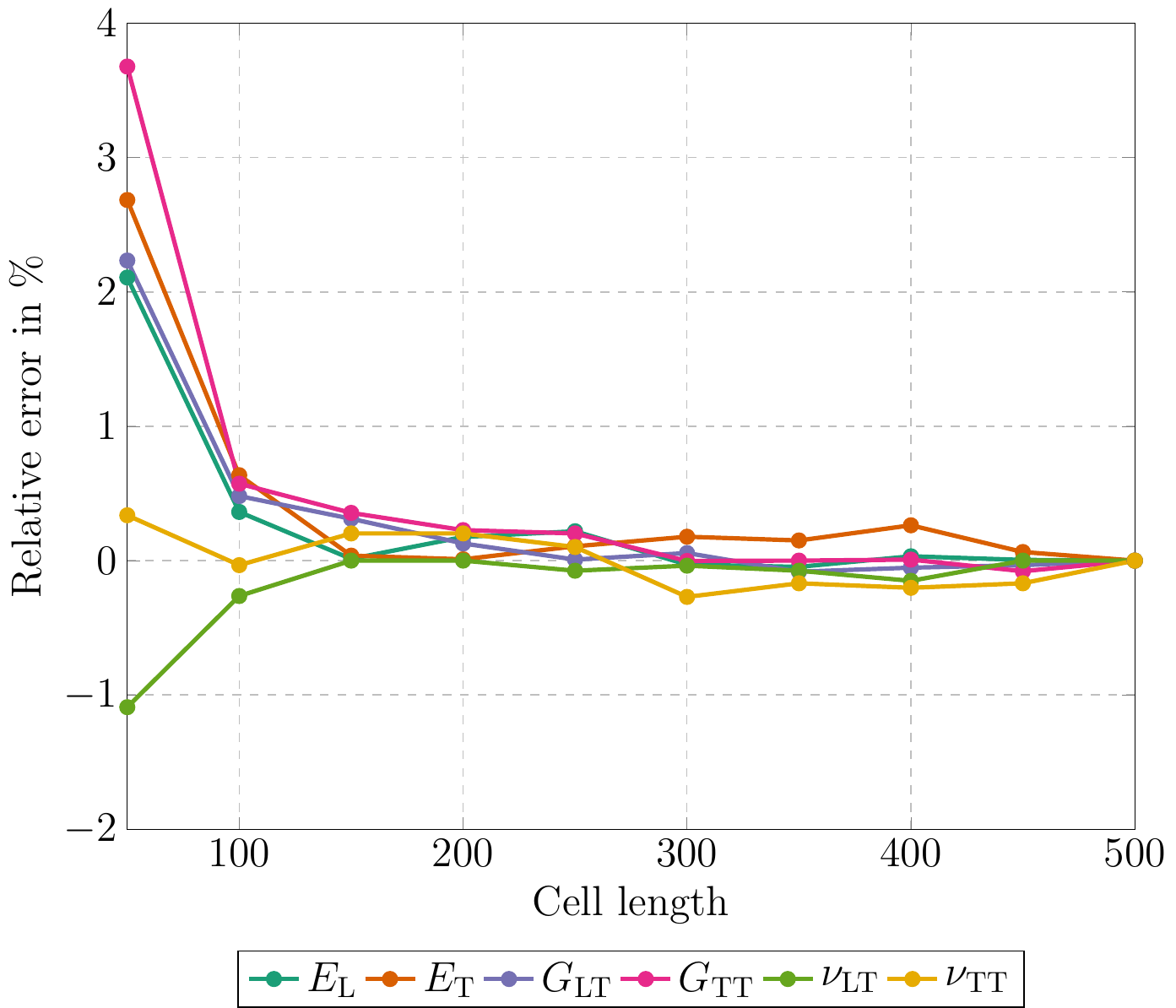}
        \caption{Cell length}
    \end{subfigure}
    \hfill
    \begin{subfigure}[b]{0.32\textwidth}
        \centering
        \includegraphics[width=\textwidth, trim= 0 25 0 0, clip]{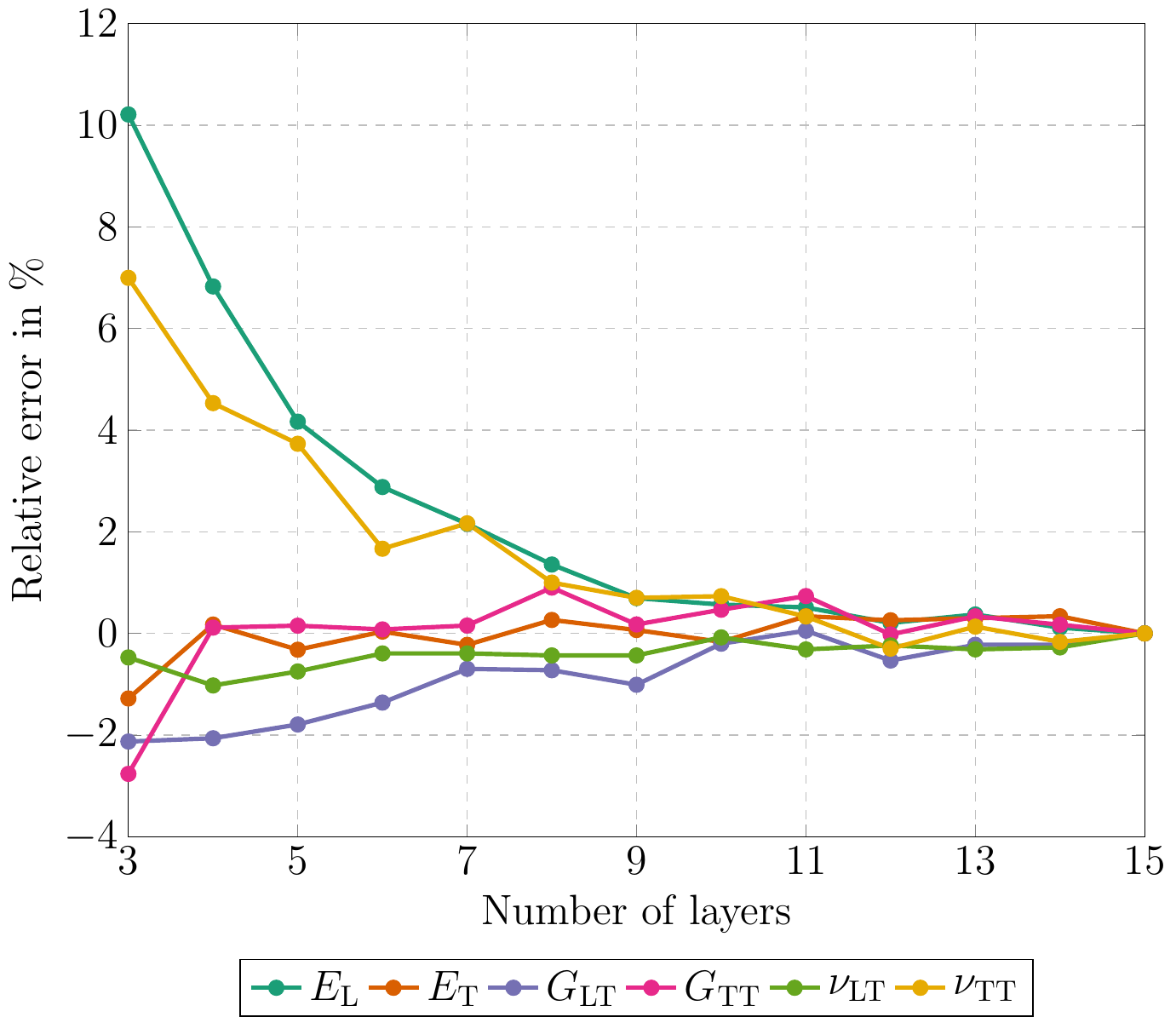}
        \caption{Number of layers}
    \end{subfigure}
    \hfill 
    \begin{subfigure}[b]{0.32\textwidth}
        \centering
        \includegraphics[width=\textwidth, trim= 0 25 0 0, clip]{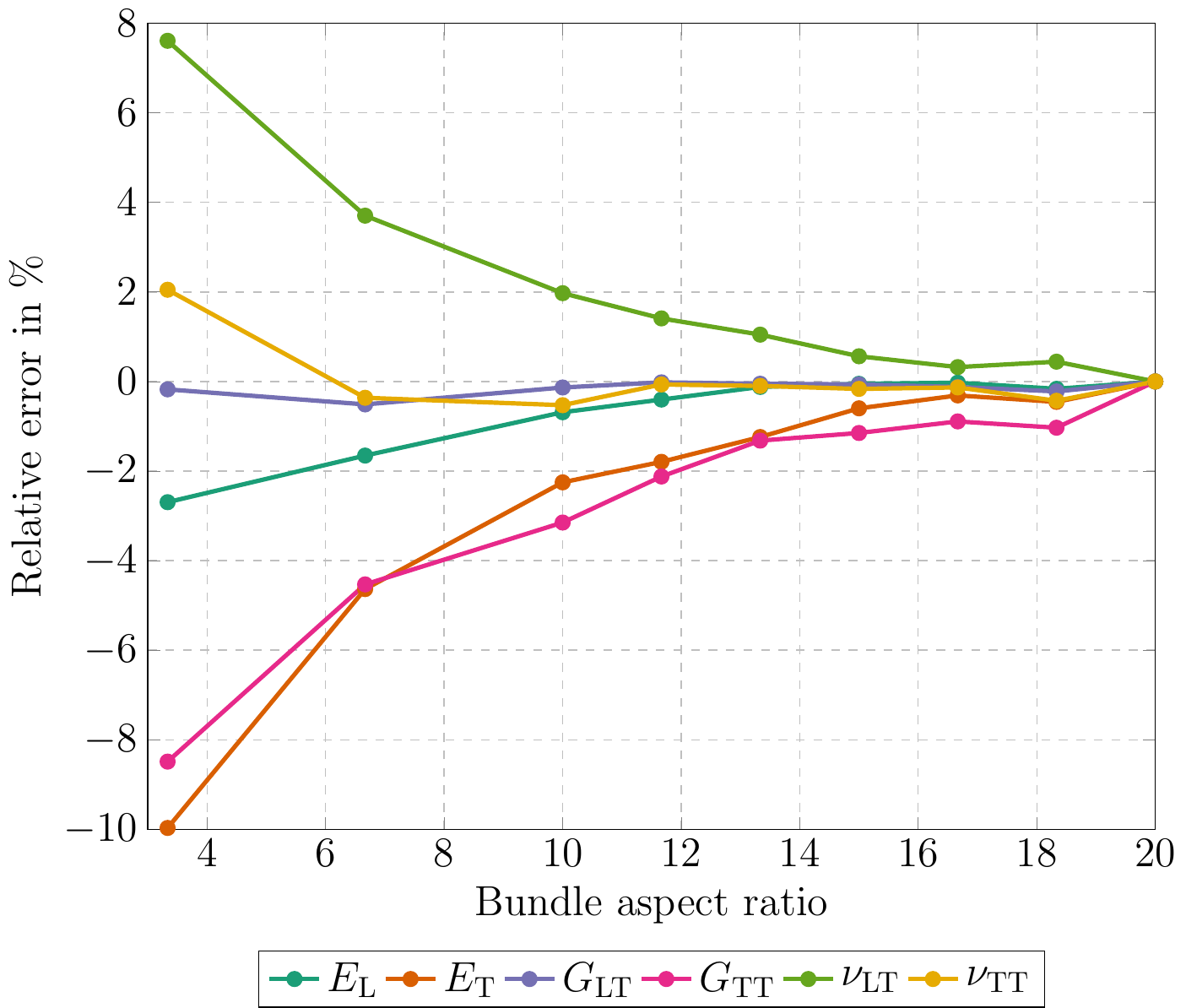}
        \caption{Bundle aspect ratio}
    \end{subfigure}
    \caption{Determination of unit cell properties based on effective transversely isotropic engineering constants}
    \label{fig:appendix:unit:cell}
\end{figure}

Using a modified random sequential adsorption approach~\cite{Feder1980} with a quasi-random Sobol' ansatz~\cite{Sobol1967} in combination with the exact closure in two dimensions~\cite{Gorthofer2020}, we generate high fidelity unit cells for SMC as introduced by Chen et al.~\cite{Chen2018generator} and extended by Görthofer et al.~\cite{Gorthofer2020}. The inescapable trade-off is the generation of unit cells that are on the one hand as small as possible to minimize the computational effort and on the other hand as large as necessary to ensure representativity. Following a procedure as presented in Görthofer et al.~\cite{Gorthofer2020} and aiming for a relative error of no more than \SI{2}{\percent} for the effective transversely isotropic engineering constants of SMC, we identify appropriate dimensions of the unit cells to be \f{500\times 500\times 50} voxels that is \f{12.5 \cdot 10^6} voxels, see Figure~\ref{fig:appendix:unit:cell}. The indices "L" and "T" of the engineering constants indicate the longitudinal and transverse direction, respectively.

\section{Computational setups}
\subsection{Direct bundle simulation}
\label{sec:dbs_computational_setup}
We ran simulations on 16-core workstations with SIMULIA Abaqus 2021 and custom subroutines VDLOAD, VEXTERNALDB, VUFIELD, VSDFIELD, VUAMP, VUVISCOSITY compiled with the INTEL ifort 16 compiler.
The simulations took \SI{15}{\hour} to \SI{30}{\hour} wall clock time depending on the number of bundles and the processor.
The data mapping described in Section \ref{sec:evaluation_mesoscale_process} is implemented as a ParaView filter (\url{https://github.com/nilsmeyerkit/paraview_map_lines}) that requires \SI{30}{\second} to \SI{60}{\second} wall clock time per specimen on a single core depending on specimen size.

\subsection{Structural simulation}
\label{sec:structure_computational_setup}
The training of the DMN and the two-scale simulations were performed on a workstation equipped with two AMD EPYC 7642 with 48 physical cores each, enabled SMT and $1024$ GB of DRAM. Sampling the linear elastic training data took about $\SI{12}{\hour}$ wall clock time where six load steps were computed in parallel on $16$ cores each. The training of the DMN took about $\SI{3}{\hour}$ wall clock time. Computing all $1024$ two-scale simulations took about $\SI{10}{\hour}$, \ie about $\SI{35}{\second}$ per simulation, with $12$ executions in parallel running on $8$ cores each.

\section{Size dependency}
\label{sec:appendix:size_dependence}
\begin{figure}[!ht]
	\centering
	\begin{subfigure}[t]{7.5cm}
		\includegraphics[width=\linewidth]{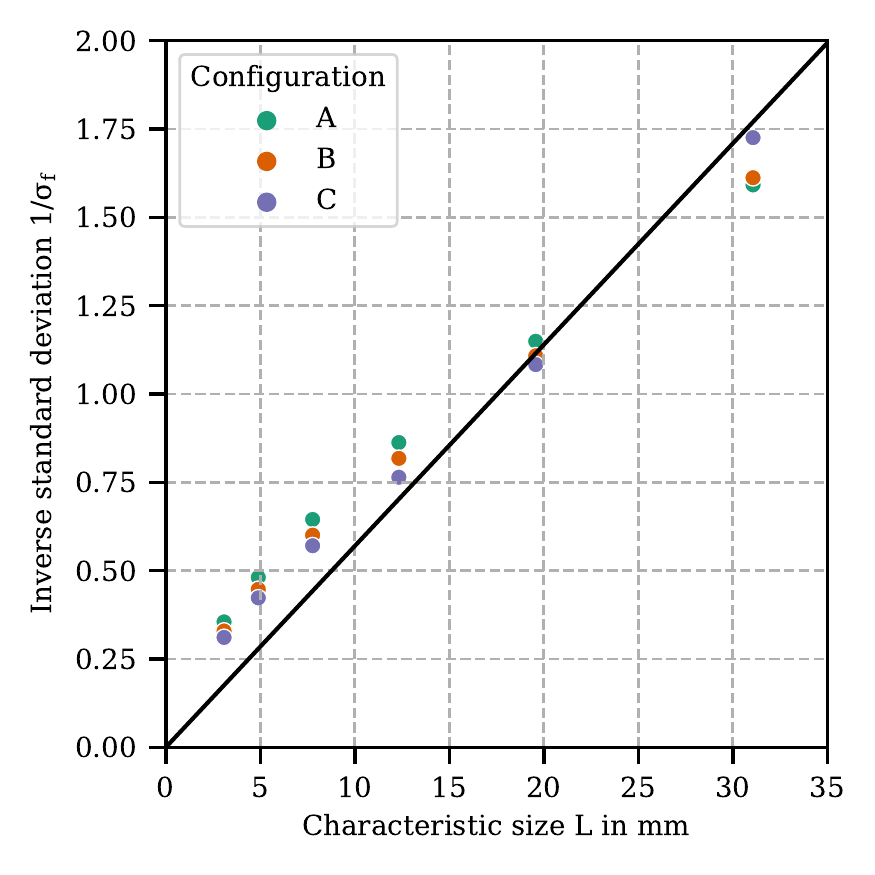}
		\caption{Inverse relation for $\sigma_f$}
	\end{subfigure}
	\begin{subfigure}[t]{7.5cm}
		\includegraphics[width=\linewidth]{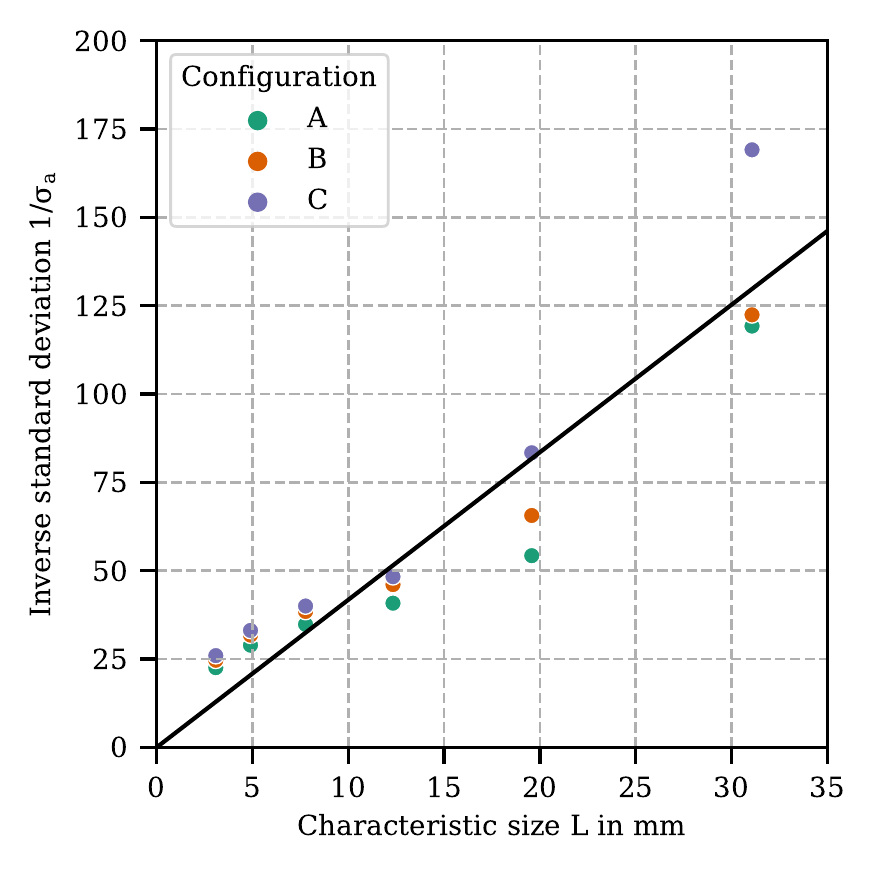}
		\caption{Inverse relation for $\sigma_a$}
	\end{subfigure}
	\caption{The scatter is approximately inversely proportional to the characteristic specimen size}
	\label{fig:lengthscale}
\end{figure}

The approximations in equation \eqref{eq:size_approximation} are plotted in Figure  \ref{fig:lengthscale} in black for comparison to evaluations of the standard deviation between different subsets of the plate for varying subset size.

\clearpage
\section{Parameters}



\begin{table}[!htbp]
    \centering
    \footnotesize
    \begin{tabular}{llr}
    \toprule
    Property                & Symbol            & Value                \\
    \midrule
    \refviscositydoc        & \refviscosity     & \SI{72}{\kilo\pascal\second}    \\
    \transsheardoc          & \transshear       & \SI{0.1}{\per\second}           \\
    \visccoefdoc            & \visccoef         & \SI{0.385}{}                           \\
    \visctdoc               & \visct            & \SI{40.73}{\celsius}            \\
    \viscadoc               & \visca            & \SI{7.94}{}                            \\
    \viscbdoc               & \viscb            & \SI{105.96}{\celsius}           \\
    \densitydoc             & \density          & \SI{1480}{\kilogram\per\meter\cubed} \\
    \conductivitydoc        & \conductivity     & \SI{0.163}{\watt\per\meter\per\kelvin} \\
    \conductancedoc         & \conductance      & \SI{403}{\watt\per\meter\squared\per\kelvin} \\
    \heatcapacitydoc        & \heatcapacity     & \SI{1530}{\joule\per\kilo\gram\per\kelvin} \\
    \refvelocitydoc         & \refvelocity      & \SI{1}{\milli\meter\per\second} \\
    \hydropowerdoc          & \hydropower       & \SI{0.6}{} \\
    \hydrofricdoc           & \hydrofric        & \SI{3.0}{\mega\newton\second\per\meter\cubed} \\
    Mass scaling factor     &                   & \SI{3e5}{} \\
    \abundledoc             & \abundle          & \SI{0.03}{\milli\meter\squared} \\ 
    \maxforcedoc            & \maxforce         & \SI{6}{\mega\newton} \\
    \bottomrule
    \end{tabular}
    \caption{Parameters for the direct bundle simulation}
    \label{tab:dbs_parameters}
\end{table}



\begin{table}[!htbp]
    \centering
    \footnotesize
    \begin{tabular}{lccc}
        \toprule
        & Young's modulus in \SI{}{\giga\pascal} & Poisson's ratio & Shear modulus in \SI{}{\giga\pascal} \\
        \midrule
        Fibers~\cite{Trauth2020Diss} & \f{E_{F}=72.00} & \f{\nu_{F\phantom{0}}=0.220} & \f{G_{F\phantom{0}}=29.51} \\
        Matrix~\cite{Trauth2020Diss} & \f{E_{M}=\phantom{0}3.45} & \f{\nu_{M\phantom{0}}=0.385} & \f{G_{M\phantom{0}}=\phantom{0}1.25} \\
        \multirow{2}{*}{Bundles} & \f{E_{L}=51.48} & \f{\nu_{TT}=0.402} & \f{G_{TT}=\phantom{0}6.63} \\
         & \f{E_{T}=18.66} & \f{\nu_{LT}=0.260} & \f{G_{LT}=\phantom{0}6.82} \\
        \bottomrule
    \end{tabular}
\caption{Elastic properties of fibers, matrix and bundles}
\label{tab:elastic:properties}
\end{table}

\begin{table}[!ht]
    \centering
    \footnotesize
    \begin{tabular}{lcccc}
        \toprule
         & Extraction tensor & \stressdamageinit in \SI{}{\mega\pascal} & \hardparam in \SI{}{\mega\pascal} & \powerexponent \\
        \midrule
        Matrix & \f{\extract_M} Eq.~\eqref{eq:extraction:matrix} & \f{36.88} & \f{213.92} & \f{1.0} \\
        \multirow{2}{*}{Bundles} & \f{\extract_{B,N}} Eq.~\eqref{eq:extraction:bundle:normal} & \f{46.03} & \f{529.00} & \f{1.0} \\
        & \f{\extract_{B,S}} Eq.~\eqref{eq:extraction:bundle:shear} & \f{44.08} & \f{283.92} & \f{1.0} \\
        \bottomrule
    \end{tabular}
\caption{Damage parameters for matrix and bundles}
\label{tab:damage:parameters}
\end{table}

\clearpage
\section{Additional plots}
\begin{figure}[!htpb]
	\centering
	\begin{subfigure}[t]{7.5cm}
		\includegraphics[width=\linewidth]{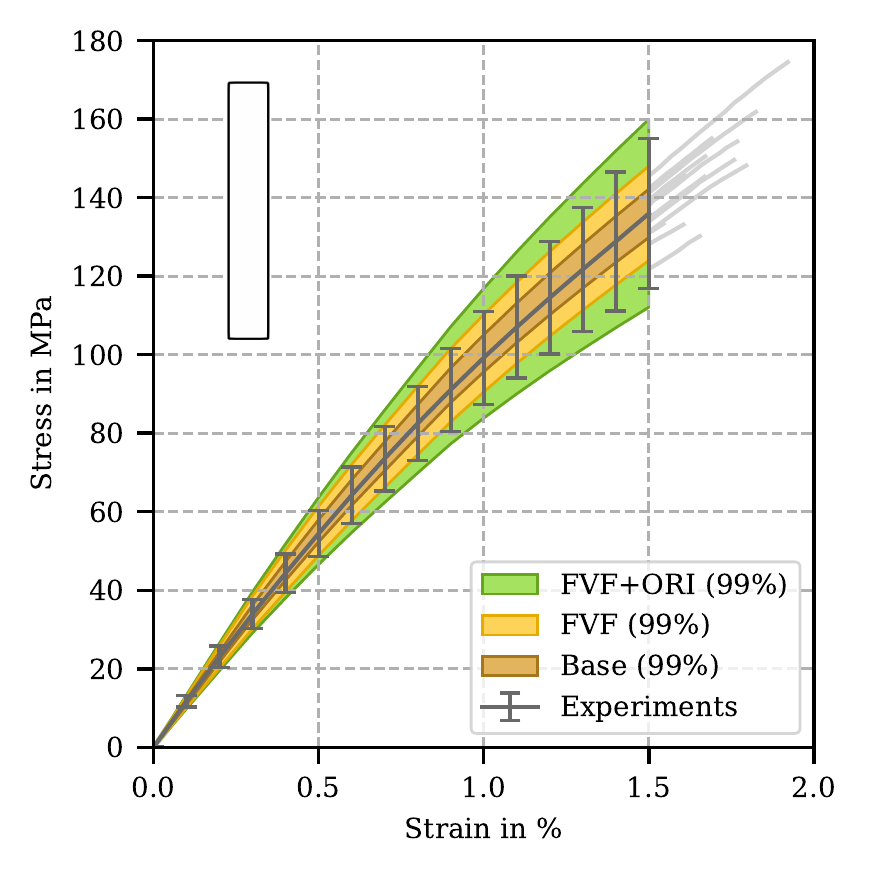}
	\end{subfigure}
	\begin{subfigure}[t]{7.5cm}
		\includegraphics[width=\linewidth]{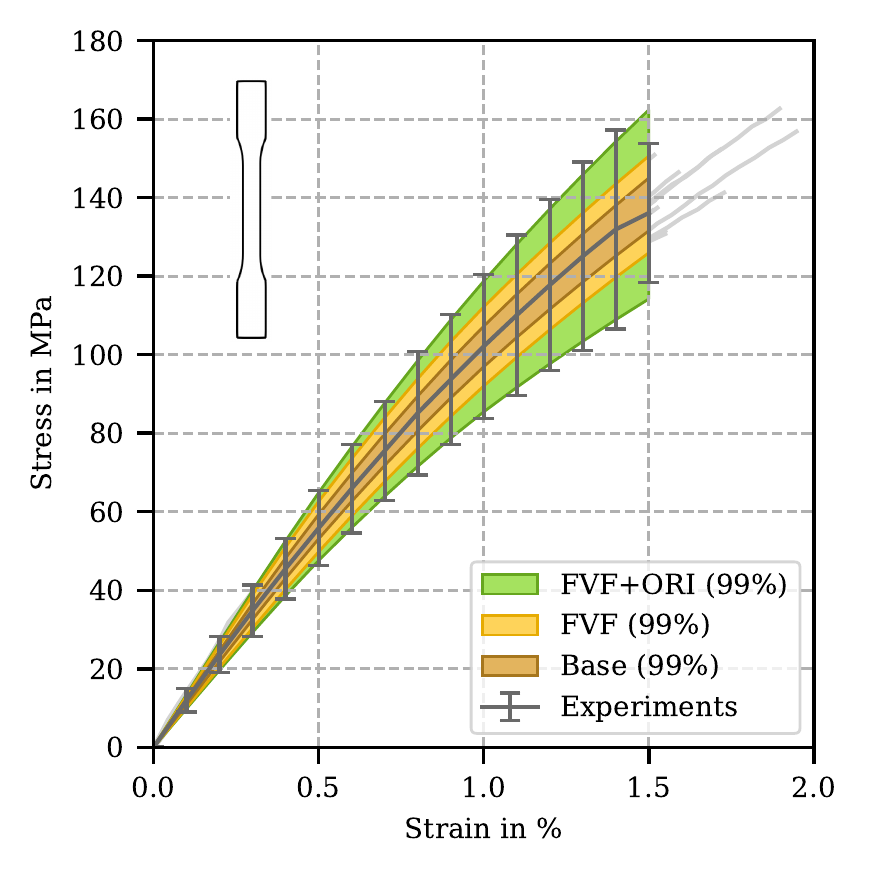}
	\end{subfigure}
	\caption{Stress-strain relations for other specimen types}
	\label{fig:stress_strain_appendix}
\end{figure}

In addition to the plots for specimens R1 and B2 in Figure~\ref{fig:stress_strain}, results for specimens R2 and B1 are given in Figure~\ref{fig:stress_strain_appendix} for completeness.

\clearpage
\bibliographystyle{bib_styles/elsarticle-num} 
\bibliography{literature}

\end{document}